# Review of Transition-Metal Diboride Thin Films

Martin Magnuson*, Lars Hultman, and Hans Högberg

Thin Film Physics Division, Department of Physics, Chemistry, and Biology (IFM)
Linköping University, SE-581 83 Linköping, Sweden.

* Electronic mail: martin.magnuson@liu.se
2021-10-08

## Abstract

We review the thin film growth, chemistry, and physical properties of Group 4-6 transition-metal diboride ($TMB_2$) thin films with $AlB_2$-type crystal structure (Strukturbericht designation C32). Industrial applications are growing rapidly as $TMB_2$ begin competing with conventional *refractory* ceramics like carbides and nitrides, including pseudo-binaries such as $Ti_{1-x}Al_xN$. The $TMB_2$ crystal structure comprises graphite-like honeycombed atomic sheets of B interleaved by hexagonal close-packed TM layers. From the C32 crystal structure stems unique properties including high melting point, hardness, and corrosion resistance, yet limited oxidation resistance, combined with high electrical conductivity. We correlate the underlying chemical bonding, orbital overlap, and electronic structure to the mechanical properties, resistivity, and high-temperature properties unique to this class of materials. The review highlights the importance of avoiding contamination elements (like oxygen) and boron segregation on both the target and substrate sides during sputter deposition, for better-defined properties, regardless of the boride system investigated. This is a consequence of the strong tendency for B to segregate to $TMB_2$ grain boundaries for boron-rich compositions of the growth flux. It is judged that sputter deposition of $TMB_2$ films is at a tipping point towards a multitude of applications for $TMB_2$ not solely as bulk materials, but also as protective coatings and electrically conducting high-temperature stable thin films.

# 1 Background

## 1.1 Introduction

Due to their technological importance, refractory and electrically conductive diborides ($TMB_2$) formed by Group 4-6 transition metals (TM): Ti, Zr, Hf, V, Nb, Ta, Cr, Mo, and W are attracting increasing research interest. This critical review is focused on Group 4-6 $TMB_2$ thin films with a predominant $AlB_2$-type crystal structure (Strukturbericht designation C32). A historical survey gives that the synthesis and characterization work on borides date back to the late 19$^{th}$ century, which resulted in the materials class known as $TMB_2$. The family of boron-containing compounds, thin films and bulk, is larger than $TMB_2$ as boron interacts with many of the elements in the periodic table to form a variety of compounds with different properties. Examples are monoborides and covalently bonded borides with high hardness, like aluminum magnesium boride (BAM) with the composition $AlMgB_{14}$, and superconducting $MgB_2$.

In section 2, the atomic sizes and the electronegativities of B and the group 4-6 TM are applied to predict the stability and chemical bonding in the C32 crystal structure to describe how the arrangement of atoms affects the property envelope shown by $TMB_2$. From binary phase diagrams, the stabilities of the Group 4-6 $TMB_2$ are considered in comparison to other competing B-containing phases of different stoichiometries. To understand properties, the underlying electronic structure, chemical bonding, and band structures are evaluated in section 3.

The progress of thin film growth of $TMB_2$ is discussed in section 4 with a focus on sputtering, as a commonly employed physical vapor deposition (PVD) technique. Other PVD techniques, such as cathodic arc-deposition, pulsed laser deposition (PLD), and e-beam evaporation as well as chemical vapor deposition (CVD) processing of $TMB_2$ films, are presented for reference. A survey of the field reveals that $TMB_2$ has been much less studied as thin film materials compared to transition metal nitrides (TMN) and transition metal carbides (TMC). Ternary and pseudobinary boride systems are also discussed here.

In section 5, we outline difficulties in controlling the level of contaminants, B/TM ratio, and microstructure in $TMB_2$ thin films that compromise their hardness, resistivity, and high-temperature stability properties. Deviations from stoichiometry and chemical purity is presented as the culprit here, and this review shows that the typical sputter-deposited Groups 4-6 $TMB_2$ film exhibits B-rich composition when grown from compound targets. This is due to mass differences between boron and the Group 4-6 TM resulting in different scattering angles when transported from the target to the substrate in the gas phase. In addition, a characteristic fine-grained microstructure is frequently reported. It too, can be explained by the strong driving force of B segregation towards grain boundaries as well as its strong reactivity towards oxygen and hydrogen, as common elements in the process ambient.

Section 6 summarizes the reviewed research studies. To enable further advancement of stoichiometric sputter deposited $TMB_2$ films, and place $TMB_2$ thin films in the position as the next generation of hard refractory and conductive functional ceramic coatings, a higher level of B-to-TM ratio control is needed. This is particularly the case for sputtering conditions pertaining to differences in mass of the metal and B species and resulting anisotropic angular distribution of their flux from the target. From the conducted research studies, we highlight the importance of retaining a strict 2B:TM composition for the growth flux in sputter-deposition of TMB$_2$ films with well-defined properties. In section 7, we suggest future research approaches with purer and more dense compound targets, HiPIMS and hybrid PVD/CVD techniques.

## 1.2   Historical exposé

The earliest work on bulk synthesis of borides depended on isolating the element boron, which was first reported in 1808 by Gay-Lussac and Thénard [1] and later Davy in 1809 [2]. The efforts were promoted by the development of the electric furnace of the arc type by Siemens in 1878-79 see *e.g.* [3], and later seen by an operational furnace by Readman in 1888. In 1894, Moissan [4] applied an electric furnace of the arc type, equipped with C cathodes to study the reactions of the TM chromium, including that to B. Moissan [4] found that the reaction product was hard and resisted acids. In two publications 1901 and 1902, Tucker and Moody [5] [6] followed in the steps of Moissan and synthesized borides from Zr, Cr, W, and Mo, using currents of 200-275 A and voltage 60-75 V between the carbon cathodes. Tucker and Moody concluded that the reaction products were crystalline, hard, of a high specific gravity, not easily attacked by acids, and having very high melting points. The synthesized compounds were determined to have the formulæ Zr$_3$B$_4$, CrB, WB$_2$, and Mo$_3$B$_4$. In addition, their studies concluded that copper and bismuth exhibit no affinity to boron [5]. In a publication from 1906, Binet du Jassoneix [7] continued by blending MoO$_2$ with amorphous B in stoichiometric amounts and reducing the oxide by H$_2$. He found that compounds containing ~20 % B were easily synthesized, while higher boron contents up to 45.6 % required higher temperatures, but at the expense of BC$_x$ precipitations in the synthesized material as well as formation of boric acid. The properties of the synthesized compounds were shown to be highly dependent on the B content judged from hardness, scratching topaz, but not corundum (Al$_2$O$_3$) and with corrosion resistance for HF and HCl, but not sulfuric acid. The visible appearance of the synthesized product containing more than 20% B was as described by Binet du Jassoneix [7]; bluish grey color, less metallic, but with no visible grains (microstructural characterization by electron microscopy was not available at the time).

Wedekind developed bulk synthesis using a vacuum oven and the parent TM, Mn, and Cr as cathodes to strike the arc and initiate boride formation with Mn [8] and Cr [9]. In a publication from 1913, [10] he summarized the work for more TMs, like Zr and described zirconium boride to be silver colored with a metallic luster and being temperature stable, but without presenting crystallographic data. The described attributes are typical for zirconium diboride, thus indicating successful synthesis of the compound. In 1936, McKenna [11] advanced the field by demonstrating a synthesis route for bulk ZrB$_2$ from heating zirconium-oxide and C with an excess of boron oxide at about 2000 °C. The resulting material consisted (in percent) of Zr 78.55 %, B 18.15 %, C 1.89%, and Si 0.03%, in total 98.62% according to chemical analysis and with a density of 5.64 g·cm$^{-3}$: the density determined for bulk ZrB$_2$ is 6.104 g·cm$^{-3}$ [12]. Characterization by x-ray diffraction (XRD) revealed a material with a hexagonal crystal structure and $c$ = 3.53 Å, $a$ = 3.15Å with a $c/a$ = 1.12. For ZrB$_2$, $c$ has later been determined to 3.53002(10) Å and $a$ = 3.16870(8) Å, which gives $c/a$ = 1.1140250 [12]. In addition, McKenna attempted to synthesize borides of tantalum and columbium (Cb), later named niobium (Nb), but with limited description of the synthesis products. Although McKenna concluded from XRD that ZrB$_2$ belongs to the family of materials with hexagonal crystal structure, no information was provided about the symmetry of the structure [11]. In 1936, Hoffmann and Jäniche [13] determined the AlB$_2$ type structure, often referred to by the Strukturbericht designation C32 or simply as α-type. This is the predominant crystal structure adopted by the TMB$_2$ formed by the Group 4-6 *d*-block elements to be discussed as thin film materials in this review. Consequently, in 1947, Ehrlich [14] found that TiB$_2$ crystallizes in the C32 structure and where Kiessling [15] in 1949 showed that $ZrB_2$ adopts the C32 structure. In a publication from 1949, Norton *et al.* [16] confirmed the crystal structure determination by Ehrlich for $TiB_2$ [14] and Kiessling [15] for $ZrB_2$ as well as expanded to the $TMB_2$ formed by the Group 5 TM, Nb, previously called Cb, Ta, and V, and determined them to be of the C32 structure. In his work from 1950, Kiessling [17] showed that the Group 6 transition-metal Cr form $CrB_2$ with C32 structure while Kiessling's earlier work from 1947 showed that Mo and W predominantly form borides of other compositions and crystal structures [18]. The most well-known composition is $TM_2B_5$ [17], often referred to as the ω-phase. By 1953, Glaser *et al.* had determined that the remaining Group 4 TM, Hf forms a $HfB_2$ with C32 crystal structure [19].

Improved synthesis methods for bulk $TMB_2$ and determination of its crystal structure made it meaningful to investigate the properties of this class of materials. Hence in 1954, Post *et al.* reported melting points for both $Mo_2B_5$ and $W_2B_5$ and their solid solutions [20]. During the 1950 and 1960s, other researchers focused on the Hall effect and the electrical conductivity of Group 4-6 $TMB_2$ [21], elastic constants of $TiB_2$ [22], hardness as a function of temperature [23] for $TiB_2$, $ZrB_2$, $HfB_2$, and $W_2B_5$, and strength, fracture mode and thermal stress resistance of $HfB_2$ and $ZrB_2$ [24]. The properties of bulk $TMB_2$ have been comprehensively reviewed by Fahrenholtz *et al.* [25].

In the 1970s, theoretical work on $TMB_2$ was initiated seen from band-structure calculations on $CrB_2$ [26] [27], $TiB_2$ [28], and $ZrB_2$ [29]. From calculations on $CrB_2$, Liu *et al.* [27] reported difficulties in determining the density of states at the Fermi level ($E_F$), which was explained by spin fluctuations. For $TiB_2$ and $ZrB_2$, the above studies found difficulties in deciding whether a charge transfer from the TM to B occurs. Ihara *et al.* [29] considered the band structure of $ZrB_2$ as a hybrid structure similar to that of graphite and zirconium metal. These early studies were pursued with insufficient accuracy in calculation of the exchange correlation and hybridization of the orbital overlaps and interstitial regions in the parent metal at the time [30]. Today's sophisticated density functional methods with higher numerical accuracy in the potentials including more atoms and shells are better suited to determine and predict properties, including magnetism and charge-transfer in $TMB_2$. DFT modelling can be utilized as a trend-giver in target-oriented experimental work. For example, Moraes *et al.* [30] used semi-automated DFT calculations across transition metal diborides and showed that point defects such as vacancies influence the phase stability that can even reverse the preference for the $AlB_2$ or $W_2B_{5-x}$ structure. Recently, they also found that $V_xW_{1-x}B_2$ exhibits ~40 GPa hardness that can be useful in demanding applications [31]. Alling *et al.* [32] observed that metastable $Al_{1-x}Ti_xB_2$ alloy could be of interest for coherent isostructural decomposition (age hardening) in thin films applications with a strong driving force for phase separation. Euchner and Mayhofer [33] also found that the ternary diboride alloys $Al_xW_{1-x}B_2$, $Ti_xW_{1-x}B_2$ and $V_xW_{1-x}B_2$ represent a new class of metastable materials that may open a large field for further investigations.

The desire to obtain materials with better defined properties resulted in the development of techniques for thin film growth. Of importance for the $TMB_2$ thin films in this review is PVD and CVD. The development of PVD began in the 1850s with the pioneering work of Groove [34] and later the more practical applications for sputter-deposited single and multi-layer metal films used as mirrors and optical coatings on telescope lenses and eyepieces were discussed in papers published in 1877 by Wright. [35] [36] Work with CVD was initiated already in the 1600s [37], but with more controlled processes developed in the late 1800s and in the early 1900s for C by Sawyer and Man [38], carbon monoxide on Ni by Mond, Langree and Quincke [39], electrical incandescing conductors by Aylesworth [40], and for Ti in 1910 by Hunter [41].

For TMB$_2$ thin films, Moers [42] developed halide-CVD in the 1930s, while sputtering was reported in the 1970s by Wheeler and Brainard [43]. In the 1980s, as the sputtering techniques were introduced, the properties of the target material were improved. In 1997, Mitterer summarized works on thin film growth of the Group 4 TMB$_2$, TiB$_2$, and ZrB$_2$, including zirconium dodecaboride (ZrB$_{12}$) coatings [44], and borides formed by the lanthanides LaB$_6$, CeB$_6$, SmB$_6$, and YB$_6$ by different sputtering techniques. Composition, structural properties, and microstructure of the investigated borides were thus directly connected to their mechanical properties. Hexaborides are formed by configuration interaction of *4f* narrow-band states of heavy-fermion elements La and Ce, LaB$_6$ [45] and CeB$_6$ [46] with charge-transfer to B. These materials exhibit a variety of interesting properties, *e.g.*, high melting points, resistance to cathode poisoning and the lowest known work function ~2.5 eV, useful as electron emitters in electron microscopes, microwave tubes, electron lithography, electron beam welding, x-ray tubes, and free electron lasers [47]. In 2015, Andrievski [48] reviewed the Group 4 TMB$_2$ both as bulk and thin films, focusing on their synthesis and the resulting microstructure and mechanical properties.

Expanding the boride family from the TM, we note that in 1970, Matkovich and Economy [49] synthesized and determined the structure of a new class of ultrahard (> 40 GPa) Al-Mg borides referred to as BAM. In BAM, boron forms a three-dimensional (3D) network of four B$_{12}$ icosahedra in the unit cell that are stabilized by the electron-donating metals. Property determination followed the successful synthesis of BAM, where one outstanding property is the hardness with a reported micro-hardness in the range 32-35 GPa and with even higher hardness values of 40-46 GPa when alloyed with 30 mol% TiB$_2$ [50]. The promising results from bulk synthesis by Cook *et al.* [50], inspired thin film growth by sputtering, see, *e.g.*, [51] [52] for recent studies.

Boron carbide (BC$_x$) discovered already in the 19th century as a by-product of reactions involving metal borides is a very stable and oxidation-resistant compound that is one of the hardest ceramic materials after diamond and cubic boron nitride (c-BN). In the 1930s, the chemical composition of boron carbide was estimated as B$_4$C [53]. Later, x-ray crystallography showed that the structure of B$_4$C is carbon deficient and highly complex, with a mixture of C-B-C chains and B$_{12}$ icosahedra consisting of a combination of the B$_{12}$C$_3$ and B$_{13}$C$_2$ (B$_{12}$CBC) units [54]. The material is commonly used in body armor systems due to its low weight, in reactors as neutron absorber and in stainless-steel claddings. The different stoichiometries ranging from B$_4$C to B$_{10.5}$C correspond to a carbon content between 20 and 8.7 at.%. Typical properties of boron carbides are high hardness (third highest Vickers hardness of 38 GPa after diamond and c-BN), corrosion resistance, and high-temperature stability. Thus, covalent materials containing connected B$_{12}$ icosahedras are boron-carbides B$_{12}$C$_3$ [55] often referred to as B$_4$C [56]. The B$_{12}$C$_3$ structure consists of eight B$_{12}$ icosahedra located at the vertex of the rhombohedral unit cell and bonded together along the long diagonal by a chain of three atoms.

In 2001, the discovery of superconductivity in bulk MgB$_2$ [57] [58] [59] further increased the interest in borides [58]. It soon turned to TMB$_2$ with C32 crystal structure. In 2001, Kaczorowski *et al.* [60] found a critical temperature (T$_c$) of 9.5 K in a powder sample of TaB$_2$. Later in 2001, V. A. Gasparov *et al.* [61] investigated polycrystalline ZrB$_2$, NbB$_2$, and TaB$_2$ pellets with over 60-90% of the theoretical mass density. From measurements of these materials, they determined T$_c$ to be 5.5 K for ZrB$_2$, while NbB$_2$ and TaB$_2$ were found to be superconducting only up to 0.37 K. Although potential superconducting materials, albeit at low T$_c$, it seems that the crystalline quality properties of the investigated TMB$_2$ bulk materials determines the measured T$_c$, which was supported from an earlier work published 1979 by Leyarovska and Leyarovski [62] that showed no superconductivity above 0.42 K when conducting measurements of pressed tablets of TMB$_2$, where TM=Ti, Zr, Hf, V, Nb, Ta, Cr, and Mo. Thus, it is apparent that to conclusively determine superconductivity in TMB$_2$ requires synthesized material of high crystal quality, suggesting thin film approaches.

Furthermore, the borides formed by Group 7 and 8 TM: Re, Os, and Ru have received considerable attention due to the high hardness reported for OsB$_2$ in 2005, [63] where a bulk piece of OsB$_2$ was found to readily scratch a polished sapphire window. The interest in these TMB$_2$ phases originates from the properties of their parent TM, where for instance Os, like diamond, shows a bulk modulus of 395-462 GPa and high valence electron density of 0.572 electrons/Å$^3$ [63]. Following the promising results with OsB$_2$ as an ultra-incompressible hard material, the attention was directed to ReB$_2$ in 2007 as Re share the previously described properties of Os with bulk modulus of 360 GPa and valence electron density of 0.4761 electrons/Å$^3$ [64]. For a ReB$_2$ pellet pressed from Re and B powders then arc-melted to form a solid metallic ingot, Chung *et al.* measured a Vickers hardness (HV) of 55.5 GPa at an applied load of 0.49 N [64]. Although promising boride materials, it should be stressed that the results on *ultrahigh hardness* in OsB$_2$ and ReB$_2$ have been debated following a technical comment by Dubrovinskaia *et al.* in 2007 [65] on the data evaluation from the mechanical testing of the ReB$_2$ sample as well as from a study by Qin *et al.* [66] in 2008 who measured a HV of only about 20 GPa for a ReB$_2$ compact synthesized at 5 GPa and 1600 °C. Thus, OsB$_2$ and ReB$_2$ need to be confirmed as superhard thin film materials.

## 2 Transition-metal diborides

### 2.1 Atomic size effects

Figure 1 shows the periodic table in which the 2$^{nd}$ row *p*-block elements B, C, and N are highlighted in yellow and the Groups 4-6 TM: Ti, Zr, Hf, V, Nb, Ta, Cr, Mo, W in light blue. The atomic radii for each element are listed above the International Union of Pure and Applied Chemistry (IUPAC) symbol determined for each TM as well as for the *p*-block elements. Atomic sizes are important when considering crystal structures from a *close-packing* model, *i.e.* packing together equal-sized spheres in 3D in a hexagonal closest packing (*hcp*) arrangement or in a cubic closest packing (*ccp*) arrangement. In the model the size of the atoms that form the basis determines the size of the interstitials that will be present between the atoms in a hcp or a ccp arrangement. Consequently, the size of the interstitials will also be decisive to what element, atomic size, that can be positioned in the hcp or ccp arrangement.

In TMB$_2$, the TM atoms will form close-packed layers, where B occupies trigonal prismatic sites, to be described in more detail in section 2.4. As observed from the trend in atomic size, the radii of the transition-metal atoms decrease from left to right in the periodic table and consequently the size of the interstitials present in the close packed structure. This is due to a more effective nuclear charge that contract the size of the electron cloud and hence the size of the atom.

Another general observation in Figure l is that the atomic radii increase when moving down columns corresponding to a given Group in the periodic table, but where Zr and Hf exhibit similar atomic radii due to what is referred to as the *lanthanide contraction* [67]. Consequently, Zr exhibits the largest size with a radius of l.60 Å, while Cr is the smallest atom with a radius of l.28 Å. When B is compared to C and N, Slater's rule for screening implies that the extension and localization of the *2p$_x$* orbitals increases with the filling by *x* = 1 to 2 and 3 electrons, respectively. However, by considering the increasing effective nuclear charge with increasing atomic number, it is evident that B is the largest atom of the three elements, exhibiting an atomic radius of 0. 88 Å compared to C (radius 0.77 Å) and N (radius 0.70 Å) [67].

**Figure 1:** Periodic table of the elements highlighting the second-row *p*-block elements and Groups 4-6 transition metals forming diborides with C32 hexagonal AlB$_2$ structure. The atomic radii and electronegativity values are from refs. [291], [72], [292], [67].

In the 1930s, Gunnar Hägg showed that hydrides, borides, carbides, nitrides, and oxides could be described in terms of metal atoms spherically packed with smaller H, B, C, N, and O atoms in octahedral or tetrahedral interstices [68]. From x-ray diffraction and considering the atomic radius ratio $r_X/r_{TM}$ between non-metals (X) and TM, Hägg empirically determined a trend in the complexity of hydride, carbide, nitride, and oxide crystal structures. According to Hägg's stability rule from 1931 [68], the structure of a non-metal/metal compound will be either *simple* or *more complex* depending on whether the atomic radii ratio is less than, or greater than 0.59. If $r_X/r_{TM}$ < 0.59, the phase will be *simple*. This rule applies to interstitial structures of carbides, nitrides, and borides formed by the transition metals of Group 4-6 in the periodic table. Thus, Häggs' rule depends on the possibility of determining atomic sizes. Here, we note that Hägg [68] applied a radius for B of 0.97 Å as derived from the phase Fe$_2$B to compare to the smaller radius of 0.88 Å that was later adopted by Hägg [67], while the radius of the C and N atoms were identical or close to our values, cf. 0.77 Å for C and 0.71 Å for N. Following in the path of Hägg, Kiessling [17] developed the stability criteria further to include more complex crystal structures among the borides.

Using Hägg's empirical rule as starting point and considering the C32 structure as *simple*, we apply the model to explain the stability of TMB$_2$. In the C32 structure, there are trigonal prismatic interstitials (voids), instead of octahedral interstitials that is common in other structures such as cubic NaCl type structure (Strukturbericht designation B1). The radius of the transition-metal atom in the C32 structure determines the size of the void for the interstitials. For $r_B/r_{TM}$ < 0.59, the model predicts a *simple* structure such as C32. However, for $r_B/r_{TM}$ ⩾ 0.59, the model predicts *more complex* crystal structures, decreasing stability, and other compositions beyond that in $TMB_2$ with decreasing stability as discussed in section 2.3.

Applying the atomic radii from Figure 1, we find that for the *3d* TM:Ti, V and Cr, the $r_B/r_{TM}$ ratio ⩾ 0.59 for all TM and increases from 0.60, to 0.66, and 0.69, and the model predicts more complex structures favored for the V-B and Cr-B binary systems, see section 2.3. For the *4d* TM: Zr, Nb and Mo, the $r_B/r_{TM}$ ratio increases from 0.55, to 0.60, and further to 0.63 and the model predicts a simple structure for $ZrB_2$, and where Mo prefers a boride with $Mo_2B_5$ composition (as mentioned in the introduction). Finally, for the *5d* TMs Hf, Ta, and W, the $r_B/r_{TM}$ ratio increases from 0.56, to 0.60, and 0.63 and the model predicts a simple structure for $HfB_2$ and where W similar to Mo forms a boride with $W_2B_5$ composition, but with a different stacking sequence compared to $Mo_2B_5$ [17]. For the stabilization of metastable structures such as $WB_2$ or $MoB_2$, Moraes *et al.* [69] [31] showed that defect engineering in terms of boron and metal vacancies is also important.

From above, the Hägg's rule predicts that $ZrB_2$ is the most stable C32 $TMB_2$ followed by $HfB_2$, while $TiB_2$ is the least stable phase in Group 4. [25] [70] $TiB_2$, $TaB_2$ and $NbB_2$ are close to the border of fulfilling Hägg stability, but less stable phases compared to $ZrB_2$ and $HfB_2$. Furthermore, the determined $r_B/r_{TM}$ ratio from Fig. 1 are identical to those presented by Kiessling [17] for Cr, Nb (Cb), Mo, and Ta, but with 0.59 for Ti, 0.54 for Zr, 0.65 for V, and 0.62 for W; while Kiessling provided no data for Hf.

When applying Hägg's rule [68] to TMC and TMN, the smaller radii of C and N compared to B will increase the number of phases with a simple B1 structure for both materials systems. For TMC:s Cr alone exhibits a $r_C/r_{TM}$ ratio ⩾ 0.59 (0.60), while all TMN show $r_N/r_{TM}$ < 0.59. Thus, from Hägg's rule transition metal borides appear to have less preference in forming a simple crystal structure as C32 when compared to TMC and TMN.

## 2.2 Electronegativity

Borides are, by definition, formed by B and an electropositive element *i.e.*, most metals. Thus, in a boride the B acts as electron acceptor and the metal acts as electron donor. As an estimate on the degree of electron transfer, the *character* of the chemical bonding as metallic, covalent, or ionic, we consider the difference in electronegativity value ($\chi$) between the electron acceptor B and the electron donating metal. A large difference in electronegative value ($\Delta\chi$) is synonymous with a high degree of electron transfer, *i.e.* ionic character in the bonding. For $\chi$ using the Pauling's scale, there is maximum $\Delta\chi$ of 3.19 in the periodic table for CsF, as the binary compound is formed from the most electronegative element in the periodic table, F with $\chi$ = 3.98 and the most electropositive and stable element, Cs with $\chi$ = 0.79. This $\Delta\chi$ value is said to correspond to a bonding with 100% ionic character and where smaller $\Delta\chi$ implies mixtures between ionic, covalent, and metallic bonding [71]. Here, we stress that this approach is a rough estimate of the chemical bonding in $TMB_2$ as the C32 crystal structure makes it necessary to not solely consider B-TM bonds, but in addition B-B bonding and TM-TM bonding, see section 2.3-2.4. However, as electron transfer is the foundation for compound formation, we apply $\Delta\chi$ to consider the stabilities of the $TMB_2$ formed by the Group 4-6 TM.

Table I lists the difference in electronegativity between B and the Group 4-6 TM, using the values presented in Figure 1 below the IUPAC symbol of each element. As can be seen, B-Hf followed by B-Zr have the largest difference in $\Delta\chi$ with 0.74 and 0.71, respectively. This is far from a bonding with ionic character, and rather corresponding to a mixture of metallic and covalent bonding with an estimated ionic character of less than 20% [71]. It is a consequence of Hf and Zr being the most electropositive TM in Group 4-6. For the other TM, $\Delta\chi$ decreases seen from B-Ta, B-Ti, B-Nb, B-V, and B-Cr. Particularly, the high $\chi$ values for Mo with $\chi = 2.20$ and W with $\chi = 2.36$ that are higher than that of B with $\chi = 2.04$ [72]. By considering the degree of electron transfer from the TM to B as an indicator of the stability of the $TMB_2$, note that $HfB_2$ and $ZrB_2$ should be the two most stable phases, while $MoB_2$ and $WB_2$ should be less stable due to their weaker B-TM bond. This is supported from the discussion above of Häggs' rule and the fact that the stronger TM-TM interaction in $TMB_2$ formed by Mo and W results in other crystal structures than C32. In section 2.4, we report difficulties in sputtering compounds like $MoB_2$ and $WB_2$ as thin films due to their lower stability than competing phases.

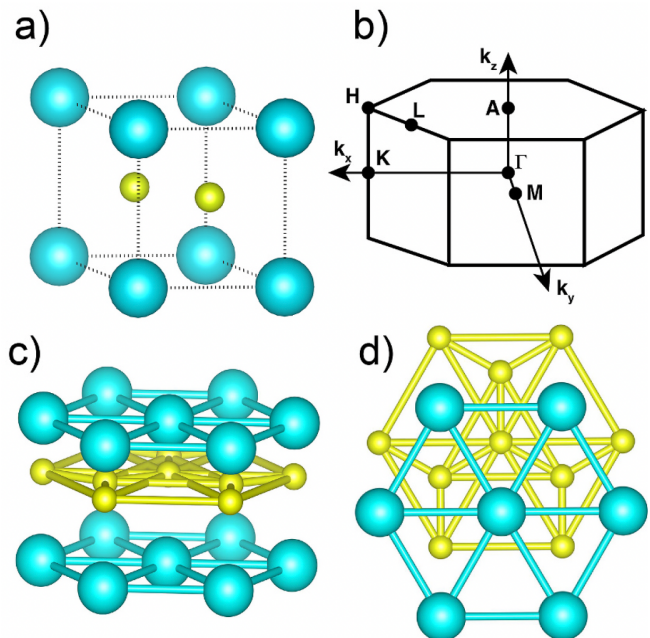

**Figure 2:** The C32 hexagonal $AlB_2$ crystal structure: (a) unit cell, (b) reciprocal space projection for the P6/mmm group that shows the major symmetry directions, (c) schematic of chemical bonds viewed from the side, and (d) from the top.

**Table I:** Electronegativity values for B and the Group 4-6 TM [72].

| $\chi$(B) | $\chi$(TM) | $\Delta\chi$ |
|---|---|---|
| 2.04 | Ti =1.54 | 0.5 |
| 2.04 | Zr=1.33 | 0.71 |
| 2.04 | Hf =1.30 | 0.74 |
| 2.04 | V=1.63 | 0.41 |
| 2.04 | Nb=1.60 | 0.44 |
| 2.04 | Ta=1.50 | 0.54 |
| 2.04 | Cr=1.66 | 0.38 |
| 2.04 | Mo=2.20 | -0.16 |
| 2.04 | W=2.36 | -0.32 |

The lower $\chi$ of B compared to C with $\chi = 2.54$ and N with $\chi = 3.04$ will result in different mixtures between ionic, metallic, and covalent bonding. Comparing Group 4-6 $TMB_2$ to TMC, and TMN, $TMB_2$ shows more metallic character, TMC bonds are of more covalent character, while TMN has the highest degree of ionic contribution [71]. Section 5, compares the hardness and resistivity of $TMB_2$ to those of carbides and nitrides.

## 2.3   The AlB$_2$ crystal structure

From our previous discussion, it is clear that all Group 4-6 TM form TMB$_2$ phases with C32 structure and where ZrB$_2$ and HfB$_2$ seems to be the two most stable phases considering Hägg's rule ($r_B/r_{TM}$ ratio) and difference in electronegative value ($\Delta\chi$) while particularly Mo and W prefer to form borides of other compositions such as TM$_2$B$_5$ and crystal structures. In this section, we describe the C32 structure in more detail as it holds the key to the properties demonstrated by the TMB$_2$, which are characterized by high hardness [25]; high melting points; good electrical and thermal conductivities, corrosion, and erosion resistance [48]; high chemical inertness and stability; as well as high wear and thermal-shock resistance [25].

Figure 2a shows the unit cell of the C32 crystal structure, with metal atoms in the basis (0,0,0) and boron atoms positioned in trigonal prismatic interstices at (1/3, 2/3, 1/2) and (2/3, 1/3, 1/2) belonging to the hexagonal crystal class and space group 191, in Hermann–Mauguin notation P6/mmm. Figure 2b shows the projection in reciprocal space in hexagonal symmetry with the major symmetry directions indicated. The symmetry points Γ, M, A, K, L, and H in the band structure give rise to prominent peak structures in the density of states (DOS) discussed in section 3. In Figures 2(c) and 2(d), the unit cell has been translated to reveal the symmetry of the layered structure characteristic to the C32 crystal structure. As can be seen, it consists of honeycomb 2D graphite-like sheets of B-atoms (*borophene*) alternating with hexagonal close-packed TM-layers. The C32 structure is often described from a stacking sequence of AHAHAH….. [17], where A corresponds to the TM layers and the *borophene* sheets to H. Furthermore, each M-atom is surrounded by six equivalent M-neighbors in metal planes and by 12 equidistant B neighbors: six above and six below. Each B has three B neighbors in the basal plane and six M-atoms out-of-plane: three above and three below. The *borophene* sheets are characteristic host sites for vacancies as well as common impurities such as oxygen and carbon and to some extent nitrogen. Vacancies and impurities within the borophene sheets may influence the properties by weakening the direct B-B bonding and destabilizes the TMB$_2$ structure. This is different from the interstitial TMC and TMN compounds with predominantly B1 crystal structure that are characterized by large homogeneity ranges, [73] where the isolated C or N atom can be removed to form a vacancy as well as possible solid solutions, where the C or N can be replaced by for instance O [74].

**Table II:** Lattice parameters *a* and *b* from [293] for TiB$_2$, [12] for ZrB$_2$, [294] for HfB$_2$, [295] for VB$_2$, [296] for NbB$_2$, [297] for TaB$_2$, [298] for CrB$_2$, [299] for MoB$_2$, and [82] for WB$_2$. The B-TM and B-B bond lengths and *c/a* ratios were determined from the lattice parameters using the structure program Visualization for Electronic and Structural Analysis (VESTA).

| Boride | TM group | a (Å) | c (Å) | c/a | -B-TM (Å) | B-B (Å) |
|---|---|---|---|---|---|---|
| TiB$_2$ | 4 | 3.03034(8) | 3.22953(14) | 1.0657295 | 2.38085(0) | 1.74957(0) |
| ZrB$_2$ | 4 | 3.16870(8) | 3.53002(10) | 1.1140250 | 2.54208(0) | 1.82945(0) |
| HfB$_2$ | 4 | 3.14245(8) | 3.47602(10) | 1.1061472 | 2.51244(0) | 1.81430(0) |
| VB$_2$ | 5 | 2.99761(9) | 3.05620(12) | 1.0195429 | 2.30875(0) | 1.73068(0) |
| NbB$_2$ | 5 | 3.11133(13) | 3.2743(2) | 1.0523855 | 2.43045(0) | 1.79633(0) |
| TaB$_2$ | 5 | 3.09803(7) | 3.22660(12) | 1.0415001 | 2.40874(0) | 1.78865(0) |
| CrB$_2$ | 6 | 2.9730(13) | 3.0709(2) | 1.0329319 | 2.30301(0) | 1.71646(0) |
| MoB$_2$ | 6 | 3.04 | 3.07 | 1.0098684 | 2.33169(0) | 1.75514(0) |
| WB$_2$ | 6 | 3.02 | 3.05 | 1.00993377 | 2.31641(0) | 1.74360(0) |

Table II lists the lattice parameters and internal bond distances in the C32 structure for the TMB$_2$ formed by the Group 4-6 TM. As can be seen, the longest *a* axis is found in ZrB$_2$, followed by HfB$_2$, NbB$_2$, TaB$_2$, MoB$_2$, TiB$_2$, WB$_2$, VB$_2$, and finally CrB$_2$. By assuming that the length of the *a* axis in TMB$_2$ corresponds to the TM-TM distance, we note from the atomic radii in Figure 1 that TiB$_2$ deviates from this hypothesis as Nb and Ta have identical atomic radii as Ti 1.46 Å and where the Mo atom is slightly smaller than the Ti atom (radius of 1.39 Å). Post *et al.* [20] explained this from the B-B distances in the *borophene* sheets, where 4*d* and 5*d* TM slightly increase the B-B distances to achieve TM-TM contaction in the *a*-axis direction, while 3*d* TM retain a "normal" B-B distance of 1.75 Å (($r_B$ + $r_B$)/2) as in TiB$_2$ or even slightly decreasing the B-B separation as for VB$_2$ and CrB$_2$, see Table I, and the B-B separation in Group 4-6 TMB$_2$. Here, WB$_2$ deviates from the *a* axis trend postulated by Post *et al.* [20] and where the authors presented no data for WB$_2$. This can be explained by the stability of WB$_2$ with C32 crystal structure as discussed in section 2.4. For the length of the *c* axis, there is a different trend where the *c* axis of TiB$_2$ is slightly larger than that of TaB$_2$ and much larger than that of MoB$_2$, but where VB$_2$ and MoB$_2$ exhibit a shorter *c* axis than CrB$_2$. For the stable TMB$_2$ phases, the *c*/*a* ratios in Table I are in the range 1.0-1.1. Norton *et al.* [16] and Kiessling [17] noted the rather regular increase in the *c*/*a* ratio that scales with the size of the metal atom. Consequently, ZrB$_2$ exhibits the largest *c*/*a* ratio and where CrB$_2$ should exhibit the smallest *c*/*a* ratio given the smallest atomic radius of 1.28 Å. Post *et al.* [20] determined a similar trend for the *c*/*a* ratios in stable TMB$_2$. From the work by Post *et al.* [20] and Table II in this review, we note the deviation for MoB$_2$ and WB$_2$ that is probably due to difficulties in determining the lattice parameters for these TMB$_2$'s. The trends in the *a*-lattice parameter and the *c*/*a* ratio as well as the B-TM and the B-B bond lengths are further developed in section 3.1.

**Table III:** Crystal structures adopted of borides with CrB$_4$ type [86], Ta$_3$B$_4$ type [75], U$_3$Si$_2$ type [84], and CuAl$_2$ type [300].

| Composition (TM$_x$B$_y$) | Crystal type structure | Crystal class | Comments |
|---|---|---|---|
| TMB$_{12}$ | UB$_{12}$ | Cubic | Borides with three-dimensional B frameworks (B$_{12}$ icosahedra) |
| TMB$_6$ | CaB$_6$ | Cubic | Borides with three-dimensional B frameworks (B$_6$ octahedra) |
| TMB$_4$ | CrB$_4$ | Orthorhombic | Borides with three-dimensional framework of tetrahedrally-coordinated B |
| TM$_2$B$_5$ | Mo$_2$B$_5$ | Trigonal | Boride with B atoms in hexagonal flat nets (H) or slightly puckered sheets (K): stacking sequence AHAKBHBKCHCKAHA |
| TM$_2$B$_5$ | W$_2$B$_5$ | Hexagonal | Boride with B atoms in hexagonal flat nets (H) or slightly puckered sheets (K): stacking sequence AHAKBHBKAHA |
| TMB$_2$ | AlB$_2$ (C32) | Hexagonal | Borides with two-dimensional hexagonal nets (H) of B atoms: stacking sequence AHAHAH |
| TM$_3$B$_4$ | Ta$_3$B$_4$ | Orthorhombic | Borides with double chain (fragments of nets) |
| TMB | CrB | Orthorhombic | Borides with boron zig-zag chains |
| TMB | FeB | Orthorhombic | Borides with boron zig-zag chains |
| TMB | MoB | Tetragonal | Borides with boron zig-zag chains |
| TM$_3$B$_2$ | U$_3$Si$_2$ | Tetragonal | Borides with B dumbbells |
| TM$_2$B | CuAl$_2$ | Tetragonal | Boride with isolated B atoms |

In addition to TMB$_2$ with C32 crystal structure, the Group 4-6 TM form borides of other compositions and crystal structures with the B atoms hosted as: 3D B$_{12}$ icosahedrons or B$_6$ octahedrons flat or puckered 2D nets or fragments of nets, 1D zig-zag chains, paired atoms or finally as individual isolated atoms depending on the stoichiometry of the phase. Table III lists the most prevalent compositions and crystal structures that are used to support the discussion on the binary phase diagrams in section 2.4. It should be emphasized that there are other more boron-rich borides than $TMB_2$ formed by the Group 4-6 TM such as $TMB_{12}$, $ZrB_{51}$ and $CrB_{41}$ as well as not listed compositions such as $V_5B_6$. For this extension of boride phases, the reader is referred to ref. [75].

For Group 4 to 6 borides, the crystal structure for borides with $TM_2B_5$ and TMB composition is worth commenting. The former composition is preferred by the TM:s Mo, and W, but differ in crystal class, since $Mo_2B_5$ is reported to be trigonal, while $W_2B_5$ is hexagonal. In his summary of crystal structures adopted by borides from 1950, Kiessling [17] described the stacking sequence for $Mo_2B_5$ and $W_2B_5$ following the terminology for hexagonal closest packing with ABABAB… and cubic closest packing with ABCABCABC… This results in a stacking sequence of AHAKBHBKCHCKAHA in $Mo_2B_5$ and AHAKBHBKAHA in $W_2B_5$ and where we previously defined the stacking sequence in $TMB_2$ with C32 crystal structure to be AHAHAHAH. Here, A, B, and C are close-packed Mo layers, where layers B and C are shifted by (a/3, 2a/3) and (2a/3, a/3), respectively, while H are planar graphite-like sheets and K are (buckled) boron rings. Thus, both the $Mo_2B_5$ and $W_2B_5$ crystal structures contain plane nets of B atoms (H) as in the C32 crystal structure, but where these nets are interleaved by puckered sheets of B atoms (K). In addition, some of the K sheets contain an additional B atom positioned in the center of the mesh. To further increase the complexity, the composition in $Mo_2B_5$ and $W_2B_5$ has been reported to concern solely a structure type and not the actual composition that is rather $MoB_{2-x}$ and $WB_{2-x}$ [76].

## 2.4 Monoborides

Similar to the $TMB_2$ materials with C32 crystal structure, all Group 4-6 form TM monoborides TMB, *albeit* with different crystal structures as reviewed by Minyaev and Hoffmann in 1991 [75]. Kiessling [17] identified at least three crystal type structures for TMB phases; CrB, FeB, and MoB, where CrB and FeB belong to the orthorhombic crystal class, while MoB is tetragonal. Common to these crystal structures are the B zig-zag chains, but where the angle between the boron atoms differs from 115-117° in CrB, to 110-112° in FeB and to 127° in MoB, [17] and with possible homogeneity ranges in the MoB type structure seen from 48.8 to 51.5 at.% B in MoB and 48.0 to 50.5 at.% B in WB [17]. Lately, the different structures of binary borides have been reviewed by, *e.g.*, Chen and Zou [77].

In the upcoming section on the binary phase diagrams for: B-Ti, B-Zr, B-Hf, B-V, B-Nb, B-Ta, B-Cr, B-Mo, and B-W, the homogeneity ranges will be discussed in $TMB_2$ with the C32 crystal structure as well as the phase order presented in Table III in connection to their respective phase diagrams.

To conclude, the group 4-6 TM borides exhibits similarities between the crystal structures determined for the same composition, although, differs in terms of symmetry, composition and homogeneity ranges. This motivates continued synthesis, exploration and theoretical work.

## 2.5 Phase diagrams and stability

Figure 3 shows binary phase diagrams for B-Zr, B-Nb, and B-Mo-B. As a general observation, the number of phases increases with increasing Group number in the periodic table [75].

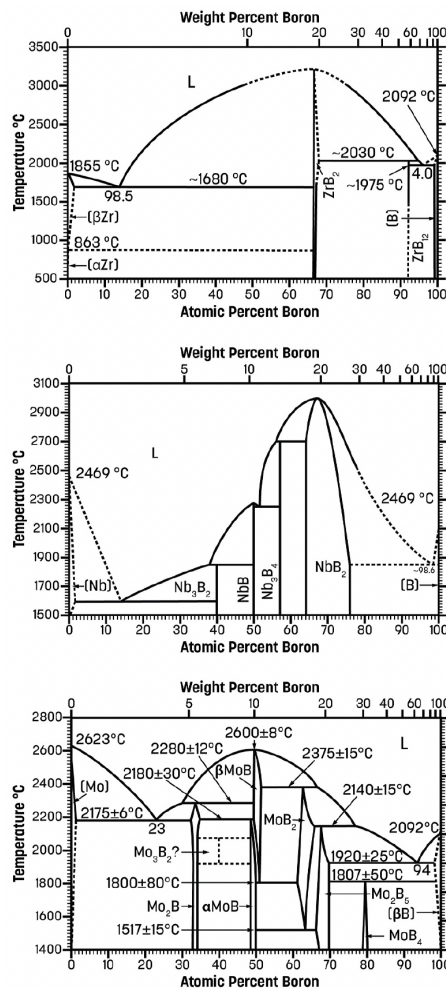

**Figure 3:** Phase diagrams of the binary systems Zr-B, Nb-B, and Mo-B. Reprinted from reference [78] with permission of AMS International.

Starting with Group 4 and the Zr-B system, it is clear from the phase diagram in Fig. 3 (top panel) that Zr forms a stable $TMB_2$ compound with C32 structure as adopted from Massalski *et al.* [78] Stable $TMB_2$ phases are also found in both the Ti-B and Hf-B systems. From the Zr-B phase diagram, the phase $ZrB_2$ is described as a line compound with a composition 33.33 at.% Zr and 66.67 at.% B. In a recent study, Engberg *et al* [79], discussed the possibly of a narrow homogeneity range in $ZrB_2$ films deposited by direct current magnetron sputtering (DCMS) investigated by atom probe tomography (ATP) in combination with cross-sectional TEM and Z-contrast STEM data, where grain stoichiometry measurements show up to 4 at.% deviations from stoichiometry. For $TiB_2$, the deviation from stoichiometry is ~1 at.% [48] and where the phase diagram published by Massalski *et al.* [78] indicates that a variation in stoichiometry is possible already at 500 °C, which is a typical temperature condition applied in PVD and CVD. The B-Hf phase diagram presented in [78] only provides data above 1400 °C and with a distinct homogeneity range from ~65 to ~68 at.% B at ~1500 °C. Andrievski [48] supports a homogeneity range in $HfB_2$ above 1400 °C from a calculated phase diagram, but where $HfB_2$ is a line compound at temperatures typical for PVD *i.e.*, below 1000 °C [48]. Although there are indications of homogeneity ranges in the Group 4 $TMB_2$ at elevated temperatures, it is evident that the C32 structure is stable and preferred by $ZrB_2$ and $HfB_2$. This is supported by our discussion in section 2.2 on the stabilities of $TMB_2$:s with C32 structure considering Hägg's rule and $\Delta\chi$.

In the B-Zr system, there is a B-rich zirconium dodecaboride $ZrB_{12}$ with $UB_{12}$-type structure [48], and reports of an ZrB monoboride with the FeB-type structure [48] not presented in the phase diagram by Massalski *et al.* [78]. Similar to Zr, Ti and Hf form monoborides, and where Hf forms also a $HfB_{12}$ phase [48]. In contrast to Zr and Hf, Ti forms $Ti_3B_4$ with $Ta_3B_4$-type structure [75].

Moving to Group 5 with TMs characterized by decreasing atomic radii and hence size of the interstitials in combination with increasing electronegativity ($\chi$) values promote the formation of phases with other stoichiometries and crystal structures. As can be seen for the B-Nb system in Figure 3 (middle panel), the number of binary phases compared to the B-Zr system have increased to at least four [78]. Similar as in the Group 4 B-TM phase diagrams, the Group 5 TMs form $TMB_2$ phases with C32 structure as present in the phase diagram for Nb-B in Fig. 3.

Differently from the Group 4 TMB$_2$, the NbB$_2$ (CbB$_2$) and TaB$_2$ phases exhibit more pronounced homogeneity ranges as reported by Kiessling [17] and later determined by Lundström [80] to extend from 65-70 at.% B in NbB$_2$ and 66-73 at.% B in TaB$_2$, but with no reports for any homogeneity range in VB$_2$, where Massalski's phase diagram [78] marks the phase as a line compound with a melting point at 2747 °C. In the B-Nb phase diagram, there is a Nb$_3$B$_4$ phase with Ta$_3$B$_4$ type structure [75], a NbB phase but with CrB type structure [17], and Nb$_3$B$_2$ with possible U$_3$Si$_2$ type structure similar to that of V$_3$B$_2$ [78]. The B-V and the B-Ta systems are similar as the B-Nb system with the formation of TM$_3$B$_4$, TMB, and TM$_3$B$_2$ phases of crystal structures previously defined for the B-Nb system. In the B-Ta phase diagram there is an additional metal-rich boride Ta$_2$B with CuAl$_2$ type structure [17], while the B-V phase diagram musters two additional phases with the composition V$_5$B$_6$ and V$_2$B$_3$ [78]. The large number of vanadium borides with different stoichiometry and crystal structures is explained by the small radius of the V atom of 1.34 Å and the relatively high $\chi$ value for V of 1.63, which results in smaller interstitials that seems to destabilize the boron sheets and where the higher $\chi$ value promotes metallic bonding. This is supported by observing the even higher number of determined phases for V's right hand neighbor Cr [75] and where we note the atomic radius for Cr of 1.28 Å with a $\chi$ value of Cr of 1.66 *i.e.*, smaller atomic radius and higher $\chi$ compared to V.

As presented in Figure 1, the atomic radii decrease and $\chi$ values increase further for the Group 6 TM. This has impact on the B-Mo phase diagram that show additional phases compared to the B-Nb phase diagram. The Group 6 TMs Cr and Mo form TMB$_2$ phases with C32 structure as seen for the Mo-B system in Fig. 3 (bottom panel), but where MoB$_2$ is only stable above 1500 °C [78], *i.e.*, higher than deposition temperatures applied in PVD. For MoB$_2$, Klesnar *et al.* [81] studied the composition of bulk MoB$_{2-x}$ and determined a narrow homogeneity range centered at 61 at.% B in the temperature range 1600-1800 °C, while in the B-Cr phase diagram [78], CrB$_2$ is illustrated as a line compound to the melting point of 2200 °C. [78] In the B-W phase diagram [78], no WB$_2$ with C32 crystal structure is present at temperatures above 1800 °C. Instead, there is a two-phase area shared by α-WB (MoB type structure) and the previously described W$_2$B$_5$ phase. However, WB$_2$ with C32 crystal structure was prepared by Woods *et al.*, in 1966 [82] by heating a boron wire in an atmosphere of WCl$_6$ and Ar and later as a sputtered thin film by Sobol in 2006 [83].

Returning to the B-Mo phase diagram, we find the Mo$_2$B$_5$ phase, α-MoB with MoB type structure, a possible Mo$_3$B$_2$ phase (U$_3$Si$_2$ type structure) [84], and a Mo-rich Mo$_2$B phase with CuAl$_2$ type structure [85]. On the B-rich side there is a MoB$_4$ phase with CrB$_4$-type structure [86] and a high temperature β-MoB phase with CrB-type structure [80]. Naturally, the B-Cr and the B-W binary phase diagrams both exhibit similarities to the B-Mo phase diagram with respect to phase distribution as seen from CrB$_4$ and WB$_4$ phases with CrB$_4$ type structure, CrB and high temperature β-WB with CrB type structure, and metal-rich phases Cr$_2$B and W$_2$B with CuAl$_2$ type structure. For Cr, there are two additional phases with Cr$_3$B$_4$ with Ta$_3$B$_4$ type structure and a phase with Cr$_5$B$_3$ composition. From the Group 4-6 phase diagrams, there is a trend where the number of phases increases when moving to the right in the periodic table and where Cr seems to form the largest number of phases ranging from B-rich CrB$_{41}$ to Cr-rich Cr$_4$B [75]. Ternary boride systems and solid solutions are presented in section 4.7.

# 3   Electronic structure of diborides

## 3.1   Chemical bonding

In section 2, we described the borides in terms of atomic size, electronegativity, and crystal structure as a basis to determine the chemical bonding in TMB$_2$. The predominant bonding consists of a mixture of (i) covalent, (ii) metallic, and (iii) ionic [25]. Generally, the borides have less directional bonds in terms of well-defined electronic orbitals involving covalent *s-p-d* hybrid configurations, compared to the stronger *p-d* hybridization in the carbides and the nitrides [87]. This has implications on their mechanical properties and properties at elevated temperatures as discussed in section 5.

Firstly, considering the *covalent* portion of B-TM bonding in TMB$_2$, it exhibits a distorted hybridization in the valence electron configuration of B. While an isolated B-atom has a valence of *2s$^2$2p$^1$*, a mixed *sp$^2$-sp$^3$* hybridized covalent bonding occurs when diborides are formed, which affects the materials properties. The strength of the covalent bonding depends on the combined effect of the energy of orbital overlaps relative to E$_F$ and the electron density in the bonding orbitals.

As shown in Table II, for Zr (the largest TM atom), the B-B distance in ZrB$_2$ of 1.83 Å exceeds the "normal" equilibrium B-B distance of 1.75 Å that has a minimum bond strain by 0.08 Å for TiB$_2$. This is due to tensile stretching in the B-B bonds as the Zr atoms are in contact with each other and give rise to the highest conductivity among the Group 4-6 TMB$_2$ [25]. In contrast, Cr and V (the smallest TM atoms) have reduced B-B distances and larger bond strengths. Among the TMB$_2$, the shortest B-B bond length, 1.72 Å, occurs for CrB$_2$ due to compressive strain on B from the TM atoms. An explanation to this is a decrease of the size of the interstices, where the boron atoms are positioned in the C32 crystal structure.

The hardness, thermal stability, and elastic properties originate from the strong B-B and B-TM bonds [88] [89] that are most significant for Group 4 TM borides (Ti, Zr, and Hf) [25]. The strengths of the covalent TM *d* – B *2p* bonding in TMB$_2$ compounds mostly connects with the bulk modulus *i.e.*, the resistance to compression, while the strength of the covalent B-B bonding and atomic separation in the *borophene* sheets, interleaved between hexagonal close-packed (hcp) structure TM atoms, reflects the hardness and the length of the *a* axis in the C32 crystal structure. As further discussed in section 5.5, the amount of covalency and directionality in the TM-B bonds and the ionic radii can, to a first approximation, explain the trend in the melting points.

Secondly, the metallic contribution to bonding (near E$_F$), increases when moving to the right in the periodic table as the *d*-bands become more filled by electrons [90]. However, the metallic bonding is partly counteracted by the amount of antibonding metal states in the vicinity of E$_F$ that cause weakening of the bond strength. As discussed in section 2.2, this phenomenon is a consequence of the increased difference in electronegativity between the metal and the B atoms.

Note that the bonding in the C32 crystal structure discussed here, for which the B atoms are positioned in 2D sheets, differs significantly from the class of materials with 3D B$_{12}$ icosahedra and UB$_{12}$ type structure commonly known as *covalent metals* that have more itinerant metallic electronic states at E$_F$. An example is ZrB$_{12}$, which has strong covalent bonding between the Zr *4d* orbitals and the B *2p* orbitals, but 3D icosahedral B-B interactions [91] [92], as there are no 2D *borophene* sheets in the structure.

Thirdly, the ionic part of the bonding in the TMB$_2$ originates from the electron transfer of two electrons from the metal atoms to each of the B atoms, involving significant charge transfer. In the simplest charge-transfer model, the TMB$_2$ can be denoted as TM$^{2+}$(B$^-$)$_2$, where the charge transfer from the metal atoms stabilizes the B-B bonding in the 2D *borophene* sheets of the C32 crystal structure.

It has been shown theoretically that in Group 4 the early TMB$_2$ compounds prefer the C32 crystal structure due to their low density-of-states (DOS) at the E$_F$ compared to competing phases such as the W$_2$B$_5$-structure [30]. According to the calculations by Moraes *et al.*, [30] early transition metal diborides (TiB$_2$, VB$_2$, etc.) tend to be chemically more stable in the AlB$_2$ structure type, whereas late transition metal diborides (WB$_2$, ReB$_2$, etc.) are preferably stabilized in the W$_2$B$_{5-x}$ structure type. However, point defects such as vacancies also significantly influence the phase stability and can even reverse the preference for the AlB$_2$ or W$_2$B$_{5-x}$ structure [30]. In reality, TMB$_2$ compounds with higher TM atomic numbers, in the same row, than VB$_2$ (*3d*), NbB$_2$ (*4d*), and TaB$_2$ (*5d*) exhibit higher *chemical stability* in the hexagonal W$_2$B$_5$ crystal structure than in the hexagonal C32 crystal structure. Thus, the theoretical calculations deviate from the observations that can be explained by B point defects such as vacancies in the buckled *borophene* layers. For HfB$_2$ (10-12 μΩ·cm), the bulk resistivity is higher than for TiB$_2$ (9-12 μΩ·cm) and ZrB$_2$ (7-10 μΩ·cm) as discussed in section 5.4.

Numerical computational methods based on the formation energy at 0 K for selected phases, including known competing phases, are often applied to predict ground-state stabilities of crystal structures. For stable structures, trends in bonding properties such as chemical orbital population and charge-transfer among atoms can be systematically compared as a function of the atomic radius of the transition-metal atom. In particular, the size of the trigonal prismatic interstitial space decreases when the size of the metal-atom decreases. This has the consequence that graphite-like B sheets are distorted from a *planar* atomic layer to a *buckled* layer. Buckling occurs when the B atoms are very close to each other and are forced to buckle as there is not enough space for them in the planar (graphite-like) structure [93]. The differences in the geometrical stabilities and the total energies of initially planar and buckled AlB$_2$ structures were theoretically analyzed by Pallas *et al.* [93]. They found that the early TM atoms (Y, Zr, Nb) prefer the planar AlB$_2$ structure, while the late TM atoms (Tc, Ru, Rh, and Pd) prefer the buckled structure where the TM-TM antibonding orbitals play an important role. In this case, MoB$_2$ is a limiting case that has no obvious preference for either planar or buckled structure and therefore form Mo$_2$B$_5$ instead of MoB$_2$. On the other hand, initially buckled structures of YB$_2$ and ZrB$_2$ became planar after geometry optimization, while buckled structures were not formed for these systems. This is due to the fact that the trigonal prismatic holes become too small to form a buckled structure. For the buckled structures, strong TM-TM antibonding orbitals are a decisive factor for the structure to form while for the planar structure, charge-transfer is more important.

The W$_2$B$_5$-type of structure is often referred to as an archetype *complex* structure in terms of stacking sequence [30] as well as phases with the orthorhombic crystal structure, which become more stable when moving to the right in the periodic table. The trend in complexity has been theoretically confirmed for both the *3d* [94] and the *4d* [93] transition-metals borides, where the r$_B$/r$_{TM}$ radius ratio <0.59 is correlated to the most energetically preferred *planar* structures. TM borides with a radius ratio equal or above 0.59 tend to form *buckled* structures [93]. However, for light atoms like B, phonon modes may also stabilize structures that is not included in 0 K ground-state calculations.

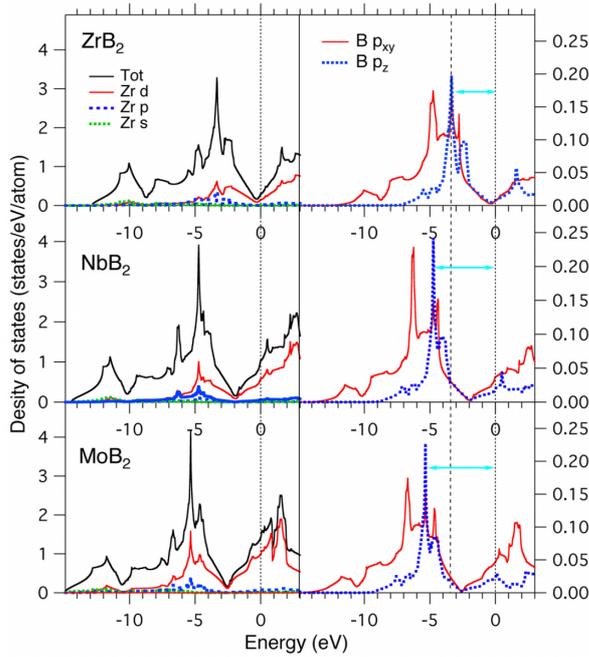

**Figure 4**: Zr, Nb, and Mo *s*, *p*, *d,* and B *s*, *p* partial densities of states for ZrB$_2$, NbB$_2$, and MoB$_2$ (left panels) and B *p$_{xy}$* and *p$_z$* states (right panels) [Author's work, unpublished]. The E$_F$ is represented by the vertical dashed line at zero energy. The energy shift of the main peak for the three systems is shown by the vertical dot/dashed line and the horizontal arrows.

The *borophene* layer in the C32 crystal structure has unique properties as it is stronger and more flexible than graphene, has low resistivity, is a superconductor, and play an important role for the properties of the TMB$_2$. In 1997, different polytypes of quasi-planar surfaces of monolayer sheets of B were theoretically predicted using *ab initio* quantum chemical and density functional theory [95]. In 2015, *borophene* was first synthesized in UHV on Ag(111) substrates using CVD by Mannix *et al.* [96] and in 2018 on Al(111) substrates [97]. In the latter case, the increased charge-transfer from Al to B in graphene-like *borophene* was found to improve the phase stability. At least four different structures of *borophene* have been identified; Besides the *corrugated* or *buckled* phase, mentioned above, with space group 2-Pmmn, the triangular lattices with space groups *b$_{12}$*, *c$_3$* [98] and the planar graphene-like phases with hexagonal honeycomb symmetry. Recently, the production of free-standing *borophene* has been shown [99].

The strength, flexibility, and resistivity properties of *borophene* can be tuned by the arrangement of vacancies and the orientation of the material [100]. We anticipate that *borophene* and borides with "built-in" *borophene* layers have potential to become important in emerging applications as anode materials in metal-ion batteries, hydrogen storage, hydrogen peroxide, catalytic reactions, and sensors.

## 3.2  Density of states

As discussed in the previous section, the electronic structure of the metal-diborides affects the properties via the chemical bonding. Figure 4 shows calculated partial density of states (DOS) for Zr, Nb, Mo *s*, *p*, *d*, and B *s*, *p* in the left panels and B *p$_{xy}$*, *p$_z$* states of ZrB$_2$, MoB$_2$, and NbB$_2$ in the right panels. For ZrB$_2$, E$_F$ is found at a DOS minimum between completely occupied bonding states and unoccupied antibonding *pd*-bands. This position of the E$_F$ located in the center of a *pseudogap*, energetically stabilizes the crystal structure in a similar way as for TiC. [87] It indicates that the isoelectronic Group 4 TiB$_2$, ZrB$_2$, and HfB$_2$ are the most stable phases [32] [101]. This observation is consistent with Hägg's rule and the trend in the difference in electronegativities between the metal and B atoms discussed in section 2.1 and section 2.2, respectively. The position of E$_F$ also influences the resistivity (see section 5.4). Comparing ZrB$_2$ with NbB$_2$ and MoB$_2$ in Figure 4, the E$_F$ for the latter two TMB$_2$:s is situated among antibonding states, which is not energetically ideal and weakens the covalent bonding.

**TABLE IV** (top): Bulk hardness in GPa, and (middle) bulk resistivity in µΩcm. Melting points °C (bottom) [73]. *TMB$_2$ with other crystal structures than C32. †TMC and TMN with other crystal structure than B1.

| Bulk hardness [GPa] | | | | | |
|---|---|---|---|---|---|
| 4 | | 5 | | 6 | |
| TiB$_2$ | 15-45 | VB$_2$ | 20.9 | CrB$_2$ | 20.5 |
| ZrB$_2$ | 22.5-23 | NbB$_2$ | 23.2 | *Mo$_2$B$_5$ | 23.0 |
| HfB$_2$ | 28 ref. [25] | TaB$_2$ | 22.6 | *W$_2$B$_5$ | 26.1 |
| TiC | 28-35 | VC | 27.2 | †Cr$_3$C$_2$ | 10-18 |
| ZrC | 25.9 | NbC | 19.6 | †Mo$_2$C | 15.5-24.5 |
| HfC | 26.1 | TaC | 16.7 | †WC | 22 (0001) |
| TiN | 18-21 | VN | 14.2 | - | - |
| ZrN | 15.8 | NbN | 13.3 | - | - |
| HfN | 16.3 | TaN | 11.0 | - | - |

| Bulk resistivity [µΩcm] | | | | | |
|---|---|---|---|---|---|
| 4 | | 5 | | 6 | |
| TiB$_2$ | 9-15 | VB$_2$ | 13 | CrB$_2$ | 18 |
| ZrB$_2$ | 7-10 | NbB$_2$ | 12 | *Mo$_2$B$_5$ | 18 |
| HfB$_2$ | 10-12 | TaB$_2$ | 14 | *W$_2$B$_5$ | 19 |
| TiC | 68 | VC | 60 | †Cr$_3$C$_2$ | 75 |
| ZrC | 43 | NbC | 35 | †Mo$_2$C | 71 |
| HfC | 37 | TaC | 25 | †WC | 22 (0001) |
| TiN | 20-25 | VN | 85 | - | - |
| ZrN | 7-21 | NbN | 58-78 | - | - |
| HfN | 33 | TaN | 135-250 | - | - |

| Bulk melting points [°C] | | | | | |
|---|---|---|---|---|---|
| 4 | | 5 | | 6 | |
| TiB$_2$ | 2980 | VB$_2$ | 2100 | CrB$_2$ | 2170 |
| ZrB$_2$ | 3040 | NbB$_2$ | 3050 | *Mo$_2$B$_5$ | 2100 |
| HfB$_2$ | 3250 | TaB$_2$ | 3200 | *W$_2$B$_5$ | 2600 |
| TiC | 3067 | VC | 2830 | †Cr$_3$C$_2$ | 1810 |
| ZrC | 3420 | NbC | 3600 | †Mo$_2$C | 2520 |
| HfC | 3928 | TaC | 3950 | †WC | 2870 (0001) |
| TiN | 2950 | VN | 2177 | - | - |
| ZrN | 2980 | NbN | ~2400 | - | - |
| HfN | 3387 | TaN | 3093 | - | - |

The density of metallic *3d*, *4d*, and *5d* states at E$_F$ increases towards the right in the periodic table as the bands become progressively more filled [88] [90]. The B-TM bonding is associated with two relatively weak *2p$_z$*-π orbitals oriented perpendicular to the B-planes while forming a strong graphite-like B network involving four *2p$_{xy}$* in-plane σ orbitals in the basal plane. For ZrB$_2$ shown in Figure 4, the states within 6 eV of E$_F$ are dominated by metallic Zr *4d* states, while B *2p* and B *2s* states are more localized at 3-4 eV and 11 eV, respectively. The right panel in Figure 4 shows that the B *2p$_{xy}$*-σ states in the basal plane of ZrB$_2$ exhibit a double-peak at 3.35 eV and 4.75 eV with a 1.4 eV peak splitting. The peak at 3.35 eV is mainly due to a flat band containing B *2p* σ states in the vicinity of the *M* (1/2,0,0) symmetry point [102] in the Brillouin zone of the reciprocal hexagonal lattice as shown in Figure 2b. In addition, there are peaks at 7.5 eV and 10 eV near the bottom of the valence band due to B *2sp* and B *2s* states, respectively. Notably, these bands only occur for the $p_{xy}$ σ states in the basal plane and are due to the strong B-B bonds in the B sheets. The out-of-plane B $2p_z$ π contribution also exhibits a double peak at 2.33 eV and 3.35 eV (1 eV peak splitting) that is due to B $2p$ – Zr $4d$ interaction. The sharp ZrB$_2$ B $2p_z$ interplanar peak located at 3.35 eV, increases to 4.74 eV for NbB$_2$ and 5.34 eV for MoB$_2$. A larger binding energy position of the B $p_z$ peak relative to E$_F$ signifies a shorter bond length and stronger bond in the $c$ direction between the B sheets and the metal layers, which affects the elasticity of the material.

For the TMB$_2$ compounds, the DOS consists of TM $spd$ hybridized states, B $sp$-hybridized states, and antibonding states above the pseudogap. The pseudogap at E$_F$ in the TMB$_2$ compounds has primarily been associated with the strong covalent interaction between the in-plane B-B $2p$-states, and less due to the covalent interplanar B-TM interaction [103].

According to *Mott's law of conductivity* in 1969 [104], the increasing number of states at the E$_F$ from ZrB$_2$ (0.2933 states/eV/atom) versus NbB$_2$ (1.0335 states/eV/atom) and MoB$_2$ (1.4241 states/eV/atom) should [104] increase the conductivity of the TMB$_2$ material due to the metal bonding and affect the stability by charge redistribution in the valence band. This rough approximation can be compared to the opposite trend in the experimental resistivity values of bulk materials in Table IV for ZrB$_2$ yielding a value of 7-10 μΩcm (8.0 μΩcm) [105] compared to 12 μΩcm [73] for NbB$_2$ and 18 μΩcm [73] for Mo$_2$B$_5$. However, Mott's law assumes isotropic scattering and carrier mobility (the Mott approximation), which differs from the experimental determination of anisotropy in carrier lifetimes. Furthermore, the higher resistivity of Mo$_2$B$_5$ can be attributed to the larger amount of *anti-bonding* states at E$_F$ in comparison to ZrB$_2$. Thus, ZrB$_2$ has the lowest resistivity: ZrB$_2$ 7-10 μΩcm, NbB$_2$ 12 μΩcm and Mo$_2$B$_5$ 18 μΩcm (see table IV for bulk materials). The respective boride's resistivity is discussed further in section 5.4.

## 3.3   Orbital overlaps and hybridization

Figure 5 schematically illustrates energy levels of the bonding and antibonding orbitals involved in the chemical bonding of diborides (middle), compared to a B sheet (left) and a transition metal (right). E$_F$ is indicated by the horizontal dashed line. The energy levels of the B sheet consist of well-defined B $2p_{xy}$ σ and B $2p_z$ π molecular orbitals, while the $sp$ and $d$ states in the transition metal are spread out in broader bands. As the TM and B atoms combine to form the C32 crystal structure, the energy positions of the hybridized bands are associated with changes in the electronic filling of the $d$-band and the covalent B $p$ – TM $d$ band.

As discussed above, the valence band of TMB$_2$ consists of B $2p$ states that reflects the 2D graphite-like layers [106] and form four in-plane $2p_{xy}$ states with σ bonds and two out-of-plane $2p_z$ states with π bonds. Foremost two in-plane B $2p_{xy}$ σ bands cross E$_F$ and contribute considerably to the density of states at E$_F$ and, hence, to the metallic-like properties. The out-of-plane B $2p_z$-π bands correspond to the weak interlayer bonding. The electrical and magnetic properties largely depend on the number of electrons in the valence band. Calculations have shown that the electron-phonon *coupling* of the σ and π bonds is highly anisotropic with a dominant role due to optical in-plane phonons [107] and phonon dispersion around the Γ-symmetry point in the Brillouin zone of ZrB$_2$ [108].

Since the discovery of bulk superconductivity in magnesium diboride, $MgB_2$ by Nagamatsu in 2001 [59] with a transition temperature of $T_c \sim 39$ K, which at that time was the highest determined for a conventional (phonon mediated, non-copper-oxide) bulk superconductor, the existence or nonexistence of superconductivity in several diborides has been a controversial issue. For example, in 2001, Gasparov *et al.* [109] claimed the discovery of a superconducting transition at 5.5 K in $ZrB_2$ polycrystalline pressed powder samples while Kaczorowski *et al.* [60] claimed a superconducting transition at 9.5 K in powder samples of $TaB_2$ and that $ZrB_2$ was not superconducting. Moreover, the difference in $T_c$ in the $TMB_2$ materials has been associated with different electron-phonon coupling and the ratio between the B *p*-states versus TM *d*-states at the $E_F$ [92]. For $MgB_2$, the obtained $T_c$ largely depend on the sample purity, in particular, it is sensitive to the amount of surface oxidation into understoichiometric $MgB_xO_y$ [110].

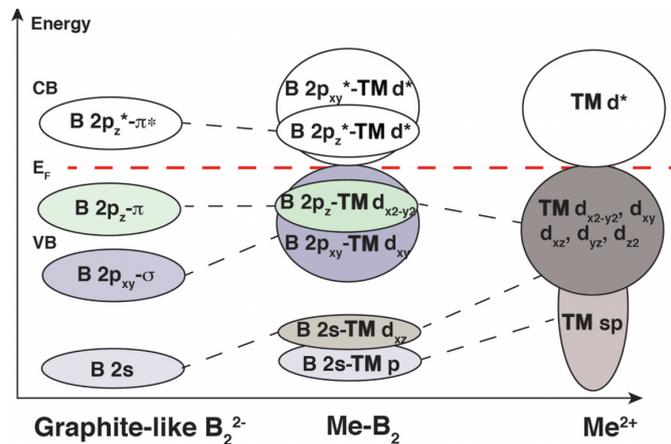

**Figure 5:** Energy levels of the bonding and antibonding orbitals involved in the chemical bonding of diborides [Author's work, unpublished].

The B layer has been identified as critical for the superconducting properties of bulk $MgB_2$ with $T_c \approx 39$ K. The Mg-Mg, Mg-B, and B-B bond lengths are 3.080, 2.502, and 1.778 Å, respectively. Raman spectroscopy has shown that for the B atoms, there are two orthogonal in-plane phonon modes around the Γ-symmetry point having $E_{2g}$ symmetry that dominates [111] [112]. The $E_{2g}$ phonon mode that changes the B-B bond length is related to the σ-bonded B $2p_{xy}$ in-plane orbitals. These phonon modes have been identified as the key factor that influences the electron-phonon interaction and the superconducting properties of $MgB_2$ [113].

A similar superconductivity mechanism as in $MgB_2$ with B/phonon interactions leading to an energy gap has been suggested [114] [115] for other borides with the C32 crystal structure. In comparison to $ZrB_2$ (0.2933 states/eV/atom), band-structure calculations show a higher DOS at $E_F$ for $MgB_2$ (0.8858 states/eV/atom with $E_F$ located at a peak of occupied states) while it is 1.0335 states/eV/atom for $NbB_2$ and 1.4241 states/eV/atom for $MoB_2$. The relatively high superconducting transition temperature $T_c$, ~39 K, for bulk $MgB_2$ has primarily been associated with the large density of B *p*-states at $E_F$, while for $ZrB_2$, the DOS at $E_F$ is dominated by metallic *4d*-states. A dominating B *2p* DOS for $MgB_2$ has also been confirmed experimentally by x-ray absorption and emission spectroscopies [116] [117]. The electronic structures of bulk $ZrB_2$ and $TiB_2$ have been experimentally investigated by photoemission spectroscopy, [29] [118] x-ray absorption spectroscopy, [119] [118] optical spectroscopy, [120] [121] [122] and point-contact spectroscopy [123].

The factors responsible for the possible superconducting properties of $TMB_2$ with C32 crystal structure have been theoretically analyzed [92] [102]. It was found that the main contributing factors to superconductivity in $MgB_2$ are the high density of $2p_{xy}$ (in-plane) B-B bonds with σ-symmetry consisting of partly occupied degenerate flat bands in the vicinity of the $E_F$ and relatively weak ionic Mg-B bonding. The degenerate and partly occupied flat B σ-bands around $E_F$ in $MgB_2$ is associated with an effective electron-phonon coupling between to the σ B-B stretching phonon mode within the B-plane and the low-frequency anharmonic $E_{2g}$ phonon mode that is advantageous for creating Cooper pairs across the small energy gaps across the $E_F$. $MgB_2$ has both σ and π electrons with different energy gaps of 2.2 and 7.1 meV, at the $E_F$ with different coherence lengths (51 nm and 13 nm) that indicate a mixture of type I and type II kind of superconductivity that has also been referred to as *type-1.5* superconductivity. [124] For comparison, Group 5 $TMB_2$, the TM - B covalent bonding changes the energy and increases the dispersion of the B σ-band away from the $E_F$ that is unfavorable for the energy gaps and phonon modes of superconductivity. For $TMB_2$ to the right of Group 4 in the periodic table, the B σ-band is completely occupied and does not have the possibility to split the degeneracy by electron-phonon coupling. A high density of TM *d*-states at $E_F$ with strong covalent TM-B bonding and hybridization is disadvantageous as the σ-band disperse away from the $E_F$. As shown in section 3.2 (Fig. 4), the *distance* of the B σ- and π-bands increase from the $E_F$ as the *d*-band is being filled from Group 4 to 6 that also affects the bands crossing the $E_F$. Thus, among the layered diborides, only $MgB_2$ is a superconductor at normal pressure without strain.

Furthermore, it has been theoretically shown that pressure-induced *strain* can reduce the covalent TM-B overlap when the lattice is compressed along the *a-b* direction or stretched along the *c* direction. Both types of strain rise the degenerate σ-band ($p_{xy}$) involved in the B-B bonds and increases the band splitting so that it nestles around the $E_F$ while elongation of the *c* axis softens the degenerate $E_{2g}$ stretching phonon that split the B σ band that is necessary to produce Cooper pairing [102]. Thus, a suitable strain can be chosen to reduce the hybridization of the TM-B bonds and increase the electron-phonon coupling. It is anticipated that extreme pressures could increase the transition temperatures of suitable materials, ideally even up towards room temperature.

Another way to control and improve the superconductivity than strain is *doping* of $TMB_2$, *e.g.*, V-doped $ZrB_2$. Renosto *et al.* [125] studied the $Zr_{0.96}V_{0.04}B_2$ alloy that exhibits a bulk superconducting transition temperature of $T_c \sim 8.7$ K as V induces superconductivity in the non-superconducting material $ZrB_2$. In 2017, Barbero *et al.* [126] also found that V-doped $ZrB_2$ and $HfB_2$ alloys $Zr_{0.96}V_{0.04}B_2$ and $Hf_{0.97}V_{0.03}B_2$ had a $T_c$ of ~ 8 K. On the other hand, doping of $MgB_2$ with $TiB_2$ [127] and $ZrB_2$ [128] can cause superconductivity by increasing the *c*-lattice parameter. This is also the case with excess B, as demonstrated for $NbB_{2+x}$ [129] [130].

A small fraction of V substitution of $TMB_2$ has several effects, that combined give rise to superconductivity. The $E_F$ moves very close to the bottom of the $d_{xz}$, $d_{yz}$ degenerated bands and obtains a flat shape along the Γ-A symmetry line. These factors change the Fermi surface due to the phonon mode involved that break the $d_{xz}$-$d_{yz}$ degeneracy, in a similar way as the $E_{2g}$ phonon mode breaks the degeneracy of the σ band in $MgB_2$. [102]

### 3.4   Bond length by x-ray spectroscopy

In the quest to understand the B-TM bonding, it is important to keep the oxygen level as low as possible as it causes unnecessary complications associated to B enrichment at surfaces and grain boundaries in the microstructure [131].

Experimentally, oxidation states and local short-range order atomic coordination symmetry have been probed with the combination of x-ray absorption near-edge spectroscopy (XANES) and extended x-ray absorption fine structure (EXAFS) spectroscopy [132] that are complementary tools to long-range order probed by x-ray diffraction (XRD). XANES provides; (i) a quantitative measure of the average oxidation state by the energy shift of the appropriate x-ray absorption edge; (ii) the amount of *p-d* hybridization in the chemical bonds, and thus the symmetry of the structure; and (iii) information about the coordination symmetry when the XANES line shape is compared to reference materials [133]. EXAFS gives quantitative information on average (i) bond lengths, (ii) coordination number (number of nearest- and next-nearest neighbors in different directions), and (iii) the mean-square disorder (Debye-Waller factor that also depends on the temperature and phonon vibrations).

For diborides, it is primarily bulk $ZrB_2$ [25] [48] that has been studied, with only a few reports on thin films [131] [134] [135]. Chu *et al.* [136] made EXAFS analysis of polycrystalline $ZrB_2$ samples synthesized from Zr and B powders by the float-zone method. Temperature-dependent measurements showed little difference between in-plane and out-of-plane vibrations in the Zr-Zr bonds. [137] Bösenberg *et al.* [138] investigated the chemical states of $ZrB_2$ and Zr powders by XANES and found that the energy of the Zr *K*-edge for $ZrB_2$ was 10 eV above that of a Zr foil reference. High-energy shifted unoccupied Zr *4s* states were also found for e-beam co-evaporated $ZrB_2$ thin films by XANES and EXAFS applied to investigate the local chemical bonding structure and the atomic distances [131]. However, the O content in these films resulted in the formation of crystalline tetragonal $ZrO_2$, which yielded longer atomic bond distances as determined by EXAFS.

Recently, a XANES and EXAFS diboride film study by Magnuson *et al.* [135] on a magnetron sputter deposited epitaxial $ZrB_2$ film using the Zr *K*-edge showed shorter B-B bonds (1.827 Å) than in a $ZrB_2$ compound target (1.833 Å) with 99.5% purity from which the film was synthesized. The Zr-B distance was also somewhat shorter (2.539 Å) in the epitaxial $ZrB_2$ thin film than in the bulk sample (2.599 Å). This can be compared to the reference value of 2.542 Å in Table II that is in between the thin film and bulk in ref. [135] The Zr-B bond distance was found to be longer in bulk than in the thin film sample due to additional superimposed Zr-O bonds resulting from contaminants, which is consistent with the observations in XANES, XPS, and XRD. The measured Zr-B value (2.539 Å) was in better agreement with the calculated bond distance from the lattice parameters of the relaxed equilibrium bulk literature value (Zr-B = 2.542 Å) [12] than that of the $ZrB_2$ compound target. Furthermore, the Zr-B bond length in the $ZrB_2$ epitaxial film was shorter than the 2.546 Å obtained by Stewart *et al.* [131] [134] for e-beam co-evaporated films from elemental Zr and B sources of 99.5% purity. Chu *et al.* [136] obtained a Zr-B bond length of 2.55 Å on single-phase polycrystalline samples prepared by the rf floating-zone method from stoichiometric mixtures of Zr and B elements compacted to pellets and arc melted. For comparison, Lee *et al.* [139] found a Zr-B bond length of 2.81 Å in magnetron sputtered $ZrB_2$ multilayers to be used in sensors. A longer Zr-B bond is likely due to additional impurities and non-directional bonds. The $ZrB_2$ film in ref. [135] exhibited superior electronic structure properties in terms of chemical bonding alignment, crystal quality and level of contaminants when compared to the $ZrB_2$ compound target from which it was synthesized. Thus, it can be anticipated that - at present level of synthesis process optimization - thin films provide better means than bulk materials to investigate and determine fundamental properties. As discussed in section 5.3, $ZrB_2$ films are more elastic than bulk $ZrB_2$, by virtue of the former's high purity.

Other diborides such as $TiB_2$, $VB_2$, and $CrB_2$ have been scarcely studied as bulk materials and even less as thin films. This is due to difficulties thus far of growing thin films with well-defined properties of these materials. Chu *et al.* [137] used EXAFS to study the bond lengths of bulk polycrystalline $TiB_2$ and found Ti-B and Ti-Ti distances of 2.37 and 3.07 Å. They also measured the bond lengths of bulk polycrystalline $VB_2$ and $CrB_2$. [140] Chartier *et al.* [141] reported EXAFS on amorphous understoichiometric thin films of $TiB_2$ deposited by dynamic ion mixing a Ti-B distance of 2.37 Å and a Ti-Ti distance of 3.03 Å in comparison to bulk TiB$_2$ with distances of 2.39 Å and 3.01 Å, respectively. Analogous to ZrB$_2$, we anticipate that the bond distances in stoichiometric single-crystal films of TiB$_2$, VB$_2$, and CrB$_2$ will be somewhat shorter and as a result the properties would be superior compared to polycrystalline materials.

# 4  Transition-metal diboride film synthesis with extension to ternary systems

There are two predominant techniques for vapor-phase growth of thin films: CVD and PVD. This review focuses on sputter deposition with DCMS, radio frequency (rf) sputtering, and pulsed techniques such as high-power impulse magnetron sputtering (HiPIMS). In addition to sputtering, PVD growth of borides has been performed with cathodic-arc evaporation as further developed in section 4.2 and exemplified by studies on the Ti-B system. In section 4.3, e-beam evaporation, and pulsed-laser deposition (PLD) are discussed, followed by the few studies on reactive sputtering of borides in section 4.4. In section 4.5, we bring attention to sputter deposition of Group 4-6 TMB$_2$ layers with C32 crystal structure and where the properties of these films are further treated in section 5.

## 4.1  Chemical vapor deposition

Vapor-phase synthesis of borides by CVD was initiated in the early 1930s, where Moers [42] investigated growth from the B precursor BBr$_3$, and the metal halides TiCl$_4$, ZrCl$_4$, HfCl$_4$, VCl$_4$, and TaCl$_5$. Moers focused on the chemistry in the CVD processes studied and provided limited amount of information on the properties of the deposited films. For instance, the phase distribution of the deposited material was characterized by x-ray diffraction (XRD), but with no conclusive evidence for growth of, *e.g.*, ZrB$_2$ or ZrB. However, the work [42] represented a stepping-stone for "halide" CVD that was later developed mainly in the USA starting from the 1960s [142] [143] [144] [145] [146] [147] and most likely inspired by the Apollo program's (1960 to 1972) demand for new coating materials such as TMB$_2$. An advantage of halide CVD was that the process chemistry resembled that developed in parallel for industrial growth of carbide and nitride coatings on cutting tools used for metal machining. Thus, experience from halide CVD of carbides and nitrides could be used when developing processes for TMB$_2$. On the other hand, halide CVD of TMB$_2$ coatings was, like CVD of TMC and TMN coatings, limited by the high temperatures required to activate and dissociate the precursors, typically well above 1000 °C, as well as parasitic etching by the hydrochloric acid generated from the chemical reaction. Under these conditions, Gannon *et al.* [142] in 1963 deposited crystalline and 110-oriented TiB$_2$ coatings on graphite sleeves at 1600 °C, but with no additional information on the composition of the grown coatings. Lowering the deposition temperature to 1400 °C, Gebhardt and Cree [143] in their work from 1965 grew crystalline stochiometric and non-porous TiB$_2$ on graphite using thermal decomposition of gaseous reactants (a pyrolysis process similar to CVD) of halide and boron trichloride in a graphite resistance-heated vacuum furnace. The B and metal content in the deposits were determined from chemical analysis including pyrohydrolysis and titration for elemental analysis of the boron and calorimetric procedures for the metal. ZrB$_2$ and HfB$_2$ coatings were also deposited, albeit as phase mixtures together with rhombohedral boron.

In 1974, Takahashi and Kamiya [144] studied growth of $TiB_2$ coatings on graphite substrates below 1200 °C, and where $TiB_2$ was deposited together with an excess of (amorphous) B. The deposition of B-rich coatings indicates a higher reactivity of the $BCl_3$ precursor compared to the applied metal halide, making it necessary to investigate alternative boron precursors to better match the reactivity to that of the metal halide. To lower the deposition temperature, reduce etching, and improve properties of the deposited coatings, Pierson and Mullendore [148] in 1980 replaced $BCl_3$ by diborane ($B_2H_6$) and grew $TMB_2$ at 500-900 °C. Stoichiometric $TiB_2$ coatings were thus deposited on graphite at 700 °C, but with B-excess at lower deposition temperatures as determined from electron microprobe analysis and induction-coupled plasma spectroscopy *i.e.*, like that reported by Gebhardt and Cree [143] for $ZrB_2$ and $HfB_2$ coatings as well as by and Takahashi and Kamiya [144] for $TiB_2$ coatings. Given the undesired etching and the high temperature processing characteristic for halide CVD described above, there was a need for developing low temperature synthesis routes, specifically at low substrate temperatures, below 600 °C. For this propose, single-source precursors such as transition-metal tetrahydroborates $TM(BH_4)_4$ have been investigated starting in the late 1980s [149] [150] [151].

Epitaxial growth of $ZrB_2$ films from the precursor $Zr(BH_4)_4$ has been demonstrated on Si(111) [152] [153], Si(001) [154], and $Al_2O_3$ [155], but at substrate temperatures of 900 to 1000 °C. These high temperatures are favorable for decomposition of the $Zr(BH_4)_4$ precursor and sufficient to activate the $H_2$ for abstraction of the excess B. These conditions promote epitaxial growth of $ZrB_2$ films of high crystal qualities as illustrated by HREM images presented by Tolle *et al.* [152] for a $ZrB_2$ film deposited on Si(111) substrate at 900 °C showing a good registry and a sharp interface between the substrate and the film as well as the *in-situ* reflection high-energy electron diffraction patterns for a $ZrB_2$ film deposited on $Al_2O_3$(0001) at 1000 °C in a study by Bera *et al.* from 2009 [155]. The process conditions required for CVD growth of epitaxial $ZrB_2$ films, in particular the choice of deposition temperature and the ultra-high vacuum (UHV) deposition conditions, inspired the development of DCMS of epitaxial $ZrB_2$ films with growth from a $ZrB_2$ compound target and on the substrates Si(111) [156], 4H-SiC(0001) [156] [157], and $Al_2O_3$(0001) [158] as well as Si(001) substrates [159].

A limitation in epitaxial growth of $ZrB_2$ from $Zr(BH_4)_4$ is low deposition rates with 0.3 nm/min on $Al_2O_3$(0001) [155] and 1.2 nm/min on Si(111) practically limiting film thicknesses to typically < 100 nm. Higher growth rates result in the growth of amorphous films on Si(111) [152]. This may be explained by difficulties at higher growth rates to remove excess B from the $Zr(BH_4)_4$ precursor in the growth zone by formation and desorption of volatile $B_2H_6$ molecules, resulting in the growth of B-rich films. This is exemplified by the study by Sung *et al.* [151] demonstrating growth of films with B/Zr ratios close to 3 in the temperature range 250-450 °C [151].

In the CVD community, there is an active development of tailored precursors for growth of specific materials systems including $TMB_2$ thin films preferably at low temperatures. Precursor development, however, is much less explored in the PVD community. It has been shown that reactive magnetron sputtering of tungsten target in krypton/trimethylboron ($B(CH_3)_3$) plasmas results in growth of W-rich 100-oriented $WC_{1-x}$ with a potential boron solid solution [160]. Moreover, there is an interesting process development in hybrid sputtering using typical CVD precursors such as $B_2H_6$, pentaborane(9) ($B_5H_9$) decaborane(14) ($B_{10}H_{14}$) as further discussed in section 7. We anticipate that a similar approach can be developed for $TMB_2$ thin films.

## 4.2  Cathodic arc deposition

The high melting points of TMB$_2$, ≥ 3000 °C generates evaporation characteristics like metals with high melting points such as W and Ta *i.e.*, the generation of solid glowing macroparticles. In 1991, Knotek *et al.* [161] studied DC arc deposition of TiB$_2$ coatings from high-pressure sintered TiB$_2$ cathodes manufactured from TiB$_2$ powder with small additions < 1 wt. % of Al and Ni as well as C and B in Ar and Ar/N$_2$ plasmas. The focus for the study was on improving the evaporation properties of ceramic cathodes by investigating cathode materials of different purities > 99% and densities between 98.7% to 99.2% of the theoretical values, resulting in specific electrical resistivities and thermal conductivities between 8.1 to 25.4 and 15.6 to 40 W/m K at 20 °C, respectively. A deposition process was found [161] that was less dependent on the properties of the cathode material, but with emission of glowing solid macroparticles and problems with controlling the movement of the arc-spot on the cathode as it often remained at a single location (immobilization) causing local overheating *i.e.*, prevailing problems in arc evaporation from ceramic cathodes. Despite these difficulties, 0001-oriented hexagonal TiB$_2$ coatings with a HV 0.05 up to 4250 were deposited on cemented carbide and high-speed steel substrates with deposition rates of 2 to 4 μm/h, but with no information provided on the composition of the coatings. The addition of N$_2$ during arc deposition appeared to stabilize the arc-spot movement, but at the expense of nitrogen uptake to form Ti-B-N coatings with reduced hardness. Here, Knotek *et al.* noted even the formation of crystalline TiN indicating a higher affinity of the metal Ti towards N compared to B in arc deposition. This is an important property of Ti that comes into play for this metal as well as the other Group 4-6 TM when TMB$_2$ is co-sputtered with N$_2$.

The pioneering work by Knotek *et al.* [161] inspired other researchers to further improve the properties of arc deposited TMB$_2$ coatings. In a publication from 1993, Tregilo *et al.* [162] investigated a pulsed arc deposition of a TiB$_2$ cathode in vacuum. From pulsing of the arc and by applying pulsed high voltage bias to the substrate, the authors sought to minimize the emission of glowing solid macroparticles and the problems with controlling the movement of the arc-spot previously described in ref. [161] The results showed that thick up to 10 μm, adherent and hard 2600 to 3300 (HV 0.025) coatings with compressive stress of 2.3 GPa were deposited on stainless steel substrates without external heating and using pulsed high voltage bias to the substrate. The authors provided no information on composition and the structural properties of their films to support that TiB$_2$ with C32 crystal structure had been deposited.

In a publication from 2015, Zhirkov *et al.* [163] investigated the influence of an external magnetic field for controlling the movement of the arc-spot on the cathode. From DC arc deposition of a TiB$_2$ cathode in vacuum, the authors showed that deposition without the external magnetic field stabilizes the conditions for the arc on the cathode. This enabled growth of smooth Ti-B coatings on MgO substrates with an approximate 1:1 Ti to B ratio as determined from quantitative analysis by XPS. The discrepancy between the composition of the cathode and the deposited film was explained by the spatial plasma distribution on the ion mass. The composition of films sputtered from TMB$_2$ sources is further discussed in section 5. To deposit TMB$_2$ as a next generation of hard coatings should be of benefit for progress in this field.

## 4.3  E-beam evaporation and pulsed-laser deposition

In a classical work from 1978, carried out under the *aegis* of the U.S.-U.S.S.R. Science and Technology Cooperative Agreement, Bunshah *et al.* [164] used electron-beam evaporation to deposit TiB$_2$ and ZrB$_2$ coatings from sintered evaporation billets manufactured from TiB$_2$ and ZrB$_2$ powders. Growth was carried out on polycrystalline Mo sheets at deposition rates ranging from 0.113 μm/min to 6.3 μm/min and temperatures from 600 to 1300 °C. Crystalline TiB$_2$ and ZrB$_2$ coatings were deposited, but with TiB$_2$ and TiB phase mixtures at a deposition rate of 6.3 μm/min, and where quantitative analysis of the phases by XRD showed TiB$_2$-rich coatings at 690 °C to 910 °C and TiB-rich coatings at higher temperatures. Due to difficulties in the quantification of the amount of B, there was no information on the composition of the coatings in, *e.g.*, ref. [164] For deposited carbide coatings, the authors used the lattice parameter to determine the carbon-to-metal ratio in the coating, which is not possible for compounds with narrow homogeneity ranges such as TiB$_2$ and ZrB$_2$. Bunshah *et al.* communicated lattice parameters for the deposited TiB$_2$ and ZrB$_2$ coatings in an earlier publication [165]. The general trend for both borides is an increasing length for the *c* axis with increasing growth temperature, while the length of the *a* axis remained relatively unchanged and close to that determined for TiB$_2$ with 3.03 Å and 3.17 Å for ZrB$_2$, see Table II. The TiB$_2$ coatings were predominantly 0001-oriented while the orientation of the ZrB$_2$ depended on the deposition rate as well as temperature with 0001-oriented coatings grown at a deposition rate of 2.14 μm/min and in the range 670 °C to 850 °C. Lower deposition rate of 0.113 μm/min resulted in $10\bar{1}0$-oriented coatings at temperatures from 600 °C to 1300 °C. The observations made in [164] and [165] on film orientation at different deposition rates and temperatures are valuable when discussing the results obtained from sputtered deposited films.

Bunshah *et al.* [164] [165], investigated the microhardness (micro-Vickers hardness under 50 g for 10 s) of the deposited TiB$_2$ and ZrB$_2$ coatings, where the general trend was increasing hardness values from ~2200 to ~3200 kg/mm$^2$ for TiB$_2$ coatings at increasing substrate temperatures, while the ZrB$_2$ coatings showed an opposite trend seen from ~3200 to ~1900 kg/mm$^2$. The hardness trend for 0001-oriented TiB$_2$ coatings is clear as higher temperatures should result in larger crystals, while the trend for coatings with a phase mixture was more unclear as TiB$_2$ due to the higher B content should exhibit a higher hardness compared to TiB. Possibly, there is a connection to stresses in the coatings given the much higher deposition rate of 6.3 μm/min applied during deposition. The trend for ZrB$_2$ can probably be explained by the change in film orientation from 0001 to $10\bar{1}0$, which is further discussed in section 5.3 and supported from nanoindentation measurements on sputtered ZrB$_2$ and TiB$_2$ films.

In 1997, Zergioti and Haidemenopoulos [166] reported PLD of a TiB$_2$ source for growth of TiB$_2$ films on Si(001) at 600 °C. Analysis of an ~50 nm thick film by transmission electron microscopy (TEM) at the zone axis <110> showed that the phase TiB$_2$ was deposited as columns between 10 to 50 nm and with an epitaxial relationship to the substrate *i.e.*, similar to epitaxial ZrB$_2$ films deposited by DCMS, see section 4.1.

The study in [166] on epitaxial growth of TiB$_2$ was followed by Zhai *et al.* [167] demonstrating epitaxial TiB$_2$ growth on Al$_2$O$_3$(0001) and SiC(0001) at 600 °C and Ferrando *et al.* [168] with epitaxial growth of TiB$_2$ on Al$_2$O$_3$(0001) at 720 °C. Later epitaxial growth of ZrB$_2$ and ScB$_2$ films were demonstrated on Al$_2$O$_3$(0001), SiC(0001), and Si(111), albeit at slightly higher substrate temperatures of 900 and 950 °C. [169] The primary motivation for epitaxial growth of TMB$_2$ by PLD thus far has been on deposit thin buffer layers < 10 nm to serve as templates for growth of superconducting MgB$_2$ layers given the higher stability of TMB$_2$ compared to MgB$_2$ at elevated temperatures. While PLD appears to be a promising technique for epitaxial growth of TMB$_2$, difficulties with upscaling and low deposition rates as well as being an expensive technique limits PLD in growth of TMB$_2$ coatings under industrial conditions.

### 4.4    Reactive sputter deposition

Growth experiments of thin-film borides by sputtering methods is *legio* and has predominantly been conducted by non-reactive sputtering and using compound targets, see section 4.5. There are several difficulties in sputter deposition of TMB$_2$ films using compound sources that are presented below in sections 4.5 and 5.1-5.6. Comparably fewer studies exist on reactive sputtering of TiB$_2$ that makes it possible to deposit films with different compositions. In fact, there are only two studies from 1989 using reactive sputtering of TiB$_2$, using B$_2$H$_6$ diluted to 6% in argon and rf diode sputtering of a Ti target by Larsson *et al.* [170] and Blom *et al.* [171]. For films deposited without external heating on vitreous carbon, oxidized silicon, and Si(001) substrates, Auger electron spectroscopy and Rutherford backscattering spectroscopy (RBS) (carbon substrates) showed that films with a 2:1 B to Ti ratio can be reactively sputtered. XRD patterns recorded from films deposited on oxidized silicon, and Si(001) substrates showed peaks from TiB$_2$ for samples heat treated for 30 min in a vacuum furnace between 400 to 900 ºC [171]. From their experimental set-up with rf diode sputtering, the authors reported substantial target poisoning, resulting in sputtering of B at low rates [170]. However, the same group showed that there is a simple solution; known as *bias sputtering* [172].

Although promising, diborane is problematic as precursor since it is both explosive and highly toxic. To handle this gas in a laboratory or an industrial environment requires rigorous safety requirements. Such routines are available in the semiconductor industry as diborane is the preferred B-source for doping Si, but comes at a high cost. Thus, it seems challenging to integrate diborane in an industrial sputtering process, but not impossible. In addition to diborane, halide BF$_3$ has been investigated for reactive sputtering of a TiB$_2$ target by rf sputtering [173]. Coatings deposited on NaCl and Si substrates exhibited a 2B:1Ti ratio, and had to be stored in a desiccator to avoid a chemical reaction with the moisture in atmosphere, otherwise resulting in etching by HF on the surface of the coating. Combined PVD deposition methods by growth of individual targets RF sputtering on B targets and DC sputtering of metal has also been shown to be effective, as discussed later.

To summarize, reactive sputtering is a promising tool for growth of compounds such as TMB$_2$. In addition to B$_2$H$_6$ there are only a few studies on the use of precursors in the sputtering process. Pentaborane (9) (B$_5$H$_9$) and decaborane (14) (B$_{10}$H$_{14}$) should be attempted as they are less poisonous than diborane, see further discussion in section 7, Outlook.

### 4.5    Sputtering from compound sources

Growth from compound sources, sintered TMB$_2$ bodies as well as TM and B compacts, is the predominant synthesis route for sputter deposited TMB$_2$ films with C32 crystal structure as initiated by Wheeler and Brainard [43] and summarized by Mitterer [44], but with a few studies where TMB$_2$ film growth have been studied by growth from elemental sources as discussed in this section. A survey of the literature shows that all Group 4-6 TMB$_2$ have been synthesized as by this technique. The most investigated TiB$_2$, ZrB$_2$, and CrB$_2$ materials systems are treated as historical markers for research on sputter deposited TMB$_2$ films, see also sections 5.3 and 5.4. In this section, we discuss important work in the field with a focus on the less investigated materials systems HfB$_2$, VB$_2$, NbB$_2$, TaB$_2$, MoB$_2$, and WB$_2$.

In 1978, Wheeler and Brainard [43] investigated 13.56 MHz rf diode sputtering of CrB$_2$, MoSi$_2$, Mo$_2$C, TiC, and MoS$_2$ hot-pressed and disk-shaped compacts, 15.2 cm in diameter. For CrB$_2$ films deposited at a sputtering power of 600 W and a bias voltage of -500 V, and without external heating, no data on the structural properties was provided, while XPS analysis of the Cr $2p$ and $B_{1s}$ peaks revealed large amounts of O in the form of $Cr_2O_3$ and $B_2O_3$, respectively. [43] The authors concluded that the origin for the O was the target material as noted in their publication: *"No other target shows as great a propensity to outgas as did the $CrB_2$"*. The properties of compound targets applied for sputtering of Group 4-6 $TMB_2$ films will be further discussed in section 5.1. In 1981, Padmanabhan and Sørensen [173] applied rf sputtering of 13.6 MHz of a $TiB_2$ compound target (99.95%) at a sputtering power of 500 W for growth of films on silicon substrates. No information from XRD or selected area electron diffraction (SAED) patterns on the structural properties of the films was communicated, but where TEM images showed the films to be relatively smooth and RBS measurements gave a B-deficient composition in the films with a Ti/B stoichiometric ratio of 0.66, *i.e.*, B/Ti ≈ 1.5. In 1981, Yoshizawa *et al.* [174] published on lattice constant measurement of 5 to 6 μm thick $TiB_2$ coatings sputtered on Mo, Cu, graphite, Ti-coated Mo, and Al-coated Mo substrates at a substrate temperature of about 200 °C, using dc cylindrical post-cathode magnetron sputtering sources. From SAED patterns, the authors concluded that polycrystalline $TiB_2$ coatings with C32 crystal structure had been deposited and where TEM and XRD showed ~5 nm fine grain sizes. In 1983 Shappirio and Finnegan [175] applied rf diode sputtering of 5 inch in diameter $TiB_2$ and $ZrB_2$ hot-pressed commercial compound targets with measured densities of 80% and 92% of the bulk values, respectively. $TiB_2$ and $ZrB_2$ films up to 800 nm thick were deposited without external substrate heating for growth on Si, thermal $SiO_2$ on Si and, Be substrates for interconnect metallization. An XRD pattern recorded from a $ZrB_2$ film revealed broad $10\bar{1}1$, $10\bar{1}0$, and $10\bar{1}2$ peaks of low intensities and where the peak distribution indicated randomly oriented films. Growth of nanocrystalline films were determined from TEM micrographs showing an average grain size of 20 nm *i.e.*, slightly larger $TiB_2$ grains than previously observed by Yoshizawa *et al.* [174] Furthermore, quantitative analysis with AES showed a B-deficient composition for both $TMB_2$ and with 6 wt.% O in $ZrB_2$ and 12 wt.% O in $TiB_2$, and where the O content was supported, but not the B/TM ratio from RBS of the films deposited on Be substrates.

Shappirio *et al.* [175] [176] evaluated the resistivities of the $TiB_2$ and $ZrB_2$ films with measured values of 500 μΩcm and 250 μΩcm, respectively, for as-deposited films. However, the influence from the substrate on the resistivity was unclear. To decrease resistivity values in the $TiB_2$ and $ZrB_2$ films, annealing by halogen lamp (rapid anneal) was performed at 1050 °C with the samples heated in flowing argon. The treatment was successful since the resistivity values were reduced to 150 μΩ·cm for $TiB_2$ and 75 μΩ·cm for $ZrB_2$. Recrystallization of $ZrB_2$ in the films was revealed by x-ray diffractograms with sharp $10\bar{1}0$, $10\bar{1}1$, and $10\bar{1}2$ peaks from $ZrB_2$ by Shapiro and Finnegan [175]. Similar to Wheeler and Brainard [43], Shappirio *et al.* [175] [176] attributed the impurity and density properties of the target material to the less favorable electrical properties of the deposited $TiB_2$ films compared to $ZrB_2$ films, where the $TiB_2$ sputtering source is said to exhibit a higher porosity that precludes elimination of contaminants even after extensive pre-sputter target cleaning [175]. The somewhat precarious condition of available $TiB_2$ targets has steered studies on $TMB_2$ as contact materials on semiconductors towards $ZrB_2$. Porous targets imply trapped gas and crystalline oxides in pure Zr sputtering target, and in the case of contaminated targets, the base pressure is less important for the sputtering process. Contaminants such as $H_2O$ (97%) in the residual gas play a role predominantly at low temperature growth without external heating, while at higher temperature the $H_2O$ and other contaminants, C and O, desorb from the growth surface. In several applications of low-temperature growth, the films strive to become two-phase systems. The use of borides as contact materials is elaborated on in section 5.4 in relation to the resistivities of TMB$_2$ films. However, the high sensitivity of a boride growth surface to O and OH was already observed by Mitterer [44] for growth of TiB$_2$ and ZrB$_2$ films. Target contamination is further developed in section 5.2.

In 1997, Mitterer [44] summarized work on depositing TiB$_2$ and ZrB$_2$ films with C32 crystal structure by sputtering in terms of stability and orientation, crystallinity as a function temperature as further developed in section 5.2. Recently, epitaxial growth of ZrB$_2$ has been demonstrated on Si(111) [156], 4H-SiC(0001) [156] [157], Al$_2$O$_3$(0001) [158], and Si(001) [159] with the substrates held at 900 °C. The orientational relationships are ZrB$_2$(0001)//Si(111) out-of-plane with two in-plane domains: ZrB$_2$[11$\bar{2}$0]//Si[10$\bar{1}$] for the majority orientation and ZrB$_2$[11$\bar{2}$0]//Si[112] for the minority orientation [156]. ZrB$_2$(0001)//4H-SiC(0001) films have the ZrB$_2$[1$\bar{1}$00]//4H-SiC[1$\bar{1}$00] in-plane relationship [157]. For ZrB$_2$(0001)//Al$_2$O$_3$(0001) layers, the two equally probable in-plane domains: ZrB$_2$[10$\bar{1}$0]//Al$_2$O$_3$[10$\bar{1}$0] and ZrB$_2$[11$\bar{2}$0]//Al$_2$O$_3$[10$\bar{1}$0] were observed [157]. For growth on Si(001) there are two in-plane orientations where ZrB$_2$[10$\bar{1}$0] grows out-of-plane.

Beyond TiB$_2$ and ZrB$_2$, CrB$_2$ has attracted interest as a thin film material, due to good corrosion resistance from the metal Cr. The previously described work by Wheeler *et al.* [43] published in 1978 reveled rf sputtered films with high content of O as determined from XPS and the O-B and O-Cr bonds present in the films. In two publications from 1998, Oder *et al.* [177] [178] used dc sputtering of a CrB$_2$ target with unspecified properties in a deposition system held at a base pressure 7-9 x 10$^{-7}$ Torr to deposit 100-200 nm films as Ohmic contacts. Similar to Wheeler and Brainard [43] no data on the structural properties of the films were communicated, but with a disturbing O contribution observed in the RBS data for as-deposited films.

In 2006, Audronis [179] showed the effect of pulsed magnetron sputtering on the structure and mechanical properties of CrB$_2$ coatings. The substrate temperature was in the range of 110–150 °C during the deposition and since CrB$_2$ is a *line-compound*, the microstructure is probably sensitive to minor deviations from the stoichiometric composition. TEM analysis revealed two very different types of microstructures in the coatings; coatings with negatively biased substrate were significantly affected by high-energy ion bombardment, while coatings were not affected as the substrate was allowed to float; coatings with neagative bias contained single crystal nanocolumns of CrB$_2$ with (0001) planes oriented along the growth direction with random rotation along the axis. On the contrary, coatings with floating bias had a randomly oriented polyrystalline microstructure with 3-5 nm crystal size with a mixture of grain shapes with a majority of different columnar shapes. Recently, Dorri *et al.* [180] achieved epitaxial growth of close-to-stoichiometric CrB$_2$ films sputter-deposited from a CrB$_2$ target onto Al$_2$O$_3$ with B-rich inclusions at overstoichometric composition.

Following the observations by Wheeler and Brainard [43] on the porosity properties of Cr targets seems to have steered most of the early work on CrB$_2$ to pulsed sputtering techniques. Pulsed deposition techniques such as HiPiMS are further developed in section 4.6. As a complement to Ti and Zr, Oder *et al.* [177] studied CrB$_2$ using DCMS as discussed in section 5.3.

HfB$_2$ is less studied for thin film TMB$_2$ in Group 4. In 1988, Lee *et al.* [181] deposited HfB$_2$ films by rf magnetron co-sputtering from Hf and B targets as well as rf diode sputtering from a pressed powder HfB$_2$ target on glass, Si and fused quartz substrates without external heating. The purpose of their study was to investigate resistivities of the deposited coatings with different boron and oxygen contents, see section 5.3 for details. From SAED patterns, the authors concluded that HfB$_x$ films of C32 crystal structure were deposited, but with diffuse rings characteristic of a film with fine-grained microstructure. The O content in the films varied depending on if O$_2$ was added to the growth flux or not and with a lowest value of ~0.4% O as determined by SIMS. Even such small O content increases the resistivity of the film and results in a lower the B/Hf ratio. In 1994 Herr *et al.* [182] studied the mechanical properties of thick 7.1 to 10.8 μm HfB$_2$ coatings deposited by rf sputtering of a HfB$_2$ compound target (of unspecified purity) on steel substrates pre-seeded with a 1 μm Ti nucleation layer. Growth was carried out without external heating and varying the rf bias between -40, -30, -20, 10, 0, and +20 V and at pressures of 0.5, 1.0, 1.5, and 2.0 Pa. GIXRD patterns showed 0001-oriented HfB$_2$ with C32 crystal structure for all conditions and with the highest intensities recorded for the 001 peak at -40 and +20 V bias. SEM images revealed a glasslike microstructure for the sputtered HfB$_2$ films, but with no information on the B/Hf ratio and the level of contaminants in the films.

In a series of publications starting from 2009, Ukrainian researchers [183] [184] [185] [186] [187] have deposited HfB$_2$ films with C32 crystal structure from rf magnetron sputtering of hot-sintered powder HfB$_2$ targets in an Ar atmosphere. Growth of HfB$_2$ films in the C32 structure was reported. [184] [186] Dub *et al.* [183] and Goncharov *et al.* [184] [185] investigated their mechanical properties, and Agulov *et al.* [187] examined the thermal stability of HfB$_2$. From the stability criteria, atomic size and electronegativity, HfB$_2$ is the most stable diboride as discussed in section 2.3 and 2.4.

Compared to HfB$_2$, VB$_2$ is even less investigated as a thin film TMB$_2$. In 1988, Kolawa *et al.* [188] applied rf sputtering of a B target covered to ~20% with narrow V strips to deposit 100 nm thick VB$_x$ films on Si and oxidized Si substrates as a potential diffusion barrier for Al in integrated circuits. During growth, the forward sputtering power was varied between 80 to 400 W to study films of different composition. RBS measurements showed VB$_x$ with increasing B content at increasing forward sputtering powers, ranging from $x$ = 1.5±0.3 at 80 W to $x$ = 2.7±0.3 at 400 W, but with no information on the level of contaminants in the films. SAED patterns confirmed the VB$_2$ phase with C32 crystal structure for all investigated compositions and where TEM micrographs revealed an average grain size of 2 nm for a VB$_{2.0}$ film that decreased further at increasing B content to form an almost amorphous film. In 1997, Martin *et al.* [189] characterized the optical and electrical properties of 0.3 μm thick TiB$_2$, ZrB$_2$, VB$_2$, NbB$_2$, and TaB$_2$ films deposited by dc sputtering from compound sources on glass substrates without external heating. The authors provided limited information on the composition and structural properties of the deposited TMB$_2$ films, but with an important notation on the rapid oxidation of the VB$_2$ films what prevented four-point van der Pauw resistivity measurements.

In 2001, Ignatenko *et al.* [190], investigated rf sputtering of a VB$_2$ compound target on NaCl(001), Si(111), and glass substrates. X-ray diffraction and SAED patterns showed that 0.2 till 3 μm thick strongly 0001-textured VB$_2$ coatings with C32 crystal structure were formed in a wide range of sputtering powers 150-600 W, substrate temperatures 140-420 °C, Ar pressures 0.09 to 0.37 Pa, and substrate bias voltages 0 to -65 V. In addition to VB$_2$, the VB phase was frequently deposited as a minority phase and with competing crystalline V$_2$O$_5$, V$_2$O$_3$, and B$_2$O$_3$ phases grown predominantly at sputtering powers above 360 W. To deposit oxide-free vanadium diboride films, Ignatenko *et al.* [190] suggested a combination of *low* sputtering power, low bias voltage, and improved vacuum conditions in the sputtering chamber, while on the other hand, high structural perfection of the films should gain from *high* sputtering powers and high substrate bias as well as higher substrate temperatures. In a study from 2005 Goncharov *et al.* [191], confirmed the phase distribution of rf sputtered VB$_2$ films previously communicated by Ignatenko *et al.*, [190]. In addition, results from TEM on the microstructure of the films were presented with nanocrystalline growth of grains 20 to 50 nm in size and where increasing the substrate temperature to 250-300 °C increased the grain size to 100 nm and above. SIMS and mass spectrum measurements provided further evidence for the formation of oxides in the deposited $VB_2$ films. Thus, the results from several studies presented above leads us to conclude that the high affinity of V towards O is problematic when depositing $VB_2$ films with well-defined properties by rf sputtering.

When moving from V to Nb the number of studies reduces further. As mentioned in a publication from 1997, Martin *et al.* [189] sputtered $NbB_2$ films, but with no information on their phase distribution. In 2014, Nedfors *et al.* [192] [193] deposited $NbB_{2-x}$ films by DCMS in a UHV system from a 50 mm in diameter $NbB_2$ compound target (99.5%) on Si(001), $Al_2O_3$ Ni-plated bronze, and polished 316L stainless steel substrates precoated with a ~50 nm Nb/NbC adhesion layer. At a constant sputtering current of 150 mA and a target-to-substrate distance of 15 cm growth was carried out at a substrate temperature of 300 °C and an Ar pressure of 0.4 Pa. X-ray diffractograms, θ/2θ and GI displayed clear peaks, in the GI diffractogram $10\bar{1}1$, 0001, 0002, $10\bar{1}2$, and $10\bar{1}0$ peaks, from weakly textured $NbB_2$ with C32 crystal structure. This result was supported from SAED patterns with discernable 0001, $10\bar{1}0$, $10\bar{1}1$, 0002, and $11\bar{2}0$ rings. The TEM micrograph showed that the $NbB_{2-x}$ film consisted of 5-10 nm columnar grains elongated in the growth direction. Quantitative analysis by ERDA revealed a composition of 35 at.% Nb, 63 at.% B and < 2 at.% C, corresponding to $NbB_{1.8}$, and where a homogeneity range for $NbB_2$ with C32 crystal structure agrees with our previous discussion in section 2.4 and with Lundström. [80]

In 2003, Lin and Lee [194], rf sputtered a 7.5 cm in diameter $TaB_2$ target at powers in the range 100-300 W, in a HV deposition chamber for growth films on Si(001) wafers without external heating, an Ar pressure of 5 mTorr, and at a substrate to target distance of 7.5 cm. The glancing angle x-ray diffractogram recorded showed clear 0001 and $10\bar{1}1$ peaks from $TaB_2$ with C32 crystal structure as well as broader $10\bar{1}0$, 0002, $11\bar{2}0$, $11\bar{2}1$, and $10\bar{1}2$ peaks of low intensities. The SAED pattern supported the results from diffraction and the TEM micrograph revealed an average grain-size of about 12 nm. Lin and Lee provided no information on the B/Ta ratio and the level of contaminants in the deposited $TaB_2$ films. In 2006, Goncharov *et al.* [195] applied rf sputtering of a sintered powdered $TaB_2$ target with a diameter of 120 mm to deposit films at power in the range 300-600 W on steel and microcrystalline Si (Sitall) at a substrate temperature in the range 80-120 °C, a sputtering pressure of 0.32 Pa and at a substrate to target distance of 110 mm. From x-ray diffractograms the authors showed that 0001-oriented to more polycrystalline $TaB_2$ films could be deposited. TEM micrograph and SAED pattern showed that the coatings deposited were nanocrystalline with grain-sizes in the region 5-30 nm and with the largest grains at the highest applied sputtering power of 600 W.

From the B-Mo phase diagram and the discussion in section 2.4, it is evident that $MoB_2$ with C32 crystal structure is only thermodynamically stable at temperatures above 1500 °C. This fact makes the phase difficult to synthesize as bulk material, but where PVD techniques such as sputtering that operates far from thermodynamically equilibrium could favor growth of metastable phases as $MoB_2$ with C32 crystal structure. In 2005, Dearnley *et al.* [196] applied unbalanced magnetron sputtering of a hot-pressed $Mo_2B_5$ target with a diameter of 100 mm to deposit coatings on commercially produced "straight grade" cemented carbide inserts at a maximum target current of 1 A. The θ/2θ diffractogram recorded from the ~6-8 μm thick coating showed $Mo_2B_5$ $10\bar{1}1$, $10\bar{1}4$, and $10\bar{1}5$ peaks that were broad and of low intensities and with no evidence on the formation of a $MoB_2$ phase with C32 crystal structure. Quantitative analysis by glow discharge optical emission spectroscopy revealed the coating to approach the stoichiometric ratios for B and Mo in $Mo_2B_5$ and with no information provided on the level of contaminants in the coating. A SEM investigation showed a smooth and featureless fractured surface for the $Mo_2B_5$ coating.

In 2016, Malinovskis *et al.* [197] investigated DCMS of a 33/67% Mo/B target of 99.9% purity for growth on preheated (300 °C) and $Ar^+$ etched Si(001), $Al_2O_3$(001), and 316L SS substrates with a Mo seed layer. The films were deposited under UHV condition at 300 °C and grown to a thickness of about 700 nm. X-ray θ/2θ and glancing incidence diffractograms revealed $10\bar{1}1$, 0001, 0002, and $11\bar{2}0$ peaks originating from $MoB_2$ with a C32 crystal structure. SAED patterns support the growth of $MoB_2$ as a thin film material and where HRTEM images show a columnar microstructure with a column width of 10-20 nm and with elongated $MoB_2$ grains with an average grain size of 10 nm. From the study by Malinovskis *et al.* [197], we note the important observation on the properties of the target material, where elastic recoil detection analysis (ERDA) showed substantial amounts of C with 4 at.% and O with 14 at.% and with a B/Mo ratio of ~1.5. This high level of C and O in the target was reduced in the films and determined by ERDA to 1 at.% and < 2 at.%, respectively. The B/Mo ratio was measured to $MoB_{1.65}$ (62 at.% B in the film), *i.e.*, a substoichiometric $MoB_2$ phase. This agrees with the discussion in section 2.4 and the study by Klesnar *et al.* [81], that determined a narrow homogeneity range centered at 61 at.% B in the temperature range 1600-1800 °C.

$WB_2$ with C32 crystal structure was prepared by Woods *et al.* in 1966 [82], by heating a boron wire in an atmosphere of $WCl_6$ and Ar 800 °C during 30 min. In 2006, Sobol [83] applied planar dc magnetron sputtering of a stoichiometric $W_2B_5$ target in a HV system to study growth on preheated (950 °C) Si(111) substrates and polycrystalline Ta sheets positioned at 55 mm directly above the target and using substrate temperatures of 300, 500, 700, and 900 °C. The coatings were grown to a thickness in the range 4-8 μm and some coatings were annealed in a vacuum system at 900-1230 °C for 40-60 min under a residual gas pressure lower than $7 \times 10^{-4}$ Pa. From recorded XRD patterns, Sobol [83] identified the $10\bar{1}1$ peak from $WB_2$ with C32 crystal structure in coatings deposited on both applied substrates at a temperature of 300 °C, while the film deposited at 500 °C was found to be amorphous and with β-WB (CrB-type structure, see section 2.4) $11\bar{2}1$ clearly visible in the coating deposited at 900 °C. Annealing for 1 h at 900 °C of the coating containing the $WB_2$ phase, improved the crystal quality of the film seen from detectable, but still broad 0001, $10\bar{1}0$, $10\bar{1}1$, $11\bar{2}0$, $11\bar{2}1$, and $10\bar{1}2$ peaks of low intensities from $WB_2$ with C32 crystal structure *i.e.*, similar to previously observed by Shappirio and Finnegan [175] for $ZrB_2$ and $TiB_2$ films. Sobol [83] used SIMS to determine the B/W ratio in coatings deposited at 300, 500, 700, and 900 °C with the trend of decreasing B content with increasing deposition temperature in $WB_{2.17}$, $WB_{2.09}$, $WB_{1.69}$, and $WB_{1.38}$, respectively. However, no information on the level of contamination in the coating was provided. In 2013, Jiang *et al.* [198], confirmed the formation of thin film $WB_2$ with C32 crystal structure previously reported by Sobol. [83] From a planar rectangular (270 mm x 70 mm x 70 mm) $WB_2$ ($WB_2$-type structure) target with a purity of 99.5%, about 2.8 μm thick coatings were deposited in a HV system on 304 SS substrates by DCMS at sputtering power of 195 W, and pressure 0.3 Pa, with substrate bias -150 V and temperature 680 °C at a target-to-substrate distance of 50 mm. A θ/2θ diffractogram showed polycrystalline $WB_2$, and where composition analysis by electron probe x-ray microanalyzer gave 35 at.% W and 65 at.% (W/B 1:1.9), again with no information on the level of contamination in the coating. From HREM and XRD, the authors [198] report a coating with a spindle-like grain morphology over 20-60 nm.

Growth of Group 4-6 TMB$_2$ films with well-defined properties are paramount for existing and envisioned applications as well as for determining the fundamental properties of TMB$_2$. In the upcoming section, we will discuss the work conducted to determine the properties for Group 4-6 TMB$_2$ films. We also anticipate the hybrid DCMS and HiPiMS techniques as discussed in section 7.

## 4.6 Pulsed sputter deposition

Sputter deposition of TMB$_2$ thin films was initiated by using rf techniques. As previously described, poorly conducting targets were used, which probably was a consequence of improperly sintered and thus porous target quality at that time. Typical density values of the compound targets were between 70 and 85% of the theoretical densities. Following the pioneering studies by Mitterer *et al.* [44], rf sputtering became replaced by DCMS or other pulsed techniques by bipolar magnetron sputtering. For example, Audronis *et al.* [179] used blended powders of Cr and amorphous B powder mixed at an atomic ratio of 1:2 without any further compacting after the mixing the blend. To achieve the desired stoichiometry, additional pieces of solid boron were placed around the racetrack region.

Following the demonstration by Kouznetsov *et al.* in 1999 [199], the sputtering technique HiPiMS has received increased attention including growth of TMB$_2$ films. In 2005, Sevvana *et al.* [200] studied the effect of the coil different configurations, coil current (2 or 5 A) and a negative substrate bias (-100 V or unbiased) of CrB$_2$ using HiPiMS. The surface roughness in terms of surface topography was smooth with a hardness of 25-30 GPa. The hardness of the films deposited by HiPiMS was higher than for comparable DCMS films. As discussed, these values are typical for sputter deposited thin films. HiPiMS combined with a secondary coil and a negative substrate bias was found to enhance the deposition rates.

Figure 6 shows cross-sectional SEM micrographs with characteristic fine-grained (glass-like) microstructure of ZrB$_2$ from HiPIMS coatings (panels (a) to (c)), compared to a DCMS coating (panel d) grown under comparable conditions. As observed, all the HiPiMS coatings appear dense and have a smooth surface, that is an appearance often reported for HiPIMS films [201]. For the highest frequencies used in the study (700 and 900 Hz) [201], a narrow, and columnar appearance can be discerned. This observation is consistent with comparatively lower ion ratio in the deposition flux and higher deposition rate, effectively reducing adatom mobility compared to the lower frequencies used. Although differences in morphology are seen between the HiPIMS coatings as the frequency increases, these are minor when put in relation to a typical DCMS ZrB$_2$ coating grown under similar conditions, that exhibits a pronounced columnar appearance. [201] A common thread since the earliest studies performed especially at low coating temperatures, is that fine-grained coatings are produced with high O levels.

Furthermore, it has been shown that problems from excess of B can be resolved by adjusting the pulse lengths in HiPiMS, while maintaining the average power and frequency of the pulses. The film growth then becomes increasingly controlled by the ionized target atoms instead of neutral species. As the sputter ejected TM atoms have a higher ionization probability than the B atoms, the B/TM ratio decreases in the film as a function of target peak current. For example, Thörnberg *et al.* reported understoichiometric TiB$_x$ by HiPiMS [202].

Recently, there has been interesting developments in hybrid HiPiMS and dc magnetron co-sputtering [203] [215], where it has been demonstrated that Zr$_{1-x}$Ta$_x$B$_x$ thin films with x ≥ 0.2 are both hard and relatively ductile. The films showed a self-organized columnar core/shell nanostructure with crystalline hexagonal Zr-rich stoichiometric $Zr_{1-x}Ta_xB_2$ cores were surrounded by narrow dense and disordered Ta-rich shells that are B-deficient. These nanostructures combine the benefits of crystalline diboride nanocolumns that give rise to high hardness with dense metallic-glass-like shells that may promote toughness.

### 4.7 Solid solution pseudo-binary alloys and ternary borides

Phase diagrams of ternary diborides are scarce. Reports include the ternary Mo-B-C system [204] containing the binary phases: $Mo_2C$, $Mo_2B$, $MoB$, $MoB_2$, $Mo_{0.8}B_3$ ($MoB_4$), $B_{13}C_2$, and one ternary phase: $Mo_2BC$ as well as the Mo-Fe-B system [205] containing the binary phases: $FeB$, $Fe_2B$, $Mo_2B$, $MoB$, $MoB_2$, $Mo_{5.1}Fe_{7.9}$, and three ternary phases: $MoFe_2B_4$, $Mo_2FeB_2$, and $Mo_8FeB_1$. From the kinetic limitations imposed in non-equilibrium processes such as PVD, solid solutions may be expected rather than separate binary or even ternary phases. In 2015, Euchner and Mayrhofer [33] theoretically studied supersaturated solid solutions of diborides based on binary constituents that crystallize in different modifications. On the exemplary systems $Al_xW_{1-x}B_2$, $Ti_xW_{1-x}B_2$, and $V_xW_{1-x}B_2$, they predicted by DFT that these ternaries represent a new class of metastable materials, which may offer improved hardness and ductility that triggered experimental investigations of solid-solution hardening of vacancy-stabilized $Ti_xW_{1-x}B_2$ films [206].

It is reiterated that the early transition metal diborides ($TiB_2$, $ZrB_2$, etc.) are chemically more stable in the $AlB_2$ structure and the late transition metal diborides (like $WB_2$) are preferably stabilized in more complicated structures such as the $W_2B_{5-x}$. For the design of thin film materials with desired properties in applications, additional degrees of freedom are offered with ternary borides, in particular pseudo-binary systems.

Alling *et al.* [32] reported a first-principles scan of the mixing thermodynamics of all 45 alloy systems formed by all $M^1_{1-x}M^2_xB_2$ combinations of the ten binary diborides $MgB_2$, $AlB_2$, $ScB_2$, $YB_2$, $TiB_2$, $ZrB_2$, $HfB_2$, $VB_2$, $NbB_2$, and $TaB_2$, all reported to crystallize in the $AlB_2$ type structure. Several metastable solid-solution alloys, in particular $Al_{1-x}Ti_xB_2$, were identified as candidates for age-hardening due to isostructural decomposition tendency with limited lattice mismatch. A related example is $Ti_{1-x}Al_xN$ - that is the archetype in hard coating applications using age hardening. We predict that ternary diborides could mimic the TiAlN system, while in the C32 $WB_2$ structure with, *e.g.*, Ta and Ti.

Materials properties and functionality can often be improved by synthesis of multicomponent systems, *e.g.*, tuning the properties by alloying to change the microstructure. Löfvander *et al.* [207] investigated alloyed Ti-Al-B films deposited by magnetron sputtering and showed improved hardness and microstructural stability associated with precipitation of nano-scale TiB disperoids during high-temperature annealing.

Mockute *et al.* [208] [209] investigated the effect of alloying $TiB_2$ with $AlB_2$ using dc magnetron sputtering. From as-deposited $(Ti_{1-x}Al_x)B_{2+\Delta}$ solid solutions they found evidence of phase separation into $TiB_2$ and $AlB_2$ accompanied by an increase in hardness upon post annealing of the thin films. Other examples of microstructural design for solid solution hardening or age hardening of the former by phase separation into the two respective metal diborides: $VB_2$ and $WB_2$ as predicted by Moraes [31].

M. Stüber et al. [210] studied the effect of alloying on the microstructure of Al-containing magnetron sputtered TiB$_2$ thin films. Films with relatively low Al concentration (Ti$_{0.90}$Al$_{0.10}$B$_{2.22}$) and moderate deviation from the diboride stoichiometry showed high hardness due to excess B at grain boundaries and possible tissue phases. On the other hand, films with relatively high Al concentration and with only a small deviation from the diboride stoichiometry did not show any excess B allocation at grain boundaries. In particular, thin films with low Al content exhibited an additional strengthening effect that might have been associated with an oscillation of the metal-concentrations during growth by sample rotation in front of the TiB$_2$ and Al targets.

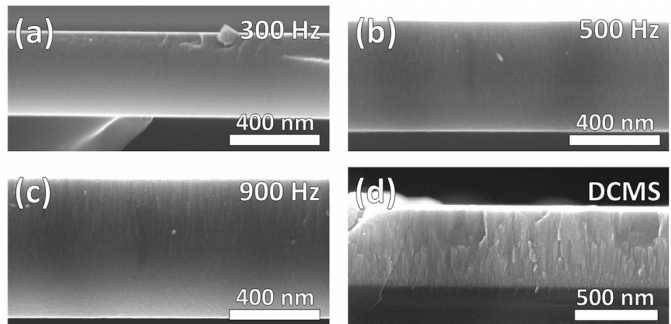

**Figure 6:** Cross sectional SEM micrographs of ZrB$_2$ films grown using HIPIMS with different pulse repetition frequencies (a–c) showing characteristic fine-grained (glass-like) microstructure, as well as for a DCMS film (d) grown under equivalent conditions. [201]

Wen et al. [211] studied solid solutions of (Hf$_{1/3}$Zr$_{1/3}$Ti$_{1/3}$)B$_2$ and (Ta$_{1/3}$Nb$_{1/3}$Ti$_{1/3}$)B$_2$ and found that both systems formed a structure with a nanocolumnar morphology. The investigation showed a promising pathway by combining theoretical analysis to predict and molten salt synthesis to fabricate multi-component solid solutions that has rarely been reported until now.

Recently, a series of studies of Ti- and Zr-based ternary diborides were conducted by Bakhit et al. [212] [213] [214] [215]. A novel strategy for tuning the B/TM ratio and increasing both hardness and toughness was exemplified for pseudobinary Zr$_{1-x}$Ta$_x$B$_y$ alloys using a hybrid HiPIMS/ DCMS sputtering technique [212]. Films synthesized with $0 \leq x \leq 0.1$ resulted in columnar stoichiometric-diboride grains encapsulated in a B-rich tissue phase, while films with $x \geq 0.2$ had a nanocolumnar structure with metal-rich boundaries. While most hard coatings suffer from brittleness, it was shown that Zr$_{1-x}$Ta$_x$B$_y$ thin films with $x \geq 0.2$ were not only hard but also tough. This was also demonstrated in nanostructured Zr$_{1-x}$Ta$_x$B$_y$ core/shell thin films [213]. These core/shell nanostructured films combined the properties of crystalline diboride nanocolumns, providing high hardness, while Ta-rich dense shells were disordered (and prefer a solid solution) similar to metallic glasses [132] [216] that give rise to increased toughness. The mechanical properties are further discussed in section 5.3.

Furthermore, increased high-temperature oxidation resistance by alloying with Al was demonstrated for Ti$_{1-x}$Al$_x$B$_y$ alloy films that exhibited a columnar structure [214], like for the TiAlN system [217]. Column boundaries of overstoichiometric TiB$_{2.4}$ were B-rich, while Ti$_{0.68}$Al$_{0.32}$B$_{1.35}$ alloys had Ti-rich columns surrounded by a Ti$_{1-x}$Al$_x$B$_y$ tissue phase that was predominantly Al rich. The kinetics of the separation into metal and B-rich domains and their respective sizes depends on diffusion and growth parameters such as substrate temperature, applied bias, and sputtering flux. Thus, the B-stoichiometry problem described for the binary metal diborides described above (see Sections 2.3, 2.5 and 4.6) remains for the ternary (pseudo-binary) systems, and becomes exacerbated and more complex to control with more metal elements involved.

In addition to improved oxidation resistance, the alloy films retained the hardness of TiB$_{2.4}$ films with low stresses and high toughness. Hybrid HfB$_2$-HiPIMS/TiB$_2$-DCMS co-sputtring without external heating was also applied to grow dense Ti$_{0.67}$Hf$_{0.33}$B$_{1.7}$ thin films that exhibited a hardness of ~41 GPa [215]. Enhanced oxidation and corrosion resistance was also observed for $Zr_{1-x}Cr_xB_y$ thin films [218] for x ≤ 0.29. However, a further decrease of Cr led to a deficiency of B and resulted in a recrystallization from the $AlB_2$-structure to an amorphous dense alloy that exhibited higher toughness and wear resistance than $ZrB_{2.19}$.

More recently, age hardening was observed in $Zr_{1-x}Ta_xB_y$ solid solution thin films for 0 ≤ x ≤ 0.3 without phase separation or spinodal decomposition [219]. The hardening > 34 GPa up to 1200 °C was explained by point-defect recovery that enhanced the chemical bond density. The absence of precipitation or spinodal decomposition (as observed by XRD, XPS, STEM, EDX, and EELS), although the $ZrB_2$-$TaB_2$ would exhibit a miscibility gap, can be explained by their high melting points (3245 °C for $ZrB_2$ and 3000 °C for $TaB_2$).

## 4.8   MAB - phases and boridene

In 2015, Ade and Hillebrecht [220] proposed inherently nanolaminated pseudobinary borides with TMB slabs separated by A-layers; MAB (M = Mo, W; A=Al, Si, Ge) and $M_2AB_2$ (M = Cr, Mn, Fe; A = Al, Si, Ge) structures, collectively known as MAB-phases [221], with MB layers for stabilization of freestanding two-dimensional (2D) *borophene* sheets, [222] [223] [224] [225] [226] in which single or double atomic A-layers are interleaved. The MAB-phases share many structural similarities with the more familiar $M_{n+1}AX_n$-phases with hexagonal crystal structure (Space group $P6_3/mmc$), in which M is an early transition metal, A is a Group 3A element and X is C or N, and *n* = 1, 2, or 3 [227] [228] [229] [87]. For the Mo-B materials system, we note theoretical studies on synthesizing the $Mo_2BC$ by sputter deposition predicted damage resistance [230] [231] [232].

$M_5SiB_2$ (T2) phases, (where M is a transition metal (like Mo and Fe) and A is an A-group element (Si, Ge, P)), were discovered by Novotny in 1957 ($Mo_5SiB_2$) [233] and Aronsson ($Fe_5SiB_2$ and $Mn_5SiB_2$). [234] T2 is an atomically layered material and crystallizes in the tetragonal $I4/mcm$ symmetry. The finding of a family of laminated quaternary metal borides, $M'_4M''SiB_2$, with out-of-plane chemical order was reported very recently by Dahlqvist *et al.* [235] Eleven chemically-ordered phases as well as 40 solid solutions (evaluated among M = Sc, Y, Ti, Zr, Hf, V, Nb, Ta, Cr, Mo, W, Mn, Fe, and Co), introducing four elements (Sc, Ti, Hf, Cr) previously not observed in these borides are predicted. The predictions are experimentally verified for $Ti_4MoSiB_2$, establishing Ti as part of the T2 boride compositional space.

In 2021, Zhou *et al.* [236] reported the experimental realization of boridene in the form of single-layer 2D molybdenum boride sheets with ordered metal vacancies, $Mo_{4/3}B_{2-x}T_z$ (where $T_z$ is F, O, or OH surface terminations), produced by selective etching of Al and Y or Sc atoms from 3D in-plane chemically ordered $(Mo_{2/3}Y_{1/3})_2AlB_2$ and $(Mo_{2/3}Sc_{1/3})_2AlB_2$ in aqueous hydrofluoric acid. The discovery of a 2D transition metal boride suggests a wealth of future 2D materials that can be obtained through the chemical exfoliation of laminated boride compounds. *Boridene* has also been named *MBene* depending on the context. [237] [238]

## 5  Property and analysis aspects for diboride thin films

### 5.1  Composition

In this section, we focus on the level of contaminants and B/TM ratio in TMB$_2$ films sputter deposited from compound targets. The early studies were concerned mainly with reducing the O and C content in the films, while more recent studies have highlighted the necessity of controlling the B/TM ratio of the sputtered species. We start by considering the properties of the compound target and continue with the gas-phase transport of the sputtered material generated from the source. Aspects on analysis methods are also raised when needed to advance the TMB$_2$ research.

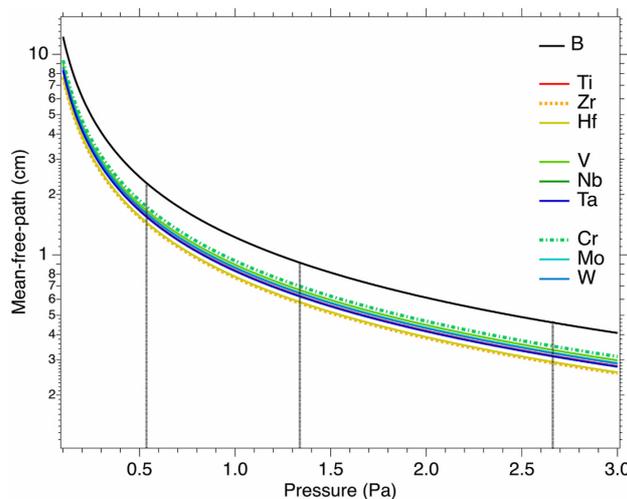

**Figure 7:** Calculated mean-free-paths of B and the group 4-6 transition metals Ti, Zr, Hf, V, Nb, Ta, Cr, Mo, and W in the pressure region 1 - 20 mTorr typically used in magnetron sputtering.

Already in 1978, Wheeler and Brainard [43] concluded that an applied CrB$_2$ compound target for deposition of CrB$_2$ films by rf sputtering was a particularly porous compact (low density). XPS showed distinguished peaks from Cr$_2$O$_3$ and B$_2$O$_3$, see section 4.5. In their work on rf sputtering of TiB$_2$ and ZrB$_2$ films from 1983, Shappirio and Finnegan [175] adhere to the previous observation of Wheeler and Brainard [43], to describe the properties of their rf sputtered TiB$_2$ films. Quantitative AES analysis suggested a B-deficient composition for both TMB$_2$ films and with 6 wt.% O in the ZrB$_2$ films and 12 wt.% O in the TiB$_2$ films, respectively. For the applied TiB$_2$ and ZrB$_2$ hot-pressed commercial compound target materials, the measured densities were 80% and 92% of the bulk values, respectively. For deposited films, Shapiro and Finnegan [175] concluded that the higher O content in their TiB$_2$ films compared to their rf sputtered ZrB$_2$ films could be explained by the lower density of the applied TiB$_2$ compound target compared to the ZrB$_2$ target. In his compilation from 1997, Mitterer [44] highlighted the importance of high-quality boride targets in sputtering. Similar to Shapiro and Finnegan [175] the target density was identified as a limitation in thin film growth, where for instance, measurements of a ZrB$_2$ target with claimed 99.5 % purity, yielded a density of 82.3 % of the theoretical value [239]. A porous target material is likely to adsorb water vapor, oxygen, nitrogen, and hydrocarbons as well as other C- and N-containing gases, thus providing a source for these contaminants to be incorporated in the deposited TMB$_2$ films, if not disturbing its nucleation and growth.

In a publication from 2015, Tengdelius *et al.* [157] applied XPS to quantify the contaminants in a ZrB$_2$ compound target of nominal 99.5% purity excluding Hf. They found high levels of ~8 at.% C and ~19 at.% O contaminations in the target even after 25 min of sputter cleaning, thus casting doubts on the manufacturer's practice for specifications. In 2016 Malinovskis *et al.* [197] noted that most boride target suppliers only give impurity levels on metals and not on *p*-elements such as O, N, and C. From ERDA measurements on a Mo/B (33%/67%) target with stated 99.9% purity, and impurity levels of analyzed elements at ppm level showed that the target contained about 4 at. % C and as much as 14 at. % O *i.e.*, similar to that previously determined from XPS analysis of a ZrB$_2$ compound target by Tengdelius *et al.* [157] Malinovskis [240] assumed that most of this C and O originated from adsorption of water and other oxygen-containing species in pores as they detected H in ERDA experiments. A scanning electron microscopy (SEM) study of the target revealed a porous microstructure, where adsorption of various species can occur. The presence of adsorbed contaminants was supported by significant outgassing in the sputter chamber from a freshly mounted target. Furthermore, the B/Mo ratio of the compound target was determined to ~1.5 *i.e.*, thereby significantly deviating from the defined ideal composition, *x*=2 of the target. In 2018, Magnuson *et al.* found that a ZrB$_2$ target with 99.5% purity contained crystalline monoclinic (m) ZrO$_2$ seen from $\bar{1}11$, 111, and 220 peaks in an x-ray diffractogram [135]. This result is different from that previously reported in 2009 by Goncharov *et. al.* [241], where XRD of hot-sintered (Ar atmosphere) HfB$_2$ target displayed no diffraction peaks from B or hafnium oxides, but where SIMS showed BO and HfO clusters. This is an indication that the driving force for forming crystalline oxides differs between the Group 4-6 TM, perhaps reflecting the affinity to O. Similar to Tengdelius *et al.* [157], XPS analysis by Magnuson *et al.* showed high amount of O (18.6 at.%) and C (7.9 at.%) in the target. With respect to B and Zr content in the resulting films, XPS determined at Zr-rich composition as judged from B/Zr = 1.29. This is consistent with the observations by Malinovskis *et al.* [197] [240] Depending on the surface roughness and the sputtering parameters (energy of the ions and time), it is likely that a very surface-sensitive technique like XPS with a probe depth of ~5 nm measures re-deposited O- and C-containing contaminants on the Ar$^+$ sputtered cleaned surfaces during the time it takes to acquire the XPS spectra. Quantitatively, is does not reflect the O and C content in the bulk of investigated samples but more of the adsorbed species. Quantitative analysis by XPS is not only limited by the high surface sensitivity, but also the large difference in sensitivity factors between different elements and dipole transition matrix probabilities from different orbitals and work functions.

During surface cleaning by Ar$^+$-etching, preferential sputtering and forward implantation are phenomena that cause surface roughening and prevents complete surface cleaning. The surface roughening causes an increased number of broken bonds, where sputtered contaminants and light elements such as O and C can easily adsorb during the acquisition time of the XPS spectra. [242] The problem of oxidation can be overcome by thin (1.5–6.0 nm) Al capping layers *in-situ* in the deposition system prior to air-exposure and loading dense and plane epitaxial films into the XPS instrument. [243] For ZrB$_2$ films, time-of-flight energy elastic recoil detection analysis (TOF-ERDA) with a probe depth of ~30-60 nm yields much lower O content than XPS, with ~1 at.% and even lower C content of ~0.4 at. % in the bulk of the investigated films. [244] The observations above suggest that the accuracy of ERDA analysis is higher than in the case of XPS. Moreover, the surface sensitivity and preferential sputtering of light atoms in XPS complicates the analysis. ERDA measurements and analysis is more complicated and specialized than XPS, but often give more accurate quantification when correctly performed.

From the studies above, it is evident that contaminants present in TMB$_2$ compound targets are released during the sputtering process and become transported together with the growth flux from the target to the substrate and condense into the growing TMB$_2$ film. Malinovski *et al.* [197] found the same trends from analyses with ERDA and XPS revealing that the final C and O contents in the films deposited by DCMS in a UHV chamber at 300 ºC were < 2 at.%. [197] Thus, in sputtering of Group 4-6 TMB$_2$ the trend is that the O content in the films decreases at high temperatures [157] [240], which complicates deposition from compound sources for applications, where low temperature growth is needed such as for semiconductor processing.

Deposition of Group 4-6 TMB$_2$ films from compound targets is further complicated by the difference in atomic masses between the constituents of the target: B with 10.81 and the TM ranging from Ti with 47.867, to W with 183.84. When emitted by the sputtering process, this mass effect results in a different angular distribution with a larger probability of lighter atoms such as boron to be sputtered along the target normal [245] as investigated by Olson *et al.* [246] For TMB$_2$ a significant deviation of the film composition in films containing elements with significant mass difference in the target has previously been observed for WB$_x$ films where, Willer [247] *et al.* in 1990 deposited W-rich films from a WB$_x$ compound target (27 at.% B). In 1997 growth of B-rich films with B/Ti ≈ 2.4 in sputtering of a TiB$_2$ compound target was reported in a publication by Mitterer [44], in 2003 by Kunc *et al.* [248] and in 2008, by Neidhardt *et al.* [249] The gas-phase transport of the sputtered species will further depend on the target-to-substrate distance and the applied pressure as these factors determine the mean-free paths (MFPs) of B and TM atoms. The difference in MFPs between B and the TM atoms during transport will result in different scattering of the atoms, which will affect the B/TM ratio of the sputtered species arriving to the substrate. The consequence of this off-set on the TMB$_2$ microstructure is discussed in the upcoming section.

Figure 7 shows calculated MFPs of B for group 4-6 transition metals Ti, Zr, Hf, V, Nb, Ta, Cr, Mo, and W, as a function of typical pressures applied during magnetron sputtering, *i.e.*, 1 - 20 mTorr. The MFPs decrease for increasing pressure according to equation (1) that describes the pressure of the MFP, where B has longer MFP than the TMs.

$$MFP = \frac{1}{\sqrt{2}} \frac{k_B T}{p} \frac{1}{\pi(r_{Ar}+r_{B,Me})^2} \:. \tag{1}$$

As an example, at typical sputtering pressures of 0.53, 1.33, and 2.67 Pa (4, 10, and 20 mTorr), the MFPs of B (0.88 Å atomic radius) are 2.28, 0.92, and 0.46 cm, respectively. For Cr and Zr, Cr is the smallest Group 4-6 TM atom (1.28 Å) while Zr is the largest atom (1.60 Å), the corresponding MFPs are 1.74, 0.70, 0.35 and 1.44, 0.58, and 0.29, respectively. Thus, the MFPs are shorter for the TM compared to B. As a result, the sputtered flux of B is classified as predominantly 'intermediate', especially at lower pressures while the sputtered TMs are more 'thermalized flow' region to the expense of energy loss of the sputtered material. [249] From Fig. 7, it is evident that scattering of B requires higher pressures and/or longer target-to-substrate distances.

In 2008, Neidhardt *et al.* [249] studied the sputtering process of Ti-B films from $Ti_xB$ (x = 0.5, 1, and 1.6) compound sources from depositions and simulations (by static TRIM emission profiles and dynamic TRIDYN calculations). Neidhardt *et al.* [249] confirmed the difference in scattering between B and Ti resulting in a B-rich composition in particular for depositions conducted along the target normal where a Ti deficiency of up to 20% was detected. From increasing the pressure or distance from 0.5 to 2 Pa (4 to 15 mTorr) and from 5 to 20 cm, respectively, the authors concluded an almost equivalent linear increase in Ti/B ratio surpassing even the target composition. In addition, films with higher Ti/B ratios were grown by off-axis depositions at lower angles 30° and 60°. The studies by Mitterer [44], Kunc *et al.* [248], and by Neidhardt *et al.* [249] show that a B-rich composition is to be expected when sputter depositing $TMB_2$ films from a compound target, but where the study in [249] also shows that the composition of the growth flux can be altered by either increasing the pressure, the target-to-substrate distance or conducting off-axis deposition to better meet a 2B:1TM composition. As discussed in section 4.6, the stoichiometry can be controlled by varying the pulse length in HiPiMS. [203]

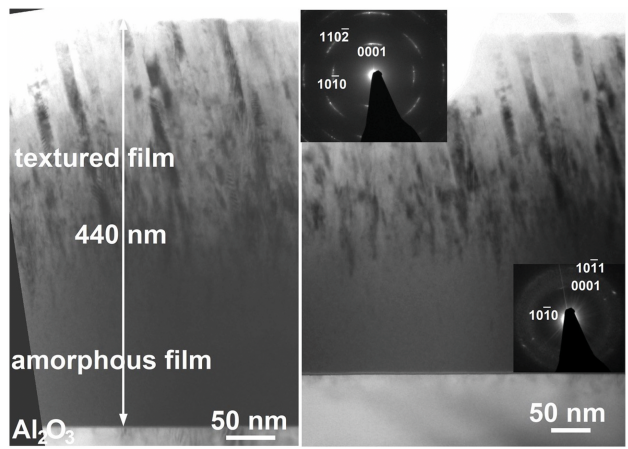

**Figure 8:** Cross-sectional TEM images of $ZrB_2$ grown without external heating (left) on $Al_2O_3$(0001) and (right) on Si(100), including corresponding SAED patterns from the film's top and bottom.

## 5.2 Microstructure

In the previous section, it was concluded that the flux of the sputtered atoms from a compound target typically is B-rich and sometimes with a high level of contaminants. In a review from 1997, Mitterer [44] summarized studies on sputter deposition of $TiB_2$ and $ZrB_2$ films. For the microstructure, he concluded that the strong directionality of the covalent B-B bonds resulted in boride coatings with a pronounced tendency to form extremely fine-grained to amorphous structures at *low substrate temperatures*.

Already in 1981, a TEM study by Yoshisava *et al.* [174] showed that sputtered polycrystalline $TiB_2$ films contain nanocrystallites with ~ 50 Å diameter when deposited on Mo, Cu, graphite, Ti-coated Mo, and Al-coated Mo substrates at 200 °C using dc cylindrical post-cathode magnetron sputtering sources. A survey of the literature shows that the microstructure reported by Yoshisava *et al.* is corroborated by numerous studies on sputter deposition of Group 4-6 $TMB_2$ films.

In 1991 Brandstetter *et al.* [250] derived a *structure diagram* for non-reactively sputtered Zr-B films using a sputtering power of 180 W and Ar pressure of 1.3 Pa and varying substrate temperature of 100-400 °C without external heating and a substrate bias from 0 to -600 V. The Zr-B coatings exhibit an amorphous structure at low substrate temperatures and high bias voltages. It was concluded that the region of crystalline films covered the complete high temperature range and extended to the low temperature and low bias voltage range. Brandstetter *et al.* [250] further concluded that the growth of crystalline films was favored by the higher mobility of the condensed atoms at higher substrate temperatures whereas an increase of bias voltage resulted in an extension of the region of amorphous film growth to *higher* substrate temperatures. In addition to the substrate temperature and bias, Brandstetter *et al.* found that the microstructure was affected by the *heat of the condensing flux*.

In a follow-up study, applying the same process conditions for growth of Zr-B films, Mitterer *et al.* [251] corroborated the observations made by Brandstetter *et al.* [250] on the evolution of the microstructure at increasing substrate temperatures and bias. In addition, the authors added results on the change in film orientations with reports of 0001-oriented $ZrB_2$ films at temperatures of 200 °C and below and bias voltages equals 200 V or lower. For films grown at high substrate temperature or low bias volage, they concluded that such basal-plane orientation in $ZrB_2$ films disappeared. A 0001 orientation in $TMB_2$ films often with a combination of a fine-grained microstructure have been frequently reported for films sputter deposited at low temperatures.

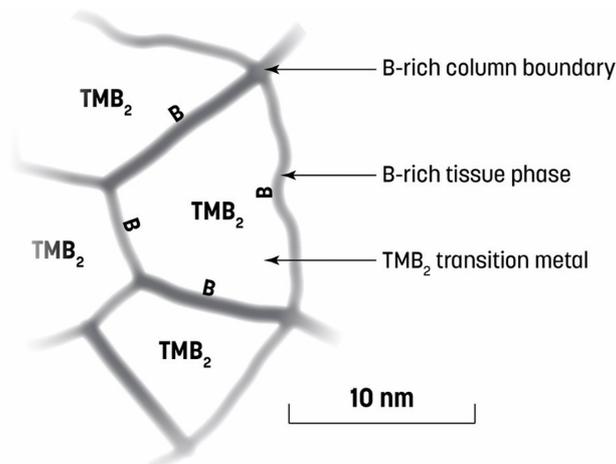

**Figure 9:** Schematic view of B-rich $TMB_2$ films adapted from Mayrhofer *et al.* [253].

The transition from amorphous to crystalline microstructure was observed by Tengdelius *et al.* [244] by the increased temperature of the heat from the condensing flux of the sputtered material without external heating of the substrate. In 2014, Tengdelius *et al.* [244], studied deposition of $ZrB_2$ films using DCMS with a compound target on 4H-SiC(0001) and Si(001) substrates, including diboride growth without external substrate heating.

Figure 8 shows TEM images of $ZrB_2$ that summarizes the results from the film growth by Tengdelius *et al.* [244]. The left panel shows TEM images of a 440 nm thick $ZrB_2$ film grown on $Al_2O_3$(0001) without external heating of the substrate. As observed, the film gradually transformed from amorphous close to the substrate to textured with a columnar structure further up in the film. Similar observations were made from cross-sectional TEM images by Sim *et al.* [252] that revealed that a sputtered $ZrB_2$ film was amorphous to about half of the total film thickness followed by a layer with a fine crystalline columnar structure that had a preferred (0001) crystallographic orientation. For comparison, the right panel shows a $ZrB_2$ film grown on a Si(001) substrate including SAED diffraction patterns with the indicated diffraction rings from the top and bottom of the film. The diffuse diffraction rings close to the substrate show amorphous structure while the elongated spots visible close to the surface indicate a significant 0001 orientation of the columnar structure. The growth process starts out as amorphous and as the deposition process proceed, the energetic flux begins to heat the substrate that enables an increased surface mobility for the film constituents. This condition favors the growth of films with higher crystalline quality and is reflected in the columnar growth behavior of the film at the film-vacuum interface. Thus, the microstructure can be controlled by the temperature of heat from the condensing flux, substrate temperature, and bias while the orientation is controlled by the substrate orientation. Similar observations as those shown in Fig. 8 were made by Sung *et al.* in 2002 applying remote-plasma CVD to produce conformal $ZrB_2$ films at low temperature [151].

In 2005, Mayrhofer *et al.* [253] developed the *structure diagram* reported by Brandstetter *et al.* for sputter-deposited films. Figure 9 schematically shows the growth-microstructure model proposed by the studies of Mayrhofer *et al.* [253] for a 3 μm thick $TiB_{2.4}$ film grown on austenitic stainless steel and Si(001) substrates at 300 °C by magnetically unbalanced magnetron sputter deposition from a stoichiometric $TiB_2$ target with 150 mm in diameter. The $TiB_2$ film composition was measured by wavelength dispersive electron probe microanalyses calibrated using a stoichiometric $TiB_2$ standard whose composition was determined by nuclear reaction analyses. The TEM studies in [253] showed an average column diameter of ~ 20 nm and a smooth surface with an average root-mean-square roughness similar to that of the polished substrate surface of 15 nm. In the structure, B segregated towards the grain boundaries. Mayrhofer's TEM studies further showed that the 20 nm diameter columns with a 0001 texture were composed of bundles of sub-columns, with an average coherence length of 5 nm. Over-stoichiometric $TiB_{2.4}$ layers with excess B exhibited a complex self-organized nanostructure with ~20 nm wide columns encapsulated in excess B and oriented along 0001 consisting of a bundle of ~5 nm diameter $TiB_2$ subcolumns separated by an ultrathin B-rich tissue phase. The combination of the nanocolumnar structure combined with the B-rich tissue phase results in superhardness.

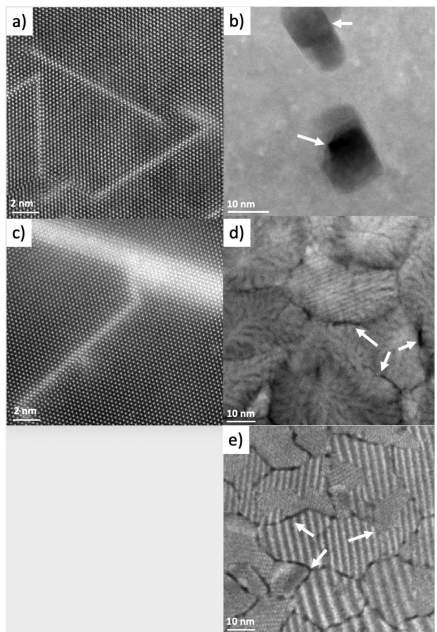

**Figure 10:** Plan-view HAADF-HRSTEM images from $CrB_{1.9}$ (a) and $CrB_{2.1}$ (b), $TiB_{1.9}$ (c) and $TiB_{2.7}$ (d), and $ZrB_2$ (e) films showing metal-rich stacking faults for the under-stoichiometric condition in (a) and (c), and B-rich inclusions (dark regions within grains or as tissue phase marked by arrows) for the over or close to stoichiometric condition in (b), (d), and (e), respectively, all images viewed along the [0001] zone axis. Images are adapted from refs. [254] [157], authors who kindly provided the original data.

A clear evidence for the line-compound nature of common $TMB_2$:s covered in this review is the behaviours of the non-matched metal in slightly understoichiometric $CrB_{1.9}$ and $TiB_{1.9}$, see Fig 10a [254] and 10b [254], respectively. For both systems, there are metal-rich stacking faults in an otherwise perfect diboride lattice. Figure 10 shows plan-view HAADF-HRSTEM images from understoichiometric $CrB_{1.9}$ (a) and overstoichiometric $CrB_{2.1}$ (b) [254], $TiB_{1.9}$ (c) and $TiB_{2.7}$ (d) [254], and $ZrB_2$ (e) [157] films showing metal-rich stacking faults for the under-stoichiometric condition in (a) and (c), and B-rich inclusions (dark regions within grains or as tissue phase marked by arrows) for the over or close to stoichiometric condition in (b), (d), and (e), respectively, all images viewed along the [0001] zone axis. Generally, deposition of $CrB_x$ films yields the same composition as a stoichiometric $CrB_2$ sputtering target, while $TiB_x$ films generated under the same conditions with $TiB_2$ results in overstoichiometric films. For overstoichiometric $CrB_x$ films, large B-rich inclusions are observed as in Fig. 10b. This is in contrast to overstoichiometric $TiB_x$ films, where a B-rich tissue phase is formed as in Fig. 10d. The B-deficiency in understoichiometric $CrB_x$ in Fig. 10a shows up as Cr-rich stacking faults in a similar way as for understoichiometric $TiB_x$ films in Fig. 10c. Generally, the understoichiometric films showed significantly better oxidation resistance compared to the overstoichiometric films with linear oxidation rates that depend on the B content.

Characterization of the mechanical properties shows that $TiB_{2.4}$ films have a high hardness of 40-50 GPa. The hardness of 0001-oriented $TMB_2$ films is evaluated in section 5.3. The microstructure model determined for $TiB_{2.4}$ films [253] has been supported for $ZrB_2$ films using ATP by Engberg *et al.* [79]. Whenever there is an excess of B, as in an investigated $ZrB_{2.5}$ film, the grain boundaries form a B-rich more or less continuous boundary network around columns consisting of a single grain or a few grains (see typical example in Figure 10d). Even for a stoichiometric $ZrB_{2.0}$ film, APT shows grain boundaries covered by a thin layer of 1–2 monolayers or incomplete < 1 monolayer B tissue phase in the stich-like semi-coherent grain boundaries (compare with Fig. 10e). This supports the notion of a strong segregation tendency for B, even at deposition temperatures as low as 300 °C [44] in PVD.

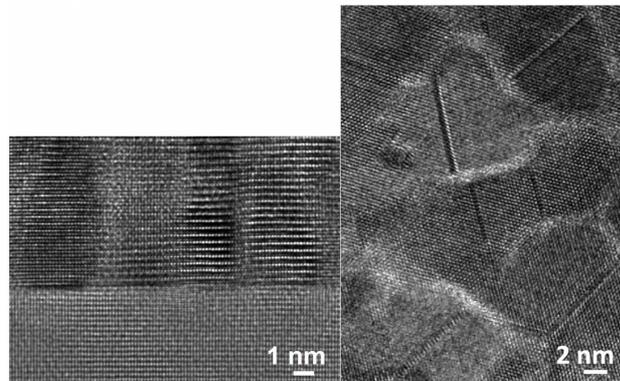

**Figure 11:** HR-TEM images of a $ZrB_2$ thin film grown on a 4H-SiC(0001) substrate. Cross-section (left panel, adapted from ref. [156]) and plane-view (right panel, author's work, unpublished).

For growth on lattice matched substrates and in the temperature region 550-900 °C, Tengdelius *et al.* [157] reported with support from pole figure measurements that epitaxial growth of $ZrB_2$ was possible on 4H-SiC (0001) above 820 °C. The crystal quality (epitaxial fraction) increased at 900 °C *i.e.*, at the highest substrate temperature studied, while lower temperature resulted in 0001-oriented films. The contribution of 0001-oriented minority domains in the film seen from decreasing intensities of the $ZrB_2$ $10\bar{1}1$ peaks in the θ/2θ diffractograms decreased with increasing substrate temperatures. From these studies, it is evident that the crystal quality of sputter-deposited $TMB_2$ films is improved by higher substrate temperatures and where Tengdelius *et al.* [157] demonstrated epitaxial growth of $ZrB_2$ on 4H-SiC(0001) substates.

The TEM images in Figure 11 shows a cross-sectional HR-TEM image of a $ZrB_2$ thin film grown on a 4H-SiC(0001) substrate right panel and a plan-view image left panel. From the images it can be concluded that the $ZrB_2$ films grow as epitaxial columns a few to some tens of nm wide parallel to the substrate that extend all the way to the film-vacuum interface orthogonal to the substrate.

More recent studies have demonstrated epitaxial growth of other $TMB_2$ materials systems than $ZrB_2$ seen from $TiB_2$ [255] and $CrB_2$ [180] on $Al_2O_3$(0001) at 900 °C, *i.e.*, similar substrate temperature previously applied by Tengdelius *et al.* for epitaxial growth on Si(111) [156], 4H-SiC(0001) [156] [157], $Al_2O_3$(0001) [158] and Schnitter *et al.* for growth on Si(001) [159]. The orientation also affects the mechanical properties to be discussed in section 5.3 and that fine-grained coatings lead to more scattering points for the phonons, which affect the resistivity, see section 5.4.

Concerning the microstructure development in thin films, the surface-mobility during nucleation and growth is important. In 1977, Bunshah *et al.* [165] reported lattice parameters for deposited $TiB_2$ and $ZrB_2$ coatings. The general trend for both $TiB_2$ and $ZrB_2$ was an increasing length for the *c* axis with increasing growth temperature, while the length of the *a* axis remained relatively unchanged and close to that determined for $TiB_2$ with 3.03 Å and 3.17 Å for $ZrB_2$, see. Table II. The $TiB_2$ coatings were predominantly 0001-oriented, while 0001- and $10\bar{1}0$-oriented $ZrB_2$ coatings were deposited with 0001 orientation at a deposition rate of 2.14 μm/min and 670 °C to 850 °C and $10\bar{1}0$ orientation at a deposition rate of 0.113 μm/min and 600 °C to 1300 °C.

The observations made by Bunshah *et al.* [164] [165] on film orientation at different deposition rates and temperatures are valuable when discussing the results obtained from sputtered deposited films. These studies showed that thinner films (~400 nm) are 0001-oriented, but with a more pronounced $10\bar{1}0$ peak compared to thicker films. On the other hand, Tengdelius *et al.* [157] showed that the $20\bar{2}0$ and $30\bar{3}0$ peaks are also visible, while diffractograms were not shown from Bunshah *et al.* However, the general results that the 0001-oriented growth gradually change to a $10\bar{1}0$-orientation with growth thickness finds support from Tengdelius *et al.* [157]. Thus, the growth parameters can be adjusted for a specific film growth orientation.

X-ray diffraction data from diborides are typically of low intensity and with quite broad peaks due to fine-grained microstructures [44]. Films deposited at low temperatures are columnar and oriented 0001, *i.e.*, the low-energy surface [256]. Higher deposition temperatures, however contra-intuitively, favor a more random or a $10\bar{1}1$ preferred orientation. Epitaxial growth relations are obtained at particular substrates for surfaces with higher Miller indices. This agrees with the microstructure evolution during growth of the parent hexagonal close-packed metals Ti and Zr [257]. Thus, at elevated temperature epitaxial growth is possible and inspires hope to eliminate hydroxides, from the residual gas of typically used vacuum systems, as recently demonstrated for $CrB_2$ by Dorri *et al.* [180].

To summarize, higher substrate temperature leads to crystallization in the films due to increased ad-atomic mobility, especially in ion-assisted PVD and desorption of contamination from the target and the residual rest gas in the vacuum system, where hydroxides on the film surface prevent desired nucleation and growth of any defined diboride. Growth at room temperature generally produces a more fine-grained structure with a preference of a 0001-orientation as 0001 surface is the low energy surface [257]. At elevated temperatures by applied substrate heating (500-900 °C), the films become $10\bar{1}1$ or 0001 oriented with epitaxial columns or epitaxial growth. At even higher temperatures (>1000 °C), the films grow with a $10\bar{1}0$ orientation, similar to e-beam deposited films, [164] [165] as discussed in sections 4.2-4.3.

## 5.3   Mechanical properties

We start by comparing the hardness values of bulk Group 4-6 $TMB_2$ to those of the corresponding TMC and TMN. For Group 4-6 $TMB_2$ films, we focus our attention to $TiB_2$, as this is the most investigated material system and to hardness and measurements performed by nanoindentation as well as the fracture toughness of Group 4 $TMB_2$.

Table 4 shows the bulk HV values for Group 4-6 $TMB_2$, TMC, and TMN, where the $TMB_2$ phases are predominantly of the C32 crystal structure, while the TMC and TMN tend to exhibit B1, but with exceptions for Group 6. Comparing the HV of the Group 4 $TMB_2$ gives for $TiB_2$, $ZrB_2$, and $HfB_2$ values of 15-45 GPa, 22.5-23, and 28 GPa, respectively. For $TiB_2$, we note the large spread in hardness from 15-45 GPa, but where a typical hardness value reported for $TiB_2$ is 25 GPa [253]. When the Group 4 $TMB_2$ bulk materials are compared to the corresponding TMC and TMN, it is clear that their hardnesses are generally lower than for the TMC, seen from 28-35 GPa for TiC, 25.9 GPa for ZrC, and with 26.1 GPa for HfC, but higher than those of the TMN with 18-21 GPa for TiN, ZrN with 15.8 GPa, and 16.3 GPa for HfN. This can be explained by the differences in chemical bonding previously discussed from the differences in the electronegativity values Δχ between the electron donor (TM) and the electron acceptor B, C or N and from theoretical calculations of the band structure in section 3.1. Generally, carbides have more directional and highly covalent bonds compared to the more metallic-like bonds in borides and the more ionic-like bonding in nitrides.

For the Group 4 $TMB_2$, Table 4 shows a small variation in hardness values for bulk materials between 22.5-28 GPa, but with the highest value for $HfB_2$. This is a consequence of the B-B and TM-B bond strength that determine the hardness for $TMB_2$ with C32 crystal structure [25] and where the bond energy increases for the heavier TM in Group 4 [73]. This trend in hardness is supported by DFT calculations on the B-B and TM-B bond strength for the heavier TM in each group as discussed in section 3.1.

Moving to Groups 5 and 6, the hardness values for bulk $TMB_2$ decreases and with no data available for bulk $MoB_2$ and $WB_2$ with C32 crystal structure, c.f. section 2.3. The decrease of hardness is a consequence of filling antibonding states above $E_F$ and change of bond strength as discussed in sections 3.2 and 3.3 and where the slightly higher hardness for $W_2B_5$ could be related to the different crystal structure. Excluding VC, Groups 5 and 6 $TMB_2$ show hardness values comparable or slightly higher than Groups 5 and 6 TMC. This suggests that the C32 crystal structure is less sensitive to changes in the TM elements and charge-transfer.

$TiB_2$ is the most investigated $TMB_2$ thin film material with respect to mechanical properties. In 1997, Mitterer [44] reviewed results from HV hardness measurements conducted on $TiB_2$ and $ZrB_2$ coatings deposited on Mo substrates by DCMS from compound targets and at substrate temperature of 300 °C. HV values for $TiB_2$ were between 3000 and 7000 HV, while reported values for $ZrB_2$ coatings were close to the measured bulk value of $ZrB_2$ with 2200 HV. From SEM images, Mitterer attributes the high HV values in $TiB_2$ coatings to a dense, extremely fine-columnar structure. From the x-ray diffractograms in his review [44], we note an 0001-orientation for the $TiB_2$ coatings. In 2001, Berger *et al*. [258] published results on low stress $TiB_2$ coatings deposited on cemented carbide substrates by DCMS from a $TiB_2$ compound target at a substrate temperature of ~340 °C. By varying the substrate bias, the highest hardness value (54 ±9 GPa) was achieved with positive bias and without external heating while negative bias and no external heating resulted in lower hardness values. Recently, $TiB_2$ thin films deposited by *nonreactive* dc sputtering from compound targets have been shown to have much higher hardnesses, 48 to 77 GPa, which are not simply due to correspondingly high residual stresses [44] [253]. However, the mechanism giving rise to this superhardness effects defined in Ref. [253] as $H\sim40$ GPa is not understood. While it has been speculated that the nanostructure and grain boundaries of the film may play an important role, these features have not been sufficiently investigated.

For thin films of $ZrB_2$, there are even fewer studies of the hardness. For epitaxial $ZrB_2$ thin films, Tengdelius *et al.* [158] reported hardness values of 45 GPa as explained by the 0001 orientation, compared to 25 GPa for polycrystalline films which is close to bulk [259]. The hardness of sputtered $HfB_2$ films is typically higher than reported for bulk materials, where Herr and Broszeit [260] reported hardness values between ~20 and ~60 GPa. The study found that the difference in hardness was a result of compressive stresses, whereby applied heat treatment could reduce the residual stresses in the film. The spread in hardness is wide and connected to the microstructure and amount of strain. Film hardness higher than bulk was also reported by Dub *et al.* [183] for a $HfB_{2.7}$ nanocrystalline thin film with a hardness of 44.0±2.4 GPa. Goncharov *et al* [185] deposited $HfB_2$ films, without external heating, 300 and 500 °C, and substrate biases of ground potential, -25, -50, and +50 V. The measured hardness values varied between 44.0±0.8 to 12.6±2.6 GPa depending on the orientation of the films. The orientation affects the hardness, where more information can be obtained if a *glide system* of the dislocations is introduced. This makes dislocation movement on the 0001 planes unlikely due to the fact that no shear force is projected directly on the basal planes when the force is applied normal to these planes. For other diboride thin films exhibiting *superhardness*, defined by a hardness ≥ 40 GPa [261], it has been suggested that the increased hardness compared to bulk is due to the small grain size in these films, thus hindering dislocation formation and motion, in combination with a B-rich tissue phase between the grains, hindering grain boundary sliding due to strongly directional bonds [253]. $ZrB_2$ films exhibit a large grain size in the z direction and it is possible that the small grain size in the *x* and *y* directions is of less importance, given that the films are 0001 oriented and that the basal glide system for $ZrB_2$ is $\{0001\}<11\bar{2}0>$. This makes dislocation movement on the 0001 planes unlikely due to the fact that no shear force is projected directly on the basal planes when the force is applied normal to these planes.

Generally, hard films have an 0001 orientation and compressive strain increase the hardness. As an example, Sobol *et al.* [83] [262] [263] deposited W-B films at substrate temperatures of 500 and 900 °C. High-temperature vacuum annealing at 1050, 1120, and 1230 resulted in crystallization into different boride phases. For the $W_2B_5$ crystal structure, the measured hardness was 26 GPa.

To summarize, the hardness values determined for sputter deposited epitaxial films with low amounts of contaminants are typically higher than what is reported for bulk materials and show a large spread compared to bulk. This can be explained by the microstructure as with a 0001-film orientation in combination with compressive strain, supported by observations of Berger *et al* [258].

The fracture toughness, as reported by Fahrenholtz *et al.* [25] of bulk $ZrB_2$ and $HfB_2$ is relatively low, typically in the range of 3.5 - 4.5 MPa√m, at both room and elevated temperatures. By adding 30 vol% of SiC to $ZrB_2$, Chamberlain *et al.* [264] were able to increase the fracture toughness from 3.5 MPa√m of pure $ZrB_2$ to 5.3 MPa√m for the $ZrB_2$-SiC composite. For $TiB_2$-SiC composites, a fracture toughness as high as 6.2 MPa√m was obtained for composites milled with WC/Co [265]. Local maxima of fracture toughness values of 7.3 MPa√m has also been reported for $TiB_2$-$ZrB_2$ composites with 20% $ZrB_2$ [266]. For comparison, the fracture toughness for understoichiometric $TiB_{1.43}$ films were measured by nanoindentation to 4.2±0.1 MPa√m, compared to overstoichiometric $TiB_{2.70}$ films at 3.1±0.1 MPa√m [267]. Thus, it has been shown that understoichiometric films could improve mechanical properties (hardness, elastic modulus, and fracture toughness) compared to corresponding overstoichiometric films. The fracture toughness was found to significantly increase for $ZrB_x$ films alloyed with Ta that increased from 4.0 MPa√m for $ZrB_{2.4}$ to 5.2 MPa√m for $Zr_{0.7}Ta_{0.3}B_{1.5}$ [212]. The increased fracture toughness was attributed to metal-rich boundaries that prevent crack propagation, while allowing for grain boundary sliding under heavy loads. The fracture toughness of diborides can also be improved by reducing the weakening boron-rich tissue phase at the grain boundaries, by alloying and in various composites. The fracture toughness of diboride films is thus comparable to bulk, let alone that the measurement methods for natural reasons are different (nanoindentation, tensile tests, etc.). For comparison with other tough ceramics, such as fine-grained nanolaminated MAX-phases ($Ti_3SiC_2$) showed a fracture toughness of 8 MPa √m, rising to 9.5 MPa√m after 1.5 mm of crack extension [268]. Coarse-grained $Ti_3SiC_2$ samples showed even higher fracture toughness with crack growth initiation between 8.5 and 11 MPa√m peaking at 14–16 MPa√m after 2.7–4 mm of crack extension.

## 5.4 Electrical resistivity

TMB$_2$ compounds are electrically conducting ceramics. Table 4 compares the resistivity values for bulk Group 4-6 TMB$_2$ to those of carbides and nitrides. The TMB$_2$ demonstrates resistivities in the range 7 to 20 μΩ·cm, which is lower than the corresponding carbides and nitrides. This is a consequence of the higher degree of metal bonding in TMB$_2$ compared to TMC and TMN. Furthermore, for bulk TMB$_2$, the resistivity generally increases from Group 4 to Group 6 and with the lowest value encountered for ZrB$_2$ followed by HfB$_2$ and TiB$_2$, but with the Group 5 TMB$_2$ demonstrating similar resistivity values as for Group 4 TMB$_2$ and with somewhat higher value for CrB$_2$. The resistivity values for bulk Mo and W borides have been measured from bulk samples consisting of Mo$_2$B$_5$ and W$_2$B$_5$, respectively. From our previous discussion on the stability of TMB$_2$ from Hägg's rule in section 2.1 and Δχ in section 2.2, we note that preference for C32 crystal structure seems to favor low resistivity values. As discussed in section 3.3, the high conductivity of ZrB$_2$ originates from the fact that the Zr atom is the largest of the Group 4-6 transition metals. Furthermore, Table 4 shows that the resistivity values of TMB$_2$ are less scattered compared to TMC and TMN, with the lowest resistivity values for ZrN followed by TiN and TaC.

In the early 1980$^{th}$, interest arose for the electrical properties of TMB$_2$ thin films, following the publications by Nelson [269] and Nicolet. [270] In 1978, Nicolet [270] authored a review paper on thin film diffusion barriers in Si-based devices focusing on issues associated with integrated circuit technology. In his review, Nicolet compared the resistivity values of the Group 4-6 TM to those of the corresponding TMB$_2$, TMC, TMN, and silicides. Nicolet highlighted the low resistivity of the TMB$_2$ and then in particular, for TiB$_2$, ZrB$_2$, and HfB$_2$, that exhibit lower resistivity values than their parent metals Ti, Zr, and Hf. This fact was originally identified by Nelson [269] in 1969.

Consequently, in the 1980s, ZrB$_2$ and TiB$_2$ films were sputtered and investigated as potential replacement of conventional refractory metal silicides in integrated circuit gates and interconnect metallization. [175] The investigated TiB$_2$ and ZrB$_2$ films were typically, rf sputtered from compound targets and where measurements of their electrical resistivity showed values 50 times higher or more than the corresponding bulk materials, with, *e.g.*, measured values for TiB$_2$ thin films of 500 μΩ·cm. [175] For such films, characterization of their phase distribution by XRD revealed broad $10\bar{1}1$ and $10\bar{1}0$ TMB$_2$ peaks of low intensities indicating a fine-grained/amorphous microstructure. AES measurements of the rf-sputtered films showed that they were B deficient and contained up to ~6 wt.% O in ZrB$_2$ and ~12 wt.% O in TiB$_2$. The level of O in the films was explained from the properties of the applied target material with a high degree of porosity in the materials, see section 5.1. From early works, we note that the rf sputtering processes were carried out without external heating of the substrate. [175] [176] [271] Applying the experiences on low-temperature sputter deposition of TMB$_2$ films summarized by Mitterer [44] it is a reasonable assumption that TiB$_2$ and ZrB$_2$ studied as diffusion barriers exhibited a fine-grained microstructure.

To lower the resistivity, post-annealing treatment was often applied to out-diffuse impurities and vacancies from underneath the surface as well as to promote recrystallization of as-deposited films. The annealing resulted in reduction of resistivity by a factor of 10 and with the lowest value of 25 μΩ·cm for a $ZrB_2$ thin film deposited by Shappiro et al. [176], but with a higher value of 130 μΩ·cm for a $TiB_2$ thin film. The experiences made from annealing of $TiB_2$ and $ZrB_2$ films suggested that $ZrB_2$ would be the preferred candidate for continued research as potential material for applications in diffusion barriers [272] [273], Ohmic contacts [274] [275] and Schottky diodes. [276] [277] Later on, aforementioned applications benefited from improved barrier properties by using $ZrB_2$ [272] with lower O content in thin films deposited at temperatures around 600 °C. [277]

The literature shows that the resistivities of other Group 4-6 $TMB_2$:s have been studied. Similar as for $TiB_2$ and $ZrB_2$ thin films, these $TMB_2$ follow the trend with much higher values compared to the bulk resistivity and where sputter process is typically carried out without external substrate heating. In 1988, Lee et al. [181] deposited $HfB_2$ films by rf magnetron co-sputtering from Hf and B targets and rf diode sputtering from a pressed powder $HfB_2$ target on glass, Si, and fused quartz substrates without external substrate heating to investigate their resistivities, discussed in section 4.5. Films deposited from a $HfB_2$ compound target showed resistivity values of ~370 μΩ·cm and contained only ~0.4% O as determined by SIMS. Lee et al. [181] also studied the influence of B and O content for $HfB_x$ films sputtered from elemental target sources. For $HfB_x$ films with $x$ between 1 and 5, the resistivity was found to fluctuate around ~20 to ~300 μΩ·cm at a constant $x$ and with little dependence on the $x$-value. Adding O during growth resulted in increasing resistivity values from ~230 to ~2200 μΩ·cm when 10% O was added to the plasma. In 1996, Wuu et al. [278], found increasing resistivity values in the range between ~230 and ~320 μΩ·cm at for 0.1 μm thick $HfB_2$ films deposited by rf sputtering on oxidized Si wafers at substrate temperatures between 100 to 400 °C and at Ar pressures between 2 to 12 mTorr. The lowest resistivity values were measured for films deposited at 400 °C and an Ar pressure of 7 mTorr or without external heating and an Ar pressure of 2 mTorr.

In 1988, Kolawa [188], together with Nicolet applied rf sputtering of a B target covered to ~20% with narrow strips of V to deposit $VB_x$ films with compositions ranging between $x$=1.5±0.3 and $x$=2.7±0.3 by varying the forward sputtering power between 80 to 400 W. Other aspects of the process conditions and the microstructure are discussed in section 4.5. The $VB_x$ films were grown to a thickness of 100 nm, where the resistivity values increased with increasing B content (increasing forward sputtering power) from 170±10 μΩ·cm for $x$ = 1.5±0.3, 190±10 μΩ·cm for $x$ = 2.0±0.3, 280±10 μΩ·cm for $x$ = 2.3±0.3, 370±10 μΩ·cm for $x$ = 2.6±0.3 to 480±10 μΩ·cm for x = 2.7±0.3. Correspondingly, the study by Kolawa et al shows that $VB_2$ films with a B rich composition exhibit higher resistivity values than stoichiometric $VB_2$ films and where V rich films exhibit lower resistivity values compared to stoichiometric $VB_2$ films. In 1997, Martin et al. [189] investigated the resistivities of $TiB_2$, $ZrB_2$, $VB_2$, $NbB_2$, and $TaB_2$ thin films deposited by DCMS from compound sources with purities better than 0.999 with growth on glass substrates and where no information on external substrate heating was provided. As the deposited $VB_2$ films were prone to oxidation following air exposure, the authors presented no resistivity values for $VB_2$. For deposited $TiB_2$ and $ZrB_2$ films, Martin et al. presented resistivities of 348 and 208 μΩ·cm, respectively. These values are in agreement with the results by Shappirio and Finnegan for rf sputtered films deposited without external substrate heating [175] with 500 μΩ·cm for $TiB_2$ and 250 μΩ·cm for $ZrB_2$ films.

In addition, Martin et al. investigated the resistivity of $NbB_2$ films, where they reported a value 552 μΩ·cm. This value is considerably higher than the 100 μΩ·cm measured by Nedfors et al. [279] [193] in 2014 for a polycrystalline $NbB_{1.8}$ film with the composition 35 at.% Nb, 63 at.% B and < 2 at.% C deposited by DCMS from a 50 mm in diameter $NbB_2$ compound target (99.5%) with a sputtering current of 150 mA on $Al_2O_3$ substrates with a ~50 nm Nb/NbC adhesion film at 300 °C in a UHV system. A TEM micrograph showed that the $NbB_{2-x}$ film consisted of thin 5-10 nm columnar grains elongated in the growth direction. From studies on the resistivities of $NbB_2$ films, we note a fine-grained microstructure in the study by Nedfors *et al.* [193] [279] and in the study by Martin *et al.* [189] with no external substrate heating.

For sputtered $TaB_2$ films, Martin *et al.* measured an even higher resistivity of 750 μΩ·cm compared to their $NbB_2$ films. In 2003 Lin and Lee [194] deposited metal rich $TaB_x$ (*x* from ~0.9 to ~1.5) films by rf magnetron sputtering from a $TaB_2$ target sputtered on Si(001) substrates without external heating and varied the bias from 0 to -200 V to change the composition in the films. $TaB_{~1.1}$ films deposited with -75 V bias by Lin and Lee [194] yielded the lowest resistivity value of ~120 μΩ·cm, where more B-rich films deposited at lower bias voltages showed slightly increasing resistivity values to about 150 μΩ·cm for a $TaB_{~1.5}$ film, i.e. a similar trend as Kolawa *et al.* [188] for $VB_2$ films. Higher applied bias voltages than -75 V resulted in increasing resistivity values to ~560 μΩ·cm at -200 V for a $TaB_{~0.9}$ film. The XRD data provided by Lin and Lee [194] showed diffractograms with broad $TaB_2$ $10\bar{1}0$ peaks of low intensities for bias voltages ≤ 75 V, with highest intensity at -75 V, and with no peaks at bias voltages above 100 V. Recorded SAED pattern showed diffuse rings, which support the broad peaks XRD. The TEM micrographs reveal a fine-grained microstructure with decreasing grain size, increasing amorphization, at increasing substrate bias voltages. In 2006, Goncharov *et al.* [195] investigated the resistivity of nanocrystalline $TaB_2$ films with grain sizes in the range 5-30 nm deposited by rf sputtering. Details on the film process conditions and properties are also discussed in section 4.5. Goncharov *et al.* [195] found that the resistivity in the deposited $TaB_2$ films decreased with increasing grain size seen from resistivity values of $5.4 \times 10^3$ μΩ·cm at a grain size of ≈ 10 nm, $3.3 \times 10^3$ μΩ·cm at grain sizes in the region of 15-20 nm, and $1.3 \times 10^3$ μΩ·cm for 25-30 grain sizes. This is an interesting observation because of the reduction of scattering points as discussed below. The studies conducted on the resistivity $TaB_2$ films described above show that a fine-grained microstructure results in high resistivity values as best illustrated in the study conducted by Goncharov *et al.* [195] on $TaB_2$ films with different grain sizes.

In 1988, Oder *et al.* [177] [178] studied the specific contact resistance ($r_C$) for $CrB_2$, $W_2B$, and $TiB_2$ films deposited on production-grade 6H-SiC substrates by DCMS with no external heating. For 100-200 nm thick layers, they measured $r_C$ values at room temperatures of $8.2 \times 10^{-5}$ Ω·cm$^2$ for $CrB_2$ and $5.8 \times 10^{-5}$ Ω·cm$^2$ for $W_2B$, while the $r_C$ for $TiB_2$ could not be accurately determined. Annealing reduced particularly the $r_C$ value at room temperature for $CrB_2$ to $1.9 \times 10^{-5}$ Ω·cm$^2$ as explained by oxygen being removed from the film. For $W_2B$, annealing has less effect compared to $CrB_2$ while $TiB_2$ annealing showed no effect at all. In a follow up study from 1988, Oder *et al.* [178] focused on the properties of $CrB_2$ thin films deposited on 6H-SiC given their superior properties as Ohmic contacts compared to $W_2B$ and $TiB_2$. Oder *et al.* [178] lowered the $r_C$ value for $CrB_2$ thin films further by applying longer annealing time. In 2009, Kiryukhantsev-Korneev *et al.* [280] deposited $CrB_2$ thin films by DCMS on Si(001) substrates with a substrate temperature in the range 250-300 °C and a bias voltage of -250 V. For 0001-oriented films, with a grain size of 60-70 nm in the 0001 direction as determined from TEM micrographs, they measured a resistivity of ~200 μΩ·cm, *i.e.* close to the resistivity values for as-deposited $ZrB_2$ films with 250 μΩ·cm as previously reported by Shappirio and Finnegan [175].

In the studies by Malinovskis *et al.* [197] [240] on $MoB_2$ films and Sobol [83] and Jiang *et al.* [198] on $WB_2$ films, there are no results on the electrical resistivity. In 1990, Willer *et al.* [247] deposited W-rich $WB_x$ films with 27 at.% B on Si and GaAs substrates pre-heated to 200 °C by rf magnetron sputtering from a compound target with unspecified composition. Resistivity values of $WB_x$ varied in the range of 140-180 μΩ·cm with varying the substrate bias between 30 to 200 V with the highest values at substrate biases higher and lower than 140 V. For the resistivities in W rich-$WB_x$ films, a similar trend is reported by Kolawa *et al* [188] for $VB_2$ films, where V-rich films exhibit lower resistivity values compared to stoichiometric $VB_2$ films.

In the studies described above, the sputter deposited Group 4-6 $TMB_2$ thin films exhibit resistivity values that are considerably higher than the corresponding bulk materials. The higher resistivity values can be explained by the fact that the deposited films are typically characterized by; i) a high level of contaminations and/or ii) a B-rich composition, and/or iii) a fine-grained microstructure typically found at low-temperature growth conditions. These properties introduce defects in the material that create scattering points for the phonons and with increased resistivity as a result. Thus, the resistivity value of an as-deposited $TMB_2$ film is a quick and easy litmus test to initially evaluate its properties. A high O content in a $TMB_2$ film will naturally increase the resistivity value by the formation of insulating B and TM oxides. An excess amount of B in $TMB_2$ films will affect the resistivity. Engberg *et al.* [79] compared the resistivity value of two ~ 400 nm thick $ZrB_2$ films deposited by DCMS on $Al_2O_3(0001)$ substrates at 900 °C. The two films showed an average composition of $ZrB_{2.5}$ and $ZrB_{2.0}$ as determined from time-of-flight energy elastic recoil detection analysis (ToF-E ERDA) and with O contents of 0.8 and 0.6 at.%, respectively. Four-point probe measurement showed a resistivity of 390 μΩ cm for the $ZrB_{2.5}$ film compared to 160 μΩ cm for the $ZrB_{2.0}$. The difference in resistivity values was explained by the variation of the microstructure between the two films. Atom probe tomography showed that both films consisted of columnar $ZrB_2$ grains with C32 crystal structure with the presence of separate disordered B-rich boundaries between the columns in the $ZrB_{2.0}$ film. In the $ZrB_{2.5}$ film, these B-rich boundaries extended to form a continuous network around the sides of the $ZrB_2$ columns *i.e.,* a similar microstructure previously reported by Mayrhofer *et al.* [253] for $TiB_{2.4}$ layers. A columnar microstructure was determined by cross-sectional and plan-view HR-TEM images presented in Figure 11 for an epitaxial $ZrB_{2.0}$ film that shows epitaxial $ZrB_2$ columns a few to some tens of nm wide, parallel to the 4H-SiC(0001) substrate that extend to the film-vacuum interface that is orthogonal to the substrate. The high resistivity value for the epitaxial $ZrB_{2.0}$ film 160 μΩ cm can be explained by the nm-size columns in the film. In 2015, Tengdelius *et al.* [157] measured a resistivity value of ~130 μΩ·cm for an epitaxial $ZrB_2$ film with similar microstructure deposited by DCMS on a 4H-SiC(0001) substrate at 900 °C. The importance of a microstructure with a minimum amount of scattering points finds further support in the study by Goncharov *et al.* [195] as well as from the higher crystal quality of CVD films compared to sputtered films. In 2009, Bera *et al.* [155] reported a resistivity value of 20 μΩ cm, still somewhat higher than the bulk value of 7-10 μΩ cm, for a stoichiometric 36 nm thick epitaxial $ZrB_2$ film deposited on an $Al_2O_3(0001)$ substrate at 1000 °C and from precursor $Zr(BH_4)_4$. In addition, a study by Nedfors *et al.* [255] from 2020, shows promising results for growing epitaxial $TiB_2$ films deposited by DCMS from a compound target on $Al_2O_3(0001)$ substrates at 900 °C. In order to control the B/Ti ratio in the films, Nedfors *et al.* [255] studied the influence of Ar pressure during growth 5 mTorr and 20 mTorr as well as the change the internal magnetic field and the strength of the central and outer magnetic poles in the magnetron. For a film with the composition $TiB_{2.09}$ as determined by ToF-ERDA, the authors determined a resistivity value of 32 μΩ cm when an Ar pressure of 20 mTorr and a stronger applied magnetic field in the outer magnetic pole. In this way, a more balanced magnetic field was achieved that influenced how the puttered electrons were ejected and increased the electron density at the sample surface. A TEM investigation showed that $TiB_2$ films without pronounced grain boundaries could be sputter deposited, contrary to the epitaxial $ZrB_2$ films on $Al_2O_3(0001)$ grown by Tengdelius *et al.* [158]. The study by Nedfors *et al.* thus supports that $TMB_2$ films with improved microstructure *i.e.,* fewer B-rich grain boundaries results in lower resistivity values.

Finally, the phonon-coupling in the borides can be compared to the nitrides, where the conductivity models of the nitrides have been developed. [281] [282] [283] Due to the lower mass of B than N, the phonon frequency is higher in borides compared to nitrides and thus the electron-phonon coupling is stronger in the borides. [284] [285] If the energy-dependent electron-phonon coupling strength is known, all the electrical transport properties can be calculated. The electronegativity of the B-cation can be compared to N as the valence electron density in the *4d - $t_{2g}$* orbitals are controlled with different TM. Kindlund *et al.* [286] studied different stress states on the material that change the charge transfer in the material and thus influence the resistivity. For comparison, research on reactively sputtered epitaxial single crystals of ZrN and HfN films deposited on MgO(001) substrates, where ZrN has a resistivity value of 12 μΩ cm [281] comparable to 7-21 μΩ cm for bulk material. HfN films show lower resistivity of 14.2 μΩ cm [287] compared to bulk seen from 33 μΩ cm.

To accurately evaluate the resistivity properties of $TMB_2$ films and correlate those to the corresponding bulk materials requires single-crystal samples as from magnetron sputter deposition so as to reduce the number of grain boundary scattering sites.

## 5.5  Properties at elevated temperatures

$TMB_2$ belongs to a class of materials known as ultrahigh-temperature refractory ceramics. As shown at the bottom of Table 4, the melting points of borides are generally slightly higher compared to the corresponding nitrides, yet somewhat lower than for the corresponding carbides. The trend in melting points can be exemplified by Group 4 and the $TMB_2$, $TiB_2$ with 2980 ºC, $ZrB_2$ with 3040 ºC, and $HfB_2$ with 3250 ºC to compare with the TMN:s TiN with 2950 ºC, ZrN with 2980 ºC, and HfN with 3387 ºC, and the TMC:s TiC with 3067 ºC, ZrC with 3420 ºC, and HfC 3928 ºC. The trend in melting points can be explained by the difference in the degree of directionality of the chemical bond between the TM atom and the B, C or N atoms. Figure 2c and 2d shows the major symmetry directions. The TM-ligand chemical bond is strongest in carbides followed by borides and nitrides. The character of the B-TM bond type can be estimated by considering the difference in electronegativity value Δχ between the electron donating TM and the electron accepting B, C or N atoms and considering the directionality of the planar TM *d* – ligand *2p*-σ and interplanar TM *d* – ligand *2p*-π covalent bonds, as previously discussed in sections 2.2 and 3.3.

The boride melting points increase within each period as described by Post *et al.* [20] for the Group 4 $TMB_2$ $TiB_2$, $ZrB_2$, and $HfB_2$. This reflects the B-TM bond strength and where the parent metals show a similar trend with Ti 1660 ºC, Zr 1850 ºC and Hf 2230 °C [73]. The B-TM bond type can be estimated by considering the difference in electronegativity between the B atom and the TM atom, as discussed in section 2.2. Applying, the electronegativity values for B and Ti, Hf, and Zr from section 2.2, we find that Δχ is highest in $HfB_2$ (Δχ = 0.74) followed by $ZrB_2$ (Δχ = 0.71) and $TiB_2$ (Δχ = 0.50). The degree of directionality (most well defined interplanar TM-B orbitals) in the chemical bond is highest in $HfB_2$ and lowest in $TiB_2$, which reflects the melting points. In their work, Post *et al.* [20] explained the trend in melting points for $TMB_2$ by the difference in ionic radii (distance between the nucleus and the outermost electron shell), where a larger radius implies a higher melting point. The trend in melting points can thus to a first approximation be explained by the amount of covalency and directionality of the TM-B bond and the ionic radii. However, other factors such as the amount of metal and ionic bonding and the B-B separation may also influence the melting points.

Moving to the right in the periodic table, the melting points decrease with increasing Group number as exemplified by the melting points of 2980 °C for $TiB_2$, 2100 °C for $VB_2$, and 2170 °C for $CrB_2$ [25]. Applying our simplified trend model, we note that moving to the right in periodic table the elements exhibit increasing $\chi$ as observed from $\chi=1.54$ for Ti, $\chi=1.63$ for V, and $\chi=1.66$ for Cr (see section 2.2.). This also implies decreasing $\Delta\chi$ values *i.e.*, decreased directionality of the covalent bonds.

In section 5.4 it was evident that annealing of as-deposited $TMB_2$ films decreases their resistivity values [175], but with no reports on high-temperature resistivity measurements on $TMB_2$ films. For the mechanical properties (H and $E_r$ measured by nanoindentation), Broitman *et al.* in 2016 [259] found that the hardness in a ~400 nm thick epitaxial $ZrB_2$ film deposited on 4H-SiC at 900 °C decreases from 47.3±0.2 GPa in measurements performed at room temperature conditions to 32.8±7.0 GPa for measurements performed at 600 °C. This means ~44% lower hardness at 600 °C compared to room temperature conditions, but still higher than the reported value for bulk $ZrB_2$ with 22.5-23 GPa [73]. In addition, Broitman *et al.* established a similar trend with decreasing hardness values for a weakly textured $ZrB_2$ film deposited on a Si(001) substrate seen from 30.8±0.2 GPa for nanoindentation performed at room temperature conditions to 24.2±7.1 GPa for nanoindentation performed at 600 °C *i.e.*, close to that of bulk $ZrB_2$ [73]. These results are consistent with what was observed for bulk material, where Koester and Noak in 1967 [23] measured the hardness of $TiB_2$, $ZrB_2$, $HfB_2$, and $W_2B_5$ samples with 98 to 100% of the theoretical densities as a function of temperature, albeit at much higher temperatures up to 1900 °C. Koester and Noak found that the Knoop hardness of $TiB_2$ $ZrB_2$, and $HfB_2$ rapidly decreased with increasing temperature seen from ~2800 kg/mm$^2$ for $TiB_2$, ~2000 kg/mm$^2$ $ZrB_2$, and ~2500 kg/mm$^2$ at room temperature testing conditions to the range of 200 to 400 kg/mm$^2$ at temperatures in the range 1600 to 1700 °C and with the highest hardness values (hot hardness) found for $ZrB_2$. The hardness of the $W_2B_5$ sample was found to decrease from ~700 kg/mm$^2$ at 1000 °C to below ~200 kg/mm$^2$ at 1600° C. Koester and Noak concluded that although the hardness decreased at higher temperatures, the investigated $TMB_2$ films exhibited higher hot hardness values compared to investigated TMC films (NbC, TiC, $TaC_{1-x}$, $HfC_{1+x}$, $(Ta_{0.8}Hf_{0.2})C_{1+x}$, $(Ta_{0.8}Hf_{0.2})C_{1+x}$, $W_2C$, and $(W_{0.65}Cr_{0.14}Re_{0.14}Ta_{0.07})C_{1+x}$).

In 2016, Sim *et al.* [252] applied a custom-designed tensile tester positioned inside a SEM to investigate the tensile strength and the elastic modulus of a thick $ZrB_2$ freestanding film up to a temperature of 1016 K. The studied $ZrB_2$ film studied was deposited by DCMS without external heating on a Si(001)/200 nm $Si_3N_4$/200 nm $SiO_2$ structure. The $ZrB_2$ film showed an average thickness of 907±26 nm, where the freestanding $ZrB_2$ tensile specimen was later fabricated from etching of the supporting Si(001)/200 nm $Si_3N_4$/200 nm $SiO_2$ substrate structure. XRD showed that the $ZrB_2$ film was 0001-oriented. The microstructure of the $ZrB_2$ film consisted of an initial amorphous layer, followed by a crystalline layer with a fine columnar structure with an average width of the grains of 12.7±3 nm. SEM images and TEM micrographs revealed no change in microstructure following annealing to 1060 K for 3 min. At room temperature, the $ZrB_2$ film exhibited a tensile strength of 1300±79 MPa, which is significantly higher than for bulk $ZrB_2$ with 350-723 MPa, see, *e.g.*, [25] and references therein. In contrast, the elastic modulus determined for the sputtered $ZrB_2$ films was 295±9 GPa, which is lower in comparison to that of the bulk material with 489 GPa [25], as explained by the microstructure of the films and in particular the degree of porosity. At 1016 K, Sim *et al.* [252] reported a tensile strength of 950±27 MPa that is still almost a factor of two higher than the value of 568 MPa communicated by Fahrenholtz *et al.* [25] for bulk $ZrB_2$ with an elastic modulus of 295±9 GPa. The study by Sim *et al.* [252] illustrates how the microstructure ranging from fine grained to amorphous and porous is characteristic for sputtered $TMB_2$ films. However, this compromises other properties such as elastic modulus, but explains why the tensile strength remains relatively unaffected considering the temperature dependent bond strength in the films that are shorter compared to the corresponding bulk material.

In a study by Padavala *et al.* from 2018 [288], the authors indirectly investigated the high temperature chemical stability of epitaxial ~400 nm thick $ZrB_2$ films deposited on 4H-SiC(0001) substrates up to temperatures of 1200°C. The $ZrB_2$ film was applied as templates for epitaxial growth of cubic boron phosphide (BP) by CVD from the precursors $PH_3$ (phosphine) and $B_2H_6$ (diborane) diluted by $H_2$. The recorded XRD patterns displayed no change in the structural properties of the $ZrB_2$ template in the temperature range 1000 to 1200 °C. High-resolution TEM micrographs showed abrupt BP/$ZrB_2$ interfaces, indicating negligible diffusion between the cubic BP film and the $ZrB_2$ template in the investigated temperature range.

A more recent study by Souqui *et al.* [289] shows that epitaxial $ZrB_2$ films retain their integrity even at 1485 °C during CVD growth of semiconducting rhombohedral BN (r-BN). Exposure to a precursor mixture of triethylboron ($B(C_2H_5)_3$)/ammonia ($NH_3$)/$H_2$ only results in surface reactions forming a few nanometer deep inclusions of a cubic (B1) $ZrB_xC_yN_z$ phase on top of the epitaxial $ZrB_2$ film as determined by x-ray $\phi$-scans and analytical microscopy [289]. Thus, epitaxial $ZrB_2$ films exhibit a good chemical resistance during CVD growth up to 1485 °C. For comparison, $TMB_2$ films that contain grain boundaries and/or segregated boron inclusions exhibit lower stability.

# 6   Summary and Concluding Remarks

Synthesis of bulk boride compounds including $TMB_2$ was initiated during "*La belle époque*" (1871 to 1914). The materials were grown from powder(s), using arc furnace techniques initially equipped with carbon cathodes and later metal cathodes. Despite this rather primitive method, these first studies gave the research community a hint on the property envelope exhibited by $TMB_2$ seen from hardness, chemical resistivity, and metallic luster. Following World War I, bulk synthesis processing for $TMB_2$ was progressed. Analysis by diffraction could soon determine that this class of materials belongs to the hexagonal crystal system and presenting lattice parameters for $ZrB_2$ close to those now determined with high precision for the phase.

After World War II, the $AlB_2$-type crystal structure (C32) was finally assigned to the Group 4-6 $TMB_2$. Further characterization of the Group 4-6 $TMB_2$, showed that the stability of the C32 crystal structure differs with $ZrB_2$ and $HfB_2$ being the most stable phases and with decreasing stabilities moving to the right in the periodic table. The binary phase diagrams illustrate the decreasing stability of the C32 crystal structure seen from an increasing number of competing phases of different stoichiometries when moving from Group 4 to 6. The crystallographic data simplified the work with property determination for bulk $TMB_2$ that was initiated during the 1950s, a work that continues today. Starting from the 1970s the experimental research on $TMB_2$ has also been supported by theoretical calculations to determine phase stabilities as well as to determine and predict their properties.

The refractory properties with high hardness and chemical stability in combination with electrical conductivity derived for TMB$_2$ from studies suggest a multitude of applications for TMB$_2$ not solely as bulk materials, but in addition as protective coatings and electrically conducting high-temperature stable thin films. Although thin film growth of TMB$_2$ by CVD was initiated already during the 1930s and with the first sputter-deposited films during 1970s, the larger interest for thin film TMB$_2$ emerged during the 1980s, where the urge for new thin film materials in the semiconductor industry catalyzed research. The enabled low-temperature growth prioritized sputtering deposition, where the properties of TiB$_2$ and ZrB$_2$ films were studied noting the low resistivity values of these TMB$_2$. The films deposited by rf sputtering from compound targets showed a fine-grained microstructure, poor crystallinity, and with a rather high content of oxygen, as an ironic effect of residual gas (hydroxides) uptake. Such sub-standard crystallinity properties of the deposited films were initially explained by a porous target material that dissolved contaminants that were subsequently incorporated in the growing film. This review shows that sputter-deposited TMB$_2$ films have continued to suffer from these conditions even today with a few promising exceptions. These sub-standard crystallinity properties occur even despite the fact that sputter-deposition as technique has been improved compared to the 1980s seen from growth by DCMS rather than rf and with HiPIMS as a new technique and where the growth is carried out from compound targets with higher densities, less amount of pores, and lower amounts of contaminants. What remains, however, is the use of conventional vacuum systems, which perhaps are not providing clean enough deposition conditions.

From the reviewed research, the importance stands out of retaining a strict 2B:TM composition for the growth flux in sputter-deposition of TMB$_2$ films with well-defined properties. This is a consequence of the strong tendency for B to segregate to the grain boundaries at B-rich compositions for the growth flux. In sputter-deposition from compound sources the offset in 2B:TM composition at the substrate originates from the mass difference and between B and the TM that during gas-phase transport causes the heavier TM atom to scatter away from the target surface normal, while the lighter B atoms preferentially travel along the target surface normal.

The typical fine-grained microstructure in sputter deposited TMB$_2$ films affect their properties and with no clear trend with respect to the choice of TM. This review found resistivity values typically 10 times or higher in thin films compared to that of the corresponding bulk materials and can be explained by the O uptake and poor crystallinity (connected to the prior), as well as off-stoichiometry. To improve the properties, the amounts of contaminants should be minimized, the B/TM ratio be close to two and the microstructure should be well-crystallized grains. The mechanical and refractory properties are less affected by the microstructure but show a large spread when compared to bulk materials. Hardness values measured by nanoindentation range from about half of that of bulk TMB$_2$ 22 to 28 GPa to values more than double that of the bulk. Regarding the hardness values, an observation presented in the literature is that 0001-oriented films display higher hardness compared to films that are more randomly oriented. This has been explained by a microstructure containing nm-sized 0001-oriented TMB$_2$ columns separated by an ultrathin B-rich tissue phase and that plasticity from dislocation glide is less efficient for deformation directed along [0001].

The high-temperature properties of TMB$_2$ films are less investigated compared to annealing of boride materials in the bulk, where the melting points increase with atomic number. As a consequence of the difference in the chemical bonding, the melting points of borides are higher compared to the corresponding nitrides, but only slightly lower than for the corresponding carbides that makes borides suitable in high-temperature applications.

We conclude that sputter deposition of $TMB_2$ films is at a tipping point for advances. As shown in this review, avoiding B segregation during all stages of growth is paramount in sputter deposition of films with improved properties, regardless of the materials system investigated. For this, we suggest further research on the properties of applied compound targets, including alternative synthesis routes such as growth from elemental targets or reactive sputtering. Research should concern the generation of sputtered species, the gas-phase transportation of sputtered species, and the condensation of the sputtered species to the substrate followed by the nucleation, initial growth and finally the continued growth of the film. The accumulated knowledge in these respective fields could be used to advance the crystalline and physical properties of the $TMB_2$ films by reassuring a B/TM ratio close to two during all stages of growth and by eliminating contaminants. It is of significance to enable interpretation of published data by that the B and O contents are measured by high precision methods that are more accurate than EDX that is not adequate. A further challenge is to master growth of contamination-free films even at reduced substrate temperature conditions, which points to a need of water-vapor less conditions like from UHV conditions, by smart ways to capture the hydroxides before interfering with the boride crystallization and managing the high B affinity to O that may hinder the necessary B-B bonding to occur in the C32 crystal structure. This points to certain elevated substrate temperature to reduce water adsorption on the growing film.

# 7 Outlook

Advances in the research and application of diborides coatings and thin films relies - more than for many other compound classes - on improved control of elemental fluxes and ratios in the used synthesis techniques. This is because of their typical line-compound nature and aptness for phase separation. More so than for well-developed contemporary carbides and nitrides, refractory transition metal diborides have potential qualities as ultra-high-temperature ceramics for extreme thermal and chemical environments. Application examples are as high-temperature electrodes, advanced nuclear fission and fusion reactors, molten metal environment, refractory crucibles, thermocouple protection tubes in steel baths and aluminum reduction cells, reinforcement fibers, solar power, aerospace, as well as in armor applications. For these, coatings and thin films are required. However, present deposition processing and eminently sputter deposited $TMB_2$ films suffer from mainly three interrelated obstacles: i) B overstoichiometry, ii) brittleness, and iii) low oxidation resistance. An advantage of the PVD technology is that it is cleaner and dry with no hazardous materials or chemical waste such as in CVD and electroplating methods.

From the synthesis methods presented in this review, we see various routes to advance deposition processing of the concerned borides:
1) For synthesis with sputtering from a *compound target*, we call for the development of *purer* and *more dense* sputtering targets with the example of $ZrB_2$. This seems doable. At present situation, oxygen contamination from the target limits the possibility to grow epitaxial $TMB_2$ thin films while oxidation of TM and B are competing reactions at the growing film surface. Carbon contamination – also inherent to present-day target fabrication - may similarly lead to competing carbide phase formation depending on the metal of the intended diboride. For corresponding coatings, segregation of contamination to column and grain boundaries would also lead to oxidation during high-temperature application. Carbon contamination can be avoided by deposition from separate purer sources as discussed in section 6.

2) It is essential to provide a strict 2:1 ratio for the B/TM flux arriving at the substrate, since the $TMB_2$ phases are predominantly line compounds and thus prone to segregation of each excess element to the growing film surface, causing renucleation, or laterally towards grain boundaries resulting in elevating or deteriorating properties such as toughness depending on the deviation from the thickness of the B-tissue phase in the case of overstoichiometry. This gives rise to inhomogeneities in the compositions and a columnar microstructure. It is a harder nut to crack. Here, a stoichiometric target is not a sufficient remedy, as there is a different angular distribution of the two concerned sputtered species. This implies finding a radial "sweet spot" off the centre of the sample holder for depositing stoichiometric $TMB_2$ or compensatory composition means, both of which would be unpractical. To adjust a B-rich composition towards a more stoichiometric composition, one may increase the Ar pressure, increase the target-substrate distance and/or change the incidence angle to the substrate away from normal.

An interesting approach to resolve the excess of B is employing HiPIMS. In this case the B/TM ratio in $TMB_x$ films can be controlled by adjusting the HiPIMS *pulse length* while maintaining the average power and pulse frequency constant. Thus, the plasma becomes dominated by $TM^+$ ions while the relative amount of $B^+$ ions are reduced. The higher amount of $TM^+$ flux compared to $B^+$ enables reduction of the B/TM ratio due to plasma steering towards the deposition surface. Consequently, the film growth can be controlled by the ionized gas atoms in HiPIMS rather than neutral species in DCMS. Due to their lower first ionization potential and larger ionization cross section, sputter-ejected TM atoms have a higher probability of being ionized than B atoms and the B/TM ratio in the films decrease as a function of target peak current.

In the case of understoichiometric compositional deviations, recent reports point to the formation of metal-rich stacking faults in the $TMB_2$ lattice to accommodate the metal that cannot find all of its B match. As far as we have found, none of the covered reference studies where the aim was to synthesize a diboride, resulted in a monoboride to compensate for an understoichiometric condition. We infer that the diboride takes precedence and that reaction coordinates (at least for thin film deposition as covered herein) go against monoboride formation. This, we suggest, should be the subject of future studies to contest.

In the case of CVD, tailored organoborane precursors hold promise to grow TM-B in a an alternative and more industrially compatible route by applying boron as a gaseous reactant and then preferably as a single precursor containing both carbon and boron. The trialkylboron triethylboron [TEB, $B(C_2H_5)_3$] is commonly used as a boron precursor in chemical vapor deposition (CVD) of boron-based thin films since this molecule is an efficient B source at temperatures below 1000 °C but fundamental studies on CVD of B-based compounds using these molecules are scarce. In a seminal study, Lewis *et al.* [290] compared trialkylboron triethylboron [TEB, $B(C_2H_5)_3$], trimethylboron (TMB, $B(CH_3)_3$) and tributylboron [TBB, $B(C_4H_9)_3$] and suggested that TEB was the most suitable for depositing $BC_x$ films with a low carbon content by CVD. On the other hand, TMB is a gas at standard pressure and temperature conditions, while TEB is a liquid and TBB is a solid, that makes TMB the easiest trialkylboron to integrate in a sputtering process while TEB does not work for PVD.

We also anticipate that B can be compensated for by more *innovative* process developments such as *hybrid* HiPIMS/DCMS by synchronizing the DCMS bias pulse with the HiPIMS cathode. The hybrid approach enables to vary the stoichiometry and thus control the microstructure and the properties of the thin films, *e.g.*, density, hardness, adhesion, toughness, and phase formation.

There is a further interesting development in hybrid reactive sputtering by sublimating precursors such as $B_2H_6$. Although diborane has been shown to be the best choice since it is gaseous (no evaporation), it is difficult to implement in production due to its high reactivity. Also, it is a toxic gas that should be diluted before use. An opportunity to work around these limitations is to utilize higher-order boron-hydrides ($B_xH_y$). Related process development to diborane is suggested for less reactive precursors of *higher order boranes* such as $B_5H_9$ that is less toxic than diborane, but more explosive. $B_{10}H_{14}$ is also less toxic but needs to be solid-phase evaporated and is thus harder to integrate into CVD-like process conditions. Solid-phase sublimation would be possible in a process like hybrid CVD-PVD, where the TM is sputtered from a solid target and the B is added (fed) from a precursor. For this purpose, $B_2H_6$ is most reactive followed by $B_5H_9$ and $B_{10}H_{14}$. Pentaborane (9) ($B_5H_9$) has fewer B and is less reactive compared to methane ($CH_4$) that is most reactive. We anticipate that this approach can be further developed for $TMB_2$ thin films. With this route, the deposition method approaches the original CVD technique that was employed for the deposition of diborides in the $1930^{th}$. Again, the CVD process becomes *hybridized* with the PVD technique, as a way towards successful $TMB_2$:s.

For the trends of properties of the materials, we foresee that thin film materials based on TM diborides with superior hardness properties will be possible to synthesize in the future. Based on the hardness vs ductility trend in bulk materials where, *e.g.*, $ZrB_2$ is harder and more ductile than $HfB_2$ while, $WB_2$ is more ductile but less hard than $ZrB_2$, we anticipate that overcoming the present difficulties in managing the deposition process in all the aforementioned aspects opens up new possibilities for the use of $TMB_2$ as thin film refractive materials in various high-temperature applications where high toughness is important. The development of diborides depends on improved crystal and electronic structures, overcoming difficulties encountered in thin-film synthesis, including properties of the sputtering source, generation of sputtered-material precursors, transport of the sputtered material, condensation on the substrate, a low level of contaminants, and control of the B/TM ratio.

Once synthesis of stoichiometric and contamination-less $TMB_2$ is provided, research can be expanded to pseudo-binary or multinary TM diborides, which may be prepared as metastable solid solutions by virtue of recoil mixing from low-energy ionized PVD deposition in combination with kinetic limitations for adatom diffusion imposed by the low deposition temperature. Such alloys will have properties worthy of exploration. With such metastable pseudo-binary (ternary) or multinary $(TM_1,TM_2,…)B_2$ alloy compounds, also *age hardening* effects may be explored from secondary phase transformations, like by spinodal decomposition or nucleation and growth of the constituent binary $TMB_2$:s. A significant character of the $TMB_2$ – unlike the corresponding nitride, carbide, and oxides, is the potential beneficial effect of the TM and B to segregate to film column boundaries and for small (~10%) deviation from stoichiometry yield semi-coherent and strongly-bonding tissue phase, respectively, which serve to strengthen the thin film material. This has been demonstrated thus far for $ZrB_2$ and $TiB_2$, but we foresee similar effects also in other $TMB_2$:s.

A further option is to deposit films from HiPIMS in order to control the preferential ejection angles and gas-phase scattering during transport between the target and the substrate. However, this synthesis route requires fine control of the sputtering parameters in terms of varying the ion flux. Carbides and nitrides are more compliant to variations in stoichiometry or oxygen substitution. As shown in this review, the ion flux (plasma heating) should be better controlled in the boride deposition process to enable applications such as conductive and corrosion resistant materials and high-temperature conductors to gain importance. For HiPIMS, the difference in the degree of ionization between the TM ions and the less ionized boron also influence the compositions of the films, where the TMs can be controlled by applying a bias voltage to the substrate.

We also note that HiPIMS is not the only solution for adapting a strict 2:1 ratio for the B/TM stoichiometry and that defected diborides with B and metal vacancies also exhibit highly interesting materials properties. Although TM diborides are hard, it may not be sufficient to prevent failure in applications that involve high stress, as hardness is typically accompanied by brittleness. To avoid failure due to cracking in applications, thin films should have high toughness, *i.e.*, both hard and ductile. Here, we project that $TMB_2$ films can be made both hard and tough using *hybrid* co-sputtering by, *e.g.*, Cr and Zr by HiPIMS and DCMS. Furthermore, by alloying the $TMB_2$ films with Al will increase the oxidation resistance with possibly minor effect of hardness. These films would have a unique combination of high hardness, toughness, wear, oxidation, and corrosion resistance. As discussed in section 3.3, recent experimental studies inspired by superconducting $MgB_2$, suggests that $ZrB_2$ becomes superconducting upon a few percent of V or Nb doping due to changes in the lattice parameters because of induced lattice strain. Although little is presently known about this strain induced effect on the electron-phonon coupling in $TMB_2$ thin films, we anticipate that continued research may lead to higher transition temperatures in layered borides.

# 8    Acknowledgments

The authors are grateful to Professor Joseph E. Greene for inspiring us to write this review. We thank Filip Wiltgren for assistance with the reproduction of the binary phase diagrams. MM acknowledges financial support from the Swedish Energy Research (no. 43606-1) and the Carl Tryggers Foundation (CTS20:272, CTS16:303, CTS14:310). LH and HH acknowledge funding from the Swedish Government Strategic Research Area in Materials Science on Advanced Functional Materials at Linköping University (Faculty Grant SFO-Mat-LiU No. 2009-00971) and the Knut and Alice Wallenberg Foundation, Project Grant (The Boride Frontier, KAW 2015.0043). HH acknowledges the Swedish Research Council (VR), contract 621-2010-3921 and the Åforsk Foundation, Grant No. 16-430. Jun Lu is acknowledged for microscopy. Claudia Schnitter and Laurent Souqui are acknowledged for assisting with translating publications written in the German language or in the French language, respectively.